\documentclass[acmlarge]{acmart}

\AtBeginDocument{%
  }

\usepackage{caption}
\usepackage{wrapfig}
\usepackage{multirow}
\usepackage{array}
\usepackage{makecell}
\usepackage{enumitem}
\usepackage{pifont}

\usepackage{booktabs}
\usepackage{tikz}
\usetikzlibrary{external, trees}
\usepackage{ textcomp }
\usepackage[utf8]{inputenc}
\usepackage{enumitem}
\usepackage{nicefrac}
\usepackage{tabularray}

\usepackage{hyperref}
\usepackage{xcolor}
\hypersetup{
    colorlinks=true,
    linkcolor=blue,
    urlcolor=magenta,
    citecolor=cyan
}
\usepackage{epstopdf}
\usepackage{makecell}
\usepackage{threeparttable}
\usepackage{longtable}
\usepackage{multirow}
\usepackage{tabularx}
\usepackage{tabulary}
\usepackage{tablefootnote}
\usepackage{array}
\newcommand{\PreserveBackslash}[1]{\let\temp=\\#1\let\\=\temp}

\newcolumntype{C}[1]{>{\PreserveBackslash\centering}p{#1}}
\newcolumntype{R}[1]{>{\PreserveBackslash\raggedleft}p{#1}}
\newcolumntype{L}[1]{>{\PreserveBackslash\raggedright}p{#1}}
\usepackage{dsfont}
\usepackage{epstopdf}
\usepackage{subfigure}
\usepackage{graphicx}
\usepackage{caption}
\usepackage{color}
\usepackage{mathtools}
\usepackage{url}
\usepackage{bm}
\usepackage{graphicx}
\usepackage{algorithm}
\usepackage{algpseudocode}
\usepackage{amsmath}

\usepackage{amsfonts}
\usepackage{subcaption}
\usepackage{multirow}
\usepackage{bm}
\usepackage[edges]{forest}
\usepackage{fontawesome5}
\usepackage{natbib}
\definecolor{hidden-draw}{RGB}{20,68,106}
\definecolor{hidden-pink}{RGB}{255,245,247}

\definecolor{mattered}{RGB}{214, 26, 60}
\definecolor{mattegreen}{HTML}{369F39}

\newcommand{\mz}[1]{\textcolor{black}{#1}}

\newcommand{\revision}[1]{\textcolor{black}{#1}}

\setcopyright{acmcopyright}
\copyrightyear{2024}
\acmYear{2024}
\acmDOI{10.1145/3690639}
\acmArticle{Pre-Print}

\begin{document}

\title{Artificial Intelligence of Things: A Survey}

\author{Shakhrul Iman Siam}

\affiliation{
  \institution{The Ohio State University}
  \country{USA}
}
\email{siam.5@osu.edu}

\author{Hyunho Ahn}
\affiliation{
  \institution{The Ohio State University}
  \country{USA}
}
\email{ahn.377@osu.edu}

\author{Li Liu}
\affiliation{
  \institution{Michigan State University}
  \country{USA}
}
\email{liuli9@msu.edu}

\author{Samiul Alam}
\affiliation{
  \institution{The Ohio State University}
  \country{USA}
}
\email{alam.140@osu.edu}

\author{Hui Shen}
\affiliation{
  \institution{The Ohio State University}
  \country{USA}
}
\email{shen.1780@osu.edu}

\author{Zhichao Cao}
\affiliation{
  \institution{Michigan State University}
  \country{USA}
}
\email{caozc@msu.edu}

\author{Ness Shroff}
\affiliation{
  \institution{The Ohio State University}
  \country{USA}
}
\email{shroff.11@osu.edu}

\author{Bhaskar Krishnamachari}
\affiliation{
  \institution{University of Southern California}
  \country{USA}
}
\email{bkrishna@usc.edu}

\author{Mani Srivastava}
\affiliation{
  \institution{University of California, Los Angeles}
  \country{USA}
}
\email{mbs@ucla.edu}

\author{Mi Zhang}
\affiliation{
  \institution{The Ohio State University}
  \country{USA}
}
\email{mizhang.1@osu.edu}

\renewcommand{\shortauthors}{Siam et al.}

\begin{abstract}

The integration of the Internet of Things (IoT) and modern Artificial Intelligence (AI) has given rise to a new paradigm known as the Artificial Intelligence of Things (AIoT).
In this survey, we provide a systematic and comprehensive review of AIoT research. 
We examine AIoT literature related to sensing, computing, and networking \& communication, which form the three key components of AIoT. In addition to advancements in these areas, we review domain-specific AIoT systems that are designed for various important application domains.
We have also created an accompanying GitHub repository, where we compile the papers included in this survey: \href{https://github.com/AIoT-MLSys-Lab/AIoT-Survey} {https://github.com/AIoT-MLSys-Lab/AIoT-Survey}. 
This repository will be actively maintained and updated with new research as it becomes available.
As both IoT and AI become increasingly critical to our society, we believe AIoT is emerging as an essential research field at the intersection of IoT and modern AI.
We hope this survey will serve as a valuable resource for those engaged in AIoT research and act as a catalyst for future explorations to bridge gaps and drive advancements in this exciting field.
\end{abstract}


\acmJournal{TOSN}
\acmMonth{8}
\acmYear{2024}
\acmDOI{10.1145/3690639}

\maketitle
\setcopyright{acmlicensed}

\section{Introduction} 
\label{intro}

The proliferation of the Internet of Things (IoT) such as smartphones, wearables, drones, and smart speakers, as well as the gigantic amount of data they capture, have revolutionized the way we work, live, and interact with the world.
Equipped with sensing, computing, networking, and communication capabilities, these devices are able to collect, analyze and transmit a wide range of data including images, videos, audio, texts, wireless signals, physiological signals from individuals and the physical world.
In recent years, advancements in Artificial Intelligence (AI), particularly in deep learning (DL)/deep neural network (DNN), foundation models, and Generative AI, have propelled the integration of AI with IoT, making the concept of \textbf{Artificial Intelligence of Things (AIoT)} a reality.
The synergy between IoT and modern AI enhances decision making, improves human-machine interactions, and facilitates more efficient operations, making AIoT one of the most exciting and promising areas that have the potential to fundamentally transform how people perceive and interact with the world.

\begin{wrapfigure}{r}{4.7cm}
\vspace{-5mm}
\includegraphics[width=4.7cm]{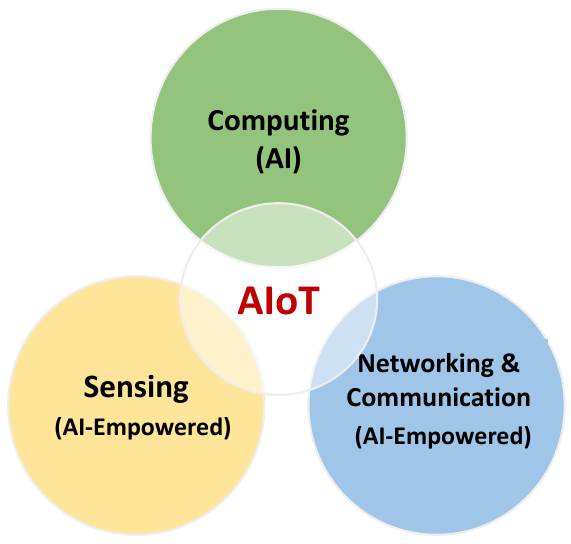} %
\vspace{-7mm}
\caption{Overview of AIoT.}
\label{fig:aiot}
\vspace{-4mm}
\end{wrapfigure}

As illustrated in Figure~\ref{fig:aiot}, at its core, AIoT is grounded on three key components: sensing, computing, and networking \& communication.
Specifically, AIoT utilizes a variety of onboard sensors such as cameras, microphones, motion, and physiological sensors to collect data from individuals and the physical world. 
The collected sensor data are processed by modern AI algorithms for a variety of tasks such as classification, localization, anomaly detection, and many others. 
Lastly, the networking \& communication component of AIoT ensures the reliable transmission of the sensor data and/or the computed outcomes to the cloud, edges or other nearby AIoT devices.
\revision{Compared to conventional IoT, the computing component of AIoT is concentrated on AI-oriented compute tasks. Moreover, the sensing and networking \& communication components of AIoT are AI empowered. It is these two key distinctions that allow AIoT to empower billions of everyday devices with breakthroughs brought by modern AI.}

Besides advancements in the three key components, domain-specific AIoT systems have been proposed and developed across a wide range of application domains.
For example, in the domain of healthcare, AIoT systems enable remote patient monitoring, facilitate  disease diagnosis on site, and act in the form of assistive technology that helps people with disabilities. 
In the domain of Augmented, Virtual, and Mixed Reality, AIoT systems enable 3D tracking to provide immersive user experiences. 
In the domain of video streaming and analytics, AIoT systems have been developed to enhance video quality and optimize video processing efficiency.
All these developed domain-specific systems demonstrate the potential of AIoT on revolutionizing a wide range of industries.

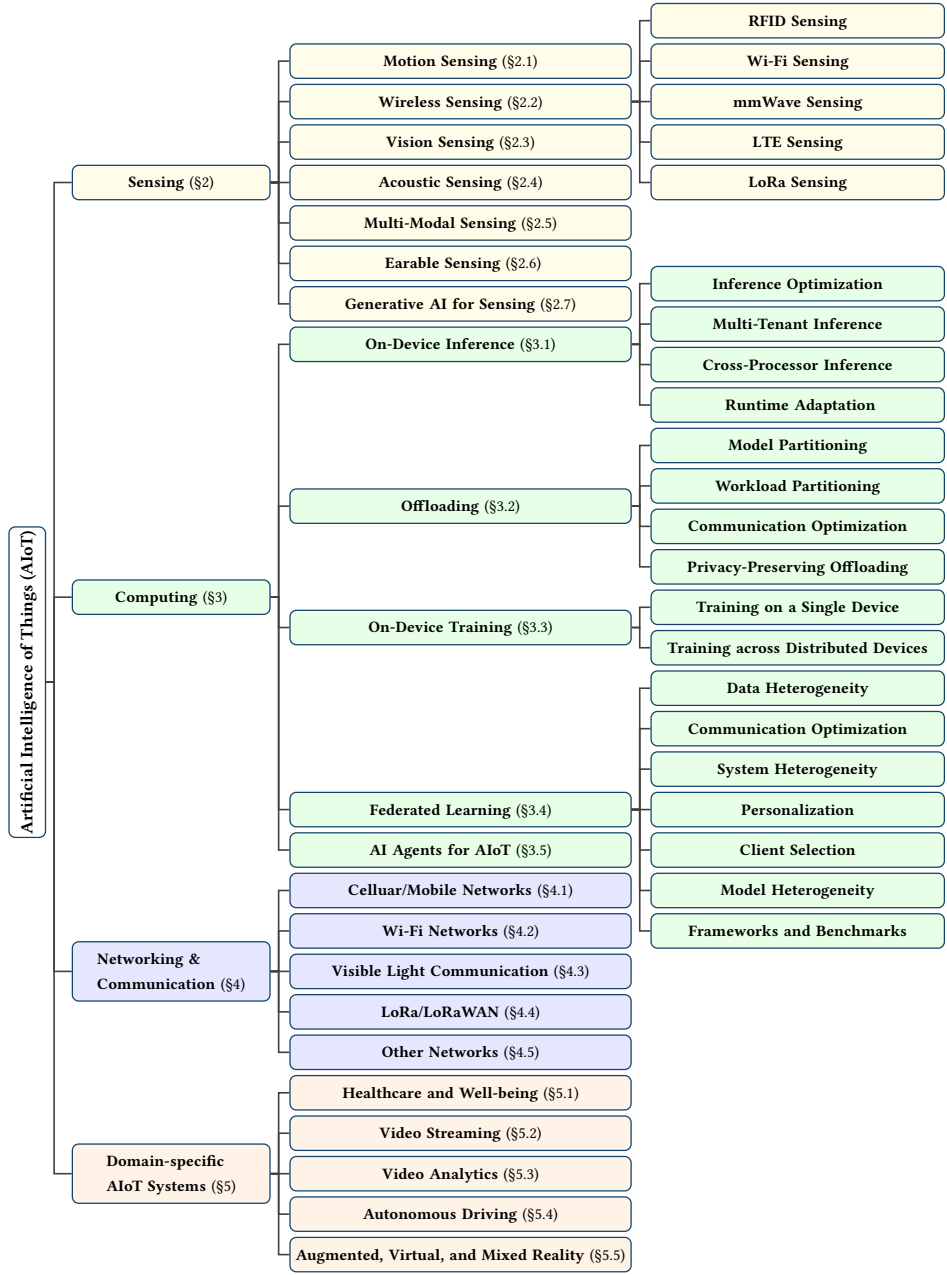
\begin{figure*}[b]
    \centering
    \resizebox{0.79\textwidth}{!}{
        \begin{forest}
            forked edges,
            for tree={
                grow=east,
                reversed=true,
                anchor=base west,
                parent anchor=east,
                child anchor=west,
                base=center,
                font=\large,
                rectangle,
                draw=hidden-draw,
                rounded corners,
                align=left,
                text centered,
                minimum width=4em,
                edge+={darkgray, line width=1pt},
                s sep=3pt,
                inner xsep=2pt,
                inner ysep=3pt,
                line width=0.8pt,
                ver/.style={rotate=90, child anchor=north, parent anchor=south, anchor=center},
            },
            where level=1{text width=12em,font=\normalsize,}{},
            where level=2{text width=21em,font=\normalsize,}{},
            where level=3{text width=18em,font=\normalsize,}{},
            where level=4{text width=16em,font=\normalsize,}{},
            where level=5{text width=17em,font=\normalsize,}{},
            [
                \textbf{Artificial Intelligence of Things (AIoT)}, ver
                [
                    \textbf{Sensing}  (\S \ref{sensing}), fill=yellow!10
                    [
                        \textbf{Motion Sensing}  (\S \ref{subsec_motion_sensing}), fill=yellow!10
                    ]
                    [
                        \textbf{Wireless Sensing}  (\S \ref{wireless_sensing}), fill=yellow!10
                        [
                            \textbf{RFID Sensing}, fill=yellow!10
                        ]
                        [
                            \textbf{Wi-Fi Sensing}, fill=yellow!10
                        ]
                        [
                            \textbf{mmWave Sensing}, fill=yellow!10
                        ]
                        [
                            \textbf{LTE Sensing}, fill=yellow!10
                        ]
                        [
                            \textbf{LoRa Sensing}, fill=yellow!10
                        ]
                    ]
                    [
                        \textbf{Vision Sensing}  (\S \ref{subsec_vision_sensing}), fill=yellow!10
                    ]
                    [
                        \textbf{Acoustic Sensing}  (\S \ref{subsec_acoustic_sensing}), fill=yellow!10
                    ]
                    [
                        \textbf{Multi-Modal Sensing}  (\S \ref{subsec_multimodal_sensing}), fill=yellow!10
                    ]
                    [
                        \textbf{ Earable Sensing}  (\S \ref{subsec_earables}), fill=yellow!10
                    ]
                    [
                        \textbf{Generative AI for Sensing}  (\S \ref{subsec_genai_sensing}), fill=yellow!10
                    ]
                ]
                [
                    \textbf{Computing}  (\S \ref{computing}), fill=green!10
                    [
                        \textbf{On-Device Inference} (\S \ref{subsec_ondevice_inference}),  fill=green!10
                        [
                        \textbf{Inference Optimization},  fill=green!10
                        ]
                        [
                        \textbf{Multi-Tenant Inference},  fill=green!10
                        ]
                        [
                        \textbf{Cross-Processor Inference},  fill=green!10
                        ]
                        [
                        \textbf{ Runtime Adaptation},  fill=green!10
                        ]
                    ]
                    [
                        \textbf{Offloading} (\S \ref{subsec_offloading}),  fill=green!10
                        [
                        \textbf{Model Partitioning},  fill=green!10
                        ]
                        [
                        \textbf{Workload Partitioning},  fill=green!10
                        ]
                        [
                        \textbf{Communication Optimization},  fill=green!10
                        ]
                        [
                        \textbf{Privacy-Preserving Offloading},  fill=green!10
                        ]
                    ]
                    [
                        \textbf{On-Device Training} (\S \ref{subsec_ondevice_training}),  fill=green!10
                        [
                        \textbf{Training on a Single Device} ,  fill=green!10
                        ]
                        [
                        \textbf{Training across Distributed Devices} ,  fill=green!10
                        ]
                    ]
                    [
                        \textbf{Federated Learning} (\S \ref{subsec_federated_learning}),  fill=green!10
                        [
                        \textbf{Data Heterogeneity},  fill=green!10
                        ]
                        [
                        \textbf{Communication Optimization},  fill=green!10
                        ]
                        [
                        \textbf{System Heterogeneity},  fill=green!10
                        ]
                        [
                        \textbf{Personalization},  fill=green!10
                        ]
                        [
                        \textbf{Client Selection},  fill=green!10
                        ]
                        [
                        \textbf{Model Heterogeneity},  fill=green!10
                        ]
                        [
                        \textbf{Frameworks and Benchmarks},  fill=green!10
                        ]
                    ]
                    [
                        \textbf{AI Agents for AIoT}  (\S \ref{subsec_genai_computing}), fill=green!10
                    ]
                ]
                [
                    \textbf{Networking \&} \\ \textbf{Communication} (\S \ref{4_Networking_and_Communication}), fill=blue!10
                    [
                        \textbf{Celluar/Mobile Networks} (\S \ref{communication-sub-cellular}),  fill=blue!10
                    ]
                    [
                        \textbf{Wi-Fi Networks} (\S \ref{subsec_wifi_networks}),  fill=blue!10
                    ]
                    [
                        \textbf{Visible Light Communication} (\S \ref{subsec_visible_light_communication}),  fill=blue!10
                    ]
                    [
                        \textbf{LoRa/LoRaWAN} (\S \ref{subsec_lora}),  fill=blue!10
                    ]
                    [
                        \textbf{Other Networks} (\S \ref{subsec_other_networks}),  fill=blue!10
                    ]
                ]
                [
                    \textbf{Domain-specific} \\ \textbf{AIoT Systems}  (\S \ref{systems}), fill=orange!10
                    [
                        \textbf{Healthcare and Well-being} (\S \ref{subsec_health}),  fill=orange!10
                    ]
                    [
                        \textbf{Video Streaming} (\S \ref{subsec_video_streaming}),  fill=orange!10
                    ]
                    [
                        \textbf{Video Analytics} (\S \ref{subsec_video_analytics}),  fill=orange!10
                    ]
                    [
                        \textbf{Autonomous Driving} (\S \ref{subsec_autonomous_driving}),  fill=orange!10
                    ]
                    [
                        \textbf{Augmented, Virtual, and Mixed Reality} (\S \ref{subsec_ar}),  fill=orange!10
                    ]
                ]
            ]
        \end{forest}
    }
    \vspace{2mm}
    \caption{Taxonomy of Artificial Intelligence of Things (AIoT) literature.}
    \label{fig:taxonomy}
    \vspace{2mm}
\end{figure*}

The overarching goal of this survey is to provide a systematic and comprehensive review of AIoT research.
As shown in Figure~\ref{fig:taxonomy}, we organize the literature of AIoT in a taxonomy consisting of four main categories: \textbf{sensing}, \textbf{computing}, \textbf{networking \& communication}, and \textbf{domain-specific AIoT systems}. 
Specifically,
\vspace{-0mm}

\tikzstyle{my-box}=[
    rectangle,
    draw=hidden-draw,
    rounded corners,
    text opacity=1,
    minimum height=1.5em,
    minimum width=5em,
    inner sep=2pt,
    align=center,
    fill opacity=.5,
    line width=0.8pt,
]
\tikzstyle{leaf}=[my-box, minimum height=1.5em,
    fill=hidden-pink!80, text=black, align=left,font=\normalsize,
    inner xsep=2pt,
    inner ysep=4pt,
    line width=0.8pt,
]

\begin{itemize}
\item \textbf{Sensing:}  
Sensing serves as the foundation of AIoT. In \S\ref{sensing}, we survey \revision{AI-empowered sensing mechanisms and techniques} in AIoT that cover research directions related to motion sensing, wireless sensing, vision sensing, acoustic sensing, multi-modal sensing, earable sensing, and \revision{Generative AI for sensing}. 

\vspace{0.5mm}
\item \textbf{Computing:}
Computing is the brain of AIoT. In \S\ref{computing}, we survey fundamental compute tasks that lie at the core of AIoT, covering topics related to on-device inference, offloading, on-device training, federated learning, and \revision{AI agents for AIoT}. 

\vspace{0.5mm}
\item \textbf{Networking \& Communication}: 
Networking and communication serve as the backbone of AIoT. In \S\ref{4_Networking_and_Communication}, we survey \revision{AI-empowered networking and communication techniques} related to a variety of networks including cellular/mobile networks, Wi-Fi networks, visible light communication, and LoRa/LoRaWAN. 

\vspace{0.5mm}
\item \textbf{Domain-specific AIoT Systems}: 
The advancements in sensing, computing, networking \& communication lay the foundation for the development of AIoT systems designed for specific application domains. In \S\ref{systems}, we survey these AIoT systems in important application domains including healthcare and well-being, video streaming and analytics, autonomous driving, as well as augmented, virtual, and mixed reality. 

\end{itemize}
\vspace{-0mm}

We have established a \textbf{GitHub repository} to organize the papers featured in the survey at \href{https://github.com/AIoT-MLSys-Lab/AIoT-Survey} {https://github.com/AIoT-MLSys-Lab/AIoT-Survey}. We will actively maintain the repository and incorporate new research as it emerges.

Although there are several surveys on topics relevant to AIoT~\cite{mohammadi_deep_2018,zhang_deep_2019,shi_communication-efficient_2020,chang_survey_2021,zhang_empowering_2021,cai_enable_2022,he_collaborative_2022,hou_trends_2023,liu_enabling_2023}, they focus on some specific aspects of AIoT. 
In contrast, this survey provides a holistic view of AIoT research. More importantly, \revision{\textit{we primarily focus on literature on sensing, computing, networking \& communication, and domain-specific AIoT systems that are built upon modern AI techniques such as DL, foundation models, and Generative AI.}}
We hope this survey along with the GitHub repository could serve as valuable resources to help researchers and practitioners gain a comprehensive understanding of AIoT research and inspire them to contribute to this important and exciting field.

\section{Sensing} 
\label{sensing}

\tikzstyle{my-box}=[
    rectangle,
    draw=hidden-draw,
    rounded corners,
    text opacity=1,
    minimum height=1.5em,
    minimum width=5em,
    inner sep=2pt,
    align=center,
    fill opacity=.5,
    line width=0.8pt,
]
\tikzstyle{leaf}=[my-box, minimum height=1.5em,
    fill=hidden-pink!80, text=black, align=left,font=\normalsize,
    inner xsep=2pt,
    inner ysep=4pt,
    line width=0.8pt,
]

\begin{figure*}[t!]
    \centering
    \resizebox{\textwidth}{!}{
        \begin{forest}
            forked edges,
            for tree={
                grow=east,
                reversed=true,
                anchor=base west,
                parent anchor=east,
                child anchor=west,
                base=center,
                font=\large,
                rectangle,
                draw=hidden-draw,
                rounded corners,
                align=left,
                text centered,
                minimum width=4em,
                edge+={darkgray, line width=1pt},
                s sep=3pt,
                inner xsep=2pt,
                inner ysep=3pt,
                line width=0.8pt,
                ver/.style={rotate=90, child anchor=north, parent anchor=south, anchor=center},
            },
            where level=1{text width=14em,font=\normalsize,}{},
            where level=2{text width=17em,font=\normalsize,}{},
            where level=3{text width=17em,font=\normalsize,}{},
            [
                \textbf{Sensing}, ver
                        [
                            \textbf{Motion Sensing}, fill=yellow!10
                            [
                            \textbf{Human Activity Recognition}, fill=yellow!10
                            [
                            Lasagana~\cite{Lasagana}{,}
                            ~\citet{deep-generative-domain}{,}
                            SenseHAR~\cite{SenseHAR}{,}
                            LIMU-BERT ~ \cite{Limu-bert}
                            , leaf, text width=31em
                            ]
                            ]
                            [
                            \textbf{Arm Tracking}, fill=yellow!10
                            [
                            ArmTroi~\cite{armtroi}{, }RTAT~\cite{RTAT},leaf, text width=11.5em
                            ]
                            ]
                        ]
                        [
                           \textbf{Wireless Sensing}, fill=yellow!10   
                           [\textbf{RFID Sensing}, fill=yellow!10
                           ]
                           [\textbf{Wi-Fi Sensing}, fill=yellow!10]
                           [\textbf{mmWave Sensing}, fill=yellow!10]
                           [\textbf{LTE Sensing}, fill=yellow!10]
                           [\textbf{LoRa Sensing}, fill=yellow!10]
                        ]
                        [
                            \textbf{Vision Sensing}, fill=yellow!10
                            [
                            \textbf{Human Activity Recognition}, fill=yellow!10
                            [
                            Mosaic~\cite{shim2023mosaic},leaf, text width=5.5em
                            ]
                            ]
                            [
                            \textbf{Image Enhancement}, fill=yellow!10
                            [
                            MobiSR~\cite{lee2019mobisr}{,}
                            Starfish~\cite{hu2020starfish}{,}
                            MicroDeblur~\cite{lee2023microdeblur}
                            ,leaf, text width=20em
                            ]
                            ]
                            [
                            \textbf{Object Detection}, fill=yellow!10
                            [
                            EagleEye~\cite{yi2020eagleeye}{,}
                            LAPD~\cite{sami2021lapd}{,}
                            LiquidHash~\cite{sun2022detecting}{,}\\
                            Mozart~\cite{xie2023mozart}{,}
                            UltraDepth~\cite{xie2021ultradepth}{,}
                            ODDS~\cite{mithun2018odds}, leaf, text width=20em
                            ]
                            ] 
                            [
                            \textbf{Eye Tracking}, fill=yellow!10
                            [
                            EMO~\cite{wu2020emo}{,}
                            GazeGraph~\cite{lan2020gazegraph}{,}
                            ASGaze~\cite{10.1145/3560905.3568544},leaf, text width=20em
                            ]
                            ]
                            [
                            \textbf{Pose Estimation}, fill=yellow!10
                            [
                            MobiPose~\cite{zhang2020mobipose}{,}
                            RoFin~\cite{zhang2023rofin},leaf, text width=12em
                            ]
                            ]
                        ]
                        [
                        \textbf{Acoustic Sensing}, fill=yellow!10
                                 [
                                 \textbf{Localization}, fill=yellow!10
                                 [
                                 DeepRange~\cite{DeepRange2020}{,}
                                 Owlet~\cite{garg2021owlet}{,}  DeepEar~\cite{DeepEar2024}, leaf, text width=21em
                                 ]
                                 ]
                                 [
                            \textbf{Movement Tracking}, fill=yellow!10
                            [
                            VSkin~\cite{sun2018vskin}{,}
                            \citet{mao2019rnn}{,}
                            ~Ipanel~\cite{yourtable2019}{,}
                            BreathListener~\cite{xu2019breathlistener}{,}\\
                            FM-Track~\cite{li2020fm}{,} 
                            SVoice~\cite{fu2022svoice}{,}
                            Experience~\cite{li2022experience},leaf, text width=27em
                            ]
                            ]
                            [
                            \textbf{Emotion Recognition}, fill=yellow!10
                            [
                            DeepEar~\cite{lane2015deepear}{,}
                            \citet{georgiev2017low}{,}
                            Mic2Mic~\cite{mathur2019mic2mic}
                            ,leaf, text width=21.5em
                            ]
                            ]
                            [
                            \textbf{Keyword and Event Detection}, fill=yellow!10
                            [
                            \citet{min2019closer}{,} 
                            SoundSieve~\cite{monjur2023soundsieve},leaf, text width=15em
                            ]
                            ]
                        ]
                        [
                        \textbf{Multi-Modal Sensing}, fill=yellow!10
                                 [
                            \textbf{Human Activity Recognition}, fill=yellow!10
                            [
                            \citet{radu2018multimodal}{,}
                            \citet{li2021low}{,}
                            \citet{leite2021optimal}{,} \\
                            VMA~\cite{hu2022vma}{,} 
                            Cosmo~\cite{ouyang2022cosmo}{,}
                            CMA~\cite{zhang2023cma},leaf, text width=22.5em
                            ]
                            ]
                                 [
                            \textbf{Human and Object Identification}, fill=yellow!10
                            [
                            XModal-ID~\cite{korany2019xmodal}{,} 
                            RF-Camera~\cite{liu2021rfid}{,}
                            Capricorn~\cite{wang2022capricorn}{,}\\
                            Vi-Fi~\cite{liu2022vi}{,}
                            RFVibe~\cite{shanbhag2023contactless},leaf, text width=23em
                            ]
                            ]
                                 [
                            \textbf{Tracking}, fill=yellow!10
                            [
                            milliEgo~\cite{lu2020milliego}{,}
                            ImmTrack~\cite{dai2023interpersonal}
                            ,leaf, text width=14em
                            ]
                            ]
                            [
                            \textbf{Localization}, fill=yellow!10
                            [
                            RFusion~\cite{boroushaki2021rfusion}{,}
                            ELF-SLAM~\cite{luo2022indoor},leaf, text width=15em
                            ]
                            ]
                            [
                            \textbf{Speech Enhancement}, fill=yellow!10
                            [
                            UltraSE~\cite{sun2021ultrase}{,}
                            Wavoice~\cite{liu2021wavoice}{,}
                            VibVoice~\cite{he2023towards},leaf, text width=19em
                            ]
                            ]    
                        ]
                        [
                            \textbf{Earable Sensing}, fill=yellow!10
                            [
                            \textbf{Facial Expression Sensing}, fill=yellow!10
                            [
                            BioFace-3D~\cite{wu2021bioface}{,}
                            FaceListener~\cite{song2022facelistener},leaf, text width=16em
                            ]
                            ]
                            [
                            \textbf{User Authentication} \\ , fill=yellow!10
                            [
                            EarGate~\cite{ferlini2021eargate}{,}
                            MandiPass~\cite{MandiPass2021}, leaf, text width=13em
                            ]
                            ]
                            [
                            \textbf{Sound Localization}, fill=yellow!10
                            [
                            ClearBuds~\cite{chatterjee2022clearbuds}{,}
                            ~\citet{zandi2022individualizing}{,}
                            ~DeepEar~\cite{DeepEar2024},leaf, text width=21em
                            ]
                            ]
                        ]
                        [
                            \textbf{Generative AI for Sensing}, fill=yellow!10
                            [
                            LLMSense~\cite{ouyang2024llmsense}{,}
                            Penetrative AI~\cite{Xu2024Penetrative}{,}
                            MEIT~\cite{Wan2024Mar}%
                            ,leaf, text width=21.5em
                            ]
                            ]
                    ]
        \end{forest}
}
    \caption{Summary of topics related to sensing.}
    \label{fig:sensing-tree}
\end{figure*}
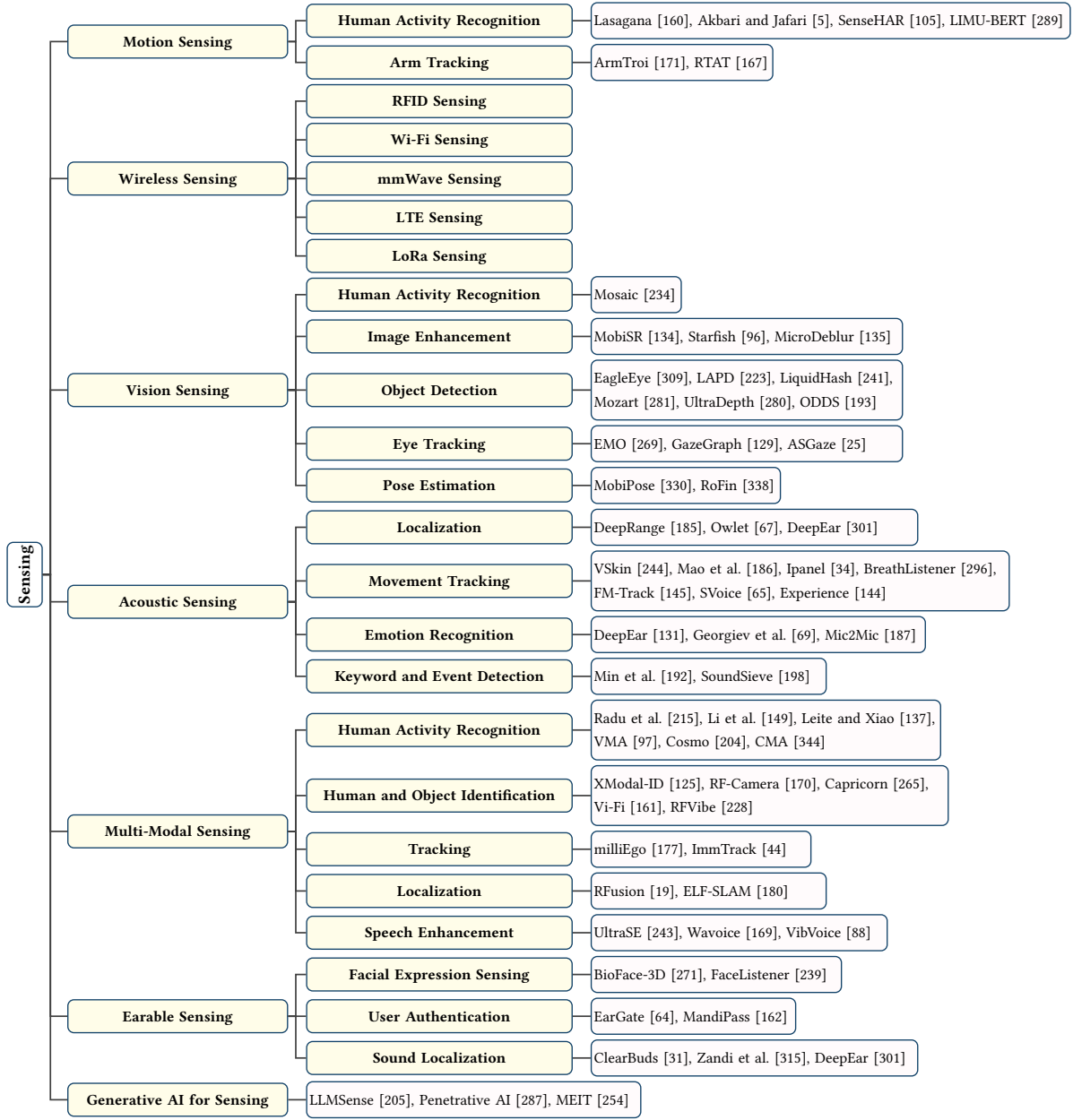

\begin{figure}[t]
\centering
\includegraphics[width=0.65\linewidth]{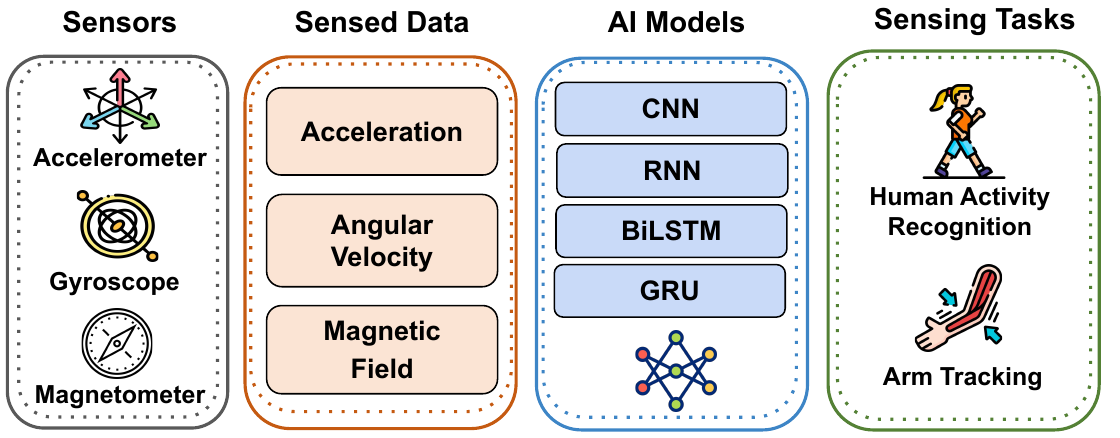}
\vspace{-2mm}
\caption{Illustration of AI-empowered motion sensing pipeline.}
\label{fig: Motion Sensing Pipeline}
\vspace{-3mm}
\end{figure}

\begin{table}[t]
\centering
\resizebox{\textwidth}{!}{
\begin{tabular}{| >{\centering\arraybackslash}m{2.8cm} | >{\centering\arraybackslash}m{1.7cm} | >{\centering\arraybackslash}m{2.3cm} | >{\centering\arraybackslash}m{1.5cm} | >{\centering\arraybackslash}m{1.6cm} | >{\centering\arraybackslash}m{3.5cm} | >{\centering\arraybackslash}m{3.7cm} |}
        \hline
        \textbf{Sensing Type} & \textbf{Contact Type} & \textbf{Computation Requirement} & \textbf{Privacy-Intrusive} & \textbf{Range} & \textbf{Advantages} & \textbf{Disadvantages}  \\
        \hline
        \textbf{Motion Sensing} & Contact & Lightweight & Low & Short & Low power, cost-effective & Require body contact, sensitive to location, limited information  \\
        \hline
        \textbf{Wireless Sensing} & Contactless & Heavy & Low & Medium to Long & Can penetrate walls, large coverage area & High computational cost, signal interference from other devices  \\
        \hline
        \textbf{Vision Sensing} & Contactless & Heavy & High & Medium to Long & Rich information, versatile applications & Privacy concerns, high computational cost, sensitive to lighting conditions and occlusions  \\
        \hline
        \textbf{Acoustic Sensing} & Contactless & Lightweight & High & Short to Medium & Rich information, versatile applications & Privacy concerns, affected by background noises  \\
        \hline
        \textbf{Multi-Modal Sensing} & Both & Heavy & Variable & Variable & Combine strengths of multiple sensors, robust & Complex integration, high computational cost  \\
        \hline
       \textbf{Earable Sensing} & Contact & Lightweight & Variable & Short & Close proximity to signal sources, versatile applications& Limited to what can be measured at the ear, sensitive to ambient noises and ear positioning  \\
        \hline
    \end{tabular}
}
\vspace{2mm}
\caption{Comparison of different sensing modalities.}
\vspace{-4mm}
\label{table:sensing_comparison}
\end{table}

\subsection{Motion Sensing}
\label{subsec_motion_sensing}
Motion sensing involves the use of motion sensors such as Inertial Measurement Unit (IMU) sensors (i.e., accelerometers, gyroscopes, and magnetometers) attached to the individuals to capture various types of motions such as arm postures, body movements, and physical activities.
As summarized in Figure~\ref{fig:sensing-tree}, depending on the sensing tasks, existing works on AI-empowered motion sensing can be grouped into two categories: human activity recognition, and arm tracking.

\vspace{1.5mm}
\noindent
\textbf{Human Activity Recognition.}
One of the most important tasks of motion sensing is human activity recognition (HAR).
Most existing HAR frameworks are limited to a few predefined activities and require prior knowledge or labeled data for supervised training. To address this limitation, \citet{Lasagana} introduce Lasagana, an unsupervised learning-based HAR framework that extracts common bases of human motions in an unsupervised manner, creating a universal multi-resolution representation for common human activities. 
\revision{Their prototype system achieves 98.9\% precision in activity classification and nearly 100\% recall with about 90\% precision in activity indexing.}
Another major limitation of existing motion sensing-based HAR frameworks is that machine learning (ML) algorithms trained on specific sensors require retraining upon any system configuration changes, such as adding a new sensor. %
To address this limitation, \citet{deep-generative-domain} propose a training scheme for the newly added sensors to identify human activities that were previously detected by existing sensors. %
As another line of research, \citet{SenseHAR} target the device heterogeneity problem, which encompasses variations in sensor types, data formats, and sampling rates, leading to lower activity recognition performance in real-life scenarios. To address this issue, they propose a DL-based HAR framework named SenseHAR, which allows sensor fusion while being robust to device heterogeneity. SenseHAR offers easy calibration for new devices, allowing seamless integration and utilization of different devices with varying sampling frequencies, sensors, and applications.
\citet{Limu-bert}  tackle the challenges of limited labeled data and device placement diversity in HAR. They propose LIMU-BERT, a lightweight DL-based HAR framework that employs self-supervised learning to extract general features from unlabeled sensor data. It adopts the key principles of the BERT framework for motion sensing and a classifier consisting of three stacked Gated Recurrent Units (GRU). The model's efficiency and ability to learn robust features make it suitable for real-time applications on mobile devices.

\vspace{1.5mm}
\noindent
\textbf{Arm Tracking.}
Another important task of motion sensing is arm tracking, which uses motion sensors to track the movements, positions, and posture of an individual's arm. %
Most arm tracking systems require attaching multiple sensors to an individual's arm, which can limit flexibility and have a negative impact on the overall user experience. To address this issue, \citet{armtroi} propose ArmTroi, a real-time 3D arm skeleton tracking system that uses a single motion sensor worn on the wrist. ArmTroi adopts an attention and recurrent neural network (RNN)-based network, which is lightweight and suitable for mobile and real-time applications.
\revision{The authors also prototype the system on LG smartwatches, Google Glass, and Samsung Galaxy S7. ArmTroi achieves real-time arm tracking with 92.7\% gesture recognition precision, and demonstrates its efficacy through fitness and gesture-based control applications.}
As another line of research, the differences among accelerometers, gyroscopes, and magnetometers of the IMU sensors as well as the heavy computation costs incurred by DL models make it challenging to leverage all of these sensors for accurate and real-time arm tracking. To address this issue, \citet{RTAT} propose RTAT, a real-time arm tracking system that utilizes a Bidirectional Long Short-Term Memory (BiLSTM)-based multitask neural network to track both the orientation and location of an arm simultaneously. RTAT also incorporates an attention mechanism to dynamically learn the importance of different IMU sensor streams to achieve high accuracy and low latency.

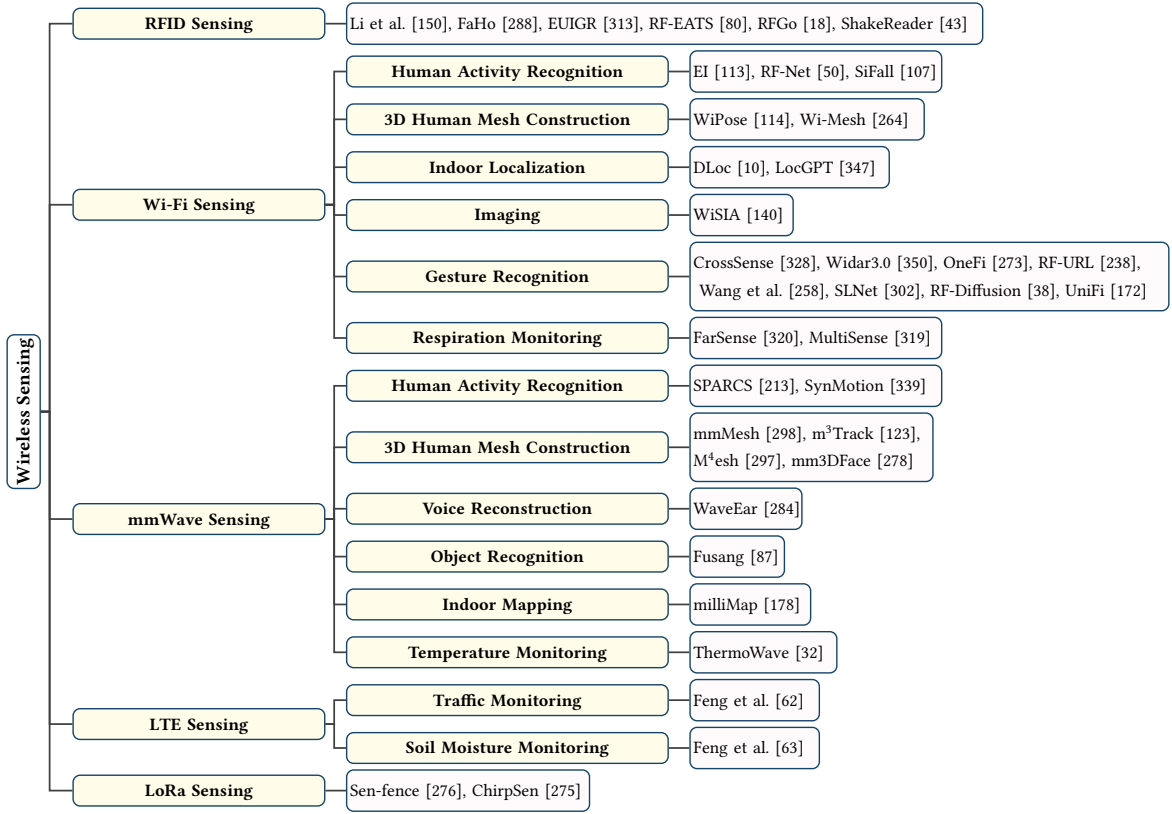
\begin{figure*}[t!]
    \centering
    \resizebox{0.98\textwidth}{!}{
        \begin{forest}
            forked edges,
            for tree={
                grow=east,
                reversed=true,
                anchor=base west,
                parent anchor=east,
                child anchor=west,
                base=center,
                font=\large,
                rectangle,
                draw=hidden-draw,
                rounded corners,
                align=left,
                text centered,
                minimum width=4em,
                edge+={darkgray, line width=1pt},
                s sep=3pt,
                inner xsep=2pt,
                inner ysep=1pt,
                line width=0.8pt,
                ver/.style={rotate=90, child anchor=north, parent anchor=south, anchor=center},
            },
            where level=1{text width=14em,font=\normalsize,}{},
            where level=2{text width=18em,font=\normalsize,}{},
            where level=3{text width=17em,font=\normalsize,}{},
            [
                \textbf{Wireless Sensing}, ver
                        [
                            \textbf{RFID Sensing}, fill=yellow!10
                            [\citet{li2016deep}{,}~ FaHo~\cite{xu2019faho}{,}~EUIGR~\cite{yu2019rfid}{,}~RF-EATS~\cite{ha2020food}{,}
                            RFGo~\cite{bocanegra2020rfgo}{,} ShakeReader~\cite{ShakeReader2021}
                            ,leaf, text width=36em]
                        ]
                        [
                            \textbf{Wi-Fi Sensing}, fill=yellow!10
                            [
                            \textbf{Human Activity Recognition}, fill=yellow!10
                            [
                            EI~\cite{jiang2018towards}{,}  
                            RF-Net~\cite{ding2020rf}{,}
                            SiFall~\cite{ji2022sifall},leaf, text width=14em
                            ]
                            ]
                            [
                            \textbf{3D Human Mesh Construction}, fill=yellow!10
                            [
                            WiPose~\cite{jiang2020towards}{,} 
                            Wi-Mesh~\cite{wang2022wi},leaf, text width=13em
                            ]
                            ]
                            [
                            \textbf{Indoor Localization}, fill=yellow!10
                            [
                            DLoc~\cite{ayyalasomayajula2020deep}{,}~LocGPT~\cite{zhao2024understanding},leaf, text width=11em
                            ]
                            ]
                            [
                            \textbf{Imaging} \\ , fill=yellow!10
                            [
                            WiSIA~\cite{li2020wi}, leaf, text width=5.5em
                            ]
                            ]
                            [
                            \textbf{Gesture Recognition}, fill=yellow!10
                            [
                            CrossSense~\cite{zhang2018crosssense}{,}
                            Widar3.0~\cite{zheng2019zero}{,}
                            OneFi~\cite{xiao2021onefi}{,} 
                            RF-URL~\cite{song2022rf}{,}\\
                            ~\citet{wang2022solving}{,}  
                            SLNet~\cite{yang2023slnet}{,} 
                            RF-Diffusion~\cite{RFDiffusion2024Chi}{,}
                            ~UniFi~\cite{liu2024unifi},leaf, text width=27em
                            ]
                            ]
                            [
                            \textbf{Respiration Monitoring}, fill=yellow!10
                            [
                            FarSense~\cite{zeng2019farsense}{,} ~MultiSense~\cite{zeng2020multisense},leaf, text width=14em
                            ]
                            ]
                        ]   
                        [
                            \textbf{mmWave Sensing}, fill=yellow!10
                            [
                            \textbf{Human Activity Recognition} 
                             , fill=yellow!10
                            [
                            SPARCS~\cite{pegoraro2022sparcs}{,}
                            SynMotion~\cite{zhang2022synthesized}, leaf, text width=14em
                            ]
                            ]
                            [
                            \textbf{3D Human Mesh Construction}, fill=yellow!10
                            [
                            mmMesh~\cite{xue2021mmmesh}{,}
                            m$^3$Track~\cite{kong2022m3track}{,}\\
                            M$^4$esh ~\cite{xue2022m4esh}{,}
                            mm3DFace~\cite{xie2023mm3dface}
                            ,leaf, text width=13.5em
                            ]
                            ]
                            [
                            \textbf{Voice Reconstruction}, fill=yellow!10
                            [
                            WaveEar~\cite{xu2019waveear},leaf, text width=6em
                            ]
                            ]
                            [
                            \textbf{Object Recognition}, fill=yellow!10
                            [
                            Fusang~\cite{he2023fusang},leaf, text width=5em
                            ]
                            ]
                            [
                            \textbf{Indoor Mapping}, fill=yellow!10
                            [
                            milliMap~\cite{lu2020see},leaf, text width=6.5em
                            ]
                            ]
                            [
                            \textbf{Temperature Monitoring}, fill=yellow!10
                            [
                        ThermoWave~\cite{chen2020thermowave},leaf, text width=8em
                            ]
                            ]
                        ]
                        [
                            \textbf{LTE Sensing}, fill=yellow!10
                            [
                            \textbf{Traffic Monitoring}, fill=yellow!10
                            [\citet{feng2021lte},leaf, text width=7em
                            ]
                            ]
                            [
                            \textbf{Soil Moisture Monitoring}, fill=yellow!10
                            [\citet{feng2022lte},leaf, text width=7em
                            ]
                            ]   
                        ]
                        [
                            \textbf{LoRa Sensing}, fill=yellow!10
                            [Sen-fence~\cite{xie2020combating}{,}
                            ChirpSen~\cite{xie2023boosting},leaf, text width=13.5em
                            ]  
                        ]       
                    ]
        \end{forest}
}
    \caption{Summary of topics related to wireless sensing.}
    \label{fig:Tree_wifi_sensing}
\end{figure*}

\subsection{Wireless Sensing} 
\label{wireless_sensing}
Wireless sensing uses wireless signals to sense individuals and objects in the environment in a contact-free manner.
As summarized in Figure \ref{fig:Tree_wifi_sensing}, based on the frequency bands wireless signals belong to, existing works on AI-empowered wireless sensing can be grouped into five categories: RFID sensing, Wi-Fi sensing, mmWave sensing, LTE sensing, and LoRa sensing.

\subsubsection{RFID Sensing}
Radio Frequency Identification (RFID) is a technology that employs an RFID tag and reader, enabling the retrieval of information from the tag using radio frequency (RF) signals emitted by the reader. By attaching RFID tags to individuals or objects, RFID can be deployed for sensing tasks such as localization, object tracking, and classification. 
\citet{li2016deep} utilize an RFID sensing system for activity recognition in the medical environment by attaching RFID tags to objects in clinical settings and recording the Received Signal Strength (RSS) from these tags. These collected data subsequently serve as the input for a convolutional neural network (CNN), enabling the recognition of activities that involve the usage of certain objects. 
While this approach effectively identifies activities using RFID, the received signal comprises both the Line of Sight (LOS) signal and multiple reflections from obstacles. This complicates the localization task, making it challenging to determine which signal accurately represents the RFID tag's location
In response to this issue, \citet{xu2019faho} introduce an algorithm that transforms the RFID signal into a hologram that encapsulates the probable location of the tag. A CNN is then employed to accurately identify the tag's actual position within this hologram. 
The accuracy of existing RFID systems is significantly impacted by the subjects and the surrounding environmental conditions. In the context of gesture recognition, some studies consider environmental variations but often neglect the impact on the user. To address this issue, \citet{yu2019rfid} develop a discriminator DNN, which identifies the user and its environment in the data. Simultaneously, it also has gesture labeling DNN, which predicts the probability of gestures. Through adversarial training of both DNNs, the gesture labeling DNN learns to create representations that are indistinguishable from the domain discriminator, resulting in a gesture recognizer that is independent of the user and environment.
\citet{ha2020food} introduces RF-EATS, a system designed to noninvasively sense food and liquids within closed containers using passive RFID tags.
The authors attach the RFID tag to the liquids and detect whether this liquid is fake or not. To manage the diversity in environmental conditions, the study employs Variational Autoencoders (VAE) to synthesize multiple samples. A classifier is then trained to distinguish counterfeit liquids using these augmented datasets.
\citet{bocanegra2020rfgo} design an RFID reader system capable of simultaneous multi-tag reading via an array of deployed antennas. To determine whether an RFID tag is within the checkout area, they also utilize a neural network, training it in a supervised manner using data captured from the reader.
\revision{Ultra-high-frequency (UHF) RFID is more appealing to retailers because it can rapidly scan multiple RFID-tagged items, substantially increasing operational efficiency; however, smartphones currently lack direct communication capabilities with UHF RFID tags. To bridge the gap, \citet{ShakeReader2021} introduce ShakeReader, a system designed to enhance interaction between smartphones and UHF RFID-tagged items without requiring hardware modifications to existing RFID systems or smartphones. ShakeReader enables users to obtain item-specific information by performing predefined gestures, such as shaking the smartphone near the RFID tag. The system utilizes a reflector polarization model to analyze the backscattered signal from the tag, which is affected by the smartphone's gestures. This model accounts for both the signal propagation and the polarization changes caused by the reflection from the smartphone, enabling the detection of specific gestures using the RFID reader even with a single tag.}

\begin{figure}[]
  \centering
  \includegraphics[width=0.8\linewidth]{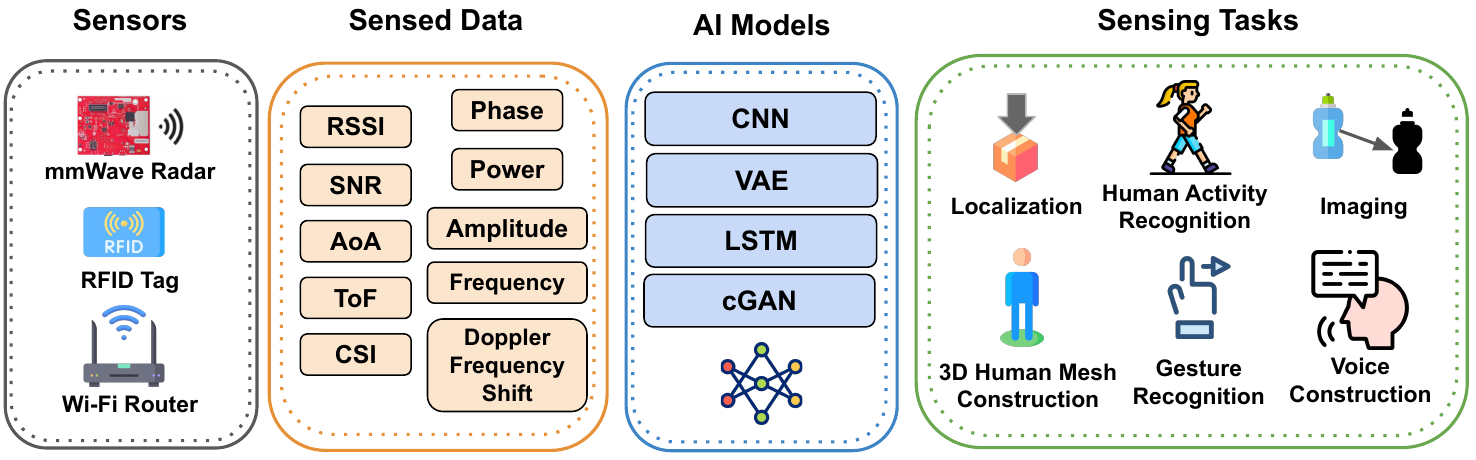}
  \caption{Illustration of AI-empowered wireless sensing pipeline.}
  \label{fig:Pipeline of Wi-Fi Sensing}
  \vspace{-3mm}
\end{figure}

\subsubsection{Wi-Fi Sensing}

\noindent Wi-Fi sensing takes advantage of the ubiquitous Wi-Fi signals and their associated hardware to detect and interpret human movements or changes in the environment.
Depending on the sensing tasks, existing works on AI-empowered Wi-Fi sensing can be grouped into the following categories.

\vspace{1.5mm}
\noindent\textbf{Human Activity Recognition.} 
One important task of Wi-Fi sensing is human activity recognition (HAR).
The major challenge in device-free human activity recognition is that wireless signals are highly influenced by the specific environment and individual characteristics of the human subject, leading to poor generalization of models across different subjects and environments.
To address this challenge, \citet{jiang2018towards} propose EI for HAR that learns domain-independent features from activity data collected in different domains. EI accepts multiple types of input signals, including Wi-Fi Channel State Information (CSI). The DL model of EI incorporates an adversarial network, including a CNN-based feature extractor, an FC-layer-based activity recognizer that predicts activity type from extracted features, and a domain discriminator that predicts the domain.
\citet{ding2020rf} present RF-Net, a metric-based meta-learning approach for one-shot human activity recognition using Wi-Fi that can perform the recognition in a new environment with only one observation per label. 
RF-Net classifies a new observation in new environments by calculating a weighted sum of all the labels in the training dataset. The weights are given by the similarity between the query observation and all the data in the support dataset of the new environment. 
Lastly, \citet{ji2022sifall} propose SiFall to formulate the fall detection problem as adaptive anomaly detection out of normal repeatable human activities instead of seeking features to characterize fall activity. %

\vspace{1.5mm}
\noindent\textbf{3D Human Mesh Construction.} 
3D human mesh construction in Wi-Fi sensing refers to the creation of three-dimensional representations of the human body using Wi-Fi signals.
\citet{jiang2020towards} present WiPose, a 3D human pose skeleton construction framework that recovers human joints on both limbs and torso of the human body using commercial Wi-Fi devices. WiPose records CSI using a single antenna transmitter with multiple distributed receivers and designs an LSTM-based deep learning model that accepts the sequence of Doppler Frequency Shift (DFS) profile transformed from non-overlapping CSI segments and outputs a series of features. The learned features from LSTM are regarded as the rotation of human body joints and fed to the forward kinematics layers to calculate the actual joint locations based on a given skeletal structure. %
\citet{wang2022wi} present Wi-Mesh, which further improves the 3D human mesh construction task with DNN based on GRU and self-attention. Wi-Mesh leverages a commodity 3-antenna transmitter and two receivers with 9 antennas in an L shape to record CSI. Received signals at the specific antenna array can be used to calculate the 2D AoA of the signal reflections based on phase shift, providing spatial information about the objects and environment. Wi-Mesh generates thirty 2D AoA spectrums per second and extracts only human images by subtracting the static components in consecutive images since the human body is moving. 
\revision{Wi-Mesh tracks way more body locations than WiPose and also outperforms WiPose with an average joint location error of 2.4cm and body vertices location error of 2.81cm, though using more complicated antenna arrays. }

\vspace{1.5mm}
\noindent\textbf{Indoor Localization.} 
Indoor localization in Wi-Fi sensing refers to the process of using Wi-Fi signals to determine the position of objects or individuals within indoor environments.
Unlike outdoor localization, which commonly relies on GPS (Global Positioning System), indoor localization requires a different set of technologies and methodologies due to challenges such as the unavailability of GPS signals indoors, multi-path reflection, and interference from walls and other structures. Thus, indoor localization remains a "last-mile" problem when forming a positioning system without blind spots. Wi-Fi has been broadly utilized to address the indoor localization problem.
\citet{ayyalasomayajula2020deep} present DLoc, a DL-based wireless localization algorithm and an automated mapping platform MapFind, which altogether forms a positioning system with a map inspired by outdoor localization services. MapFind constructs location-tagged maps of the environment and generates training data for DLoc. Together, they solve the active indoor localization scenario in which off-the-shelf Wi-Fi devices like smartphones can access a map of the environment and estimate their position by sending packets to surrounding Wi-Fi access points with respect to that map. 
While deep learning approaches for indoor localization rely on high-quality training samples and are hard to adapt to varied scenarios, \citet{zhao2024understanding} propose LocGPT, which is a specialized Generative Pre-training Transformer (GPT) variant that excels in generating profound contextual insights, to explore the underlying principles of indoor localization. The model is configured with 36 million parameters tailored for transfer learning. To facilitate the benchmarking, training, and transfer learning in indoor localization, they have established Ray, the first 3D indoor localization dataset on a scale of millions, including RFID, Wi-Fi, and BLE samples. LocGPT achieves near-par accuracy when fine-tuned with merely half the conventional dataset, which shows its superiority in transfer learning within the indoor localization domain.

\vspace{1.5mm}
\noindent\textbf{Imaging.} 
Wi-Fi imaging exploits the capabilities of Wi-Fi signals to create images of objects or humans in the environment.
\citet{li2020wi} present a Wi-Fi imaging system WiSIA that is capable of simultaneously detecting and segmenting objects and humans within the imaging plane using commodity Wi-Fi devices. WiSIA leverages two receivers with three orthogonal antennas sharing the same transmitter antenna as the imaging model on the object side to record CSI that contains the changes in the Wi-Fi signal of both amplitude and phase. 
WiSIA incorporates a conditional Generative Adversarial Network (cGAN) to refine the boundaries in an image-to-image translation fashion. WiSIA achieves 0.9 in similarity and tagging accuracy for all five tested objects which is comparable to the state-of-the-art computer vision and acoustics imaging while outperforming the state-of-the-art vision-based method in conditions with darkness or obstructions.

\vspace{1mm}
\noindent\textbf{\mz{Gesture Recognition.}} 
Wi-Fi signals can be used in the gesture recognition task by analyzing the variations in the Wi-Fi signal caused by human body movements. 
\citet{zhang2018crosssense} propose CrossSense, a system designed to improve the scalability and efficiency of WiFi-based gesture recognition. The primary challenge addressed is the need for extensive, site-specific training data collection, which is labor-intensive and impractical for large-scale deployments. CrossSense tackles this by using machine learning to generate synthetic training samples from existing measurements, allowing these samples to be effectively used across different environments.
\citet{zheng2019zero} propose Widar3.0, a Wi-Fi-based zero-effort cross-domain gesture recognition system. Widar3.0 calculates the body-coordinate velocity profile (BVP) of gestures from CSI at the lower signal level, which represents power distribution over different velocities and is unique from gesture to gesture while independent from the domain. On this basis, Widar3.0 adopts a one-fits-all model based on CNN, GRU, and dense layers that requires only one-time training but can adapt to different data domains. 
Similar to Widar3.0, OneFi \cite{xiao2021onefi} proposes to use velocity distribution which can be derived from DFS as the unique feature that describes a gesture. It adopts a backbone based on self-attention, noted as Wi-Fi Transformer, as the gesture recognition framework. To avoid model re-training, OneFi adopts a lightweight one-shot learning framework based on transductive fine-tuning and opens up a new direction for one-shot (or few-shot) learning in Wi-Fi-based gesture recognition. 
\citet{song2022rf} present RF-URL, an unsupervised representation learning framework for human gesture recognition tasks. RF-URL combines signal-processing-based RF sensing with learning-based RF sensing by using a contrastive framework. 
Experimental results indicate that RF-URL pre-training model is capable of extracting general information for gesture recognition and applying it effectively across different datasets.
\citet{wang2022solving} carry out an in-depth study on the domain variation problem in Wi-Fi-based gesture recognition task, which can alter multi-path effects and introduce noise into wireless signals. These variations, including changes in the environment, can lead to significant performance degradation in Wi-Fi sensing applications due to the resulting fluctuations in wireless signal patterns. To mitigate these effects, the authors propose a robust framework based on conformal prediction, which quantifies the similarity between testing and training data without the need for retraining or generating new features.
\citet{yang2023slnet} propose SLNet, an architecture for enhancing wireless sensing applications through the integration of deep learning and spectrogram analysis. SLNet utilizes neural networks to generate super-resolution spectrograms, addressing the limitations of traditional time-frequency uncertainty. This design improves the accuracy of Wi-Fi-based gesture recognition, human identification, fall detection, and breathing estimation tasks. Experiments demonstrate that SLNet achieves superior performance with reduced computational demands, making it suitable for practical deployment on edge devices.
\citet{RFDiffusion2024Chi} introduce RF-Diffusion, a novel approach to generating high-quality, time-series radio frequency (RF) data using diffusion models. The proposed methodology involves training RF-Diffusion with a real-world dataset to generate synthetic RF signals of the designated type. These synthetic samples are then integrated with the original dataset, and collectively employed to train the wireless sensing model. The authors highlight that RF-Diffusion when used as a data augmentation tool, leads to substantial improvements in Wi-Fi-based gesture recognition accuracy. This enhancement is attributed to the model's ability to produce diverse and high-quality RF data that enriches the training datasets of existing systems.

\vspace{1mm}
\noindent\textbf{\mz{Respiration Monitoring.}} 
Wi-Fi signals can be used for respiration monitoring by analyzing the subtle variations in wireless signals caused by the movement of a person's chest during breathing. Existing methods of respiration monitoring are limited by short sensing ranges, susceptibility to noise, and issues with phase offset stability.
\revision{To overcome these limitations,~\citet{zeng2019farsense} introduce FarSense, a system for enhancing Wi-Fi-based respiration sensing. FarSense leverages the CSI ratio from two antennas to overcome the limitations of existing methods that rely on individual CSI readings. By using the CSI ratio, FarSense cancels out most of the noise and phase offset issues, significantly extending the sensing range. The system combines the amplitude and phase information of the CSI ratio to address the "blind spots" problem and improves the sensitivity of detecting subtle respiration signals.
\citet{zeng2020multisense} present MultiSense, a system for accurately monitoring the respiration patterns of multiple individuals simultaneously using commodity Wi-Fi devices. MultiSense overcomes the challenges faced in existing methods by leveraging multiple antennas on Wi-Fi devices and modeling the multi-person respiration sensing problem as a Blind Source Separation (BSS) problem. MultiSense cancels out time-varying phase offsets and removes background static signals, allowing for robust separation and continuous monitoring of detailed respiration patterns.
}

\vspace{-2mm}
\subsubsection{mmWave Sensing}
\noindent
Millimeter Wave (mmWave) sensing refers to the use of electromagnetic waves with wavelengths in the millimeter range, typically between 30 GHz and 300 GHz frequency band, for a variety of sensing tasks.
The high frequency, short wavelength, and broadband capacity make mmWave more sensitive to minor reflection distance variations, and thus can provide finer sensing resolution. 
At the same time, mmWave has limited penetration capabilities so it can easily be attenuated or blocked by obstacles. As such, mmWave sensing often requires a direct line-of-sight between the transceivers and the sensing target. 
Depending on the sensing tasks, existing works on AI-empowered mmWave sensing can be grouped into the following categories.

\vspace{1mm}
\noindent \textbf{Human Activity Recognition.} 
The capability of mmWave signals to capture micro-motions and micro-vibrations of different human body parts makes it feasible for the task of human activity recognition (HAR). 
~\citet{pegoraro2022sparcs} introduce SPARCS for mmWave-based HAR. It focuses on extracting micro-Doppler signatures of human movement from irregular and sparse Channel Impulse Response (CIR) samples. This approach leverages the inherent sparsity of the mmWave channel to reduce sensing overhead drastically while integrating seamlessly with existing communication protocols. By formulating micro-Doppler extraction as a sparse recovery problem, SPARCS achieves high-quality human activity recognition with significantly lower overhead compared to existing methods, demonstrating its applicability and efficiency in real-world scenarios.
While research on introducing DL to mmWave-based human activity recognition achieves promising performance, collecting and labeling mmWave datasets for such tasks is difficult and expensive. To close the gap, \citet{zhang2022synthesized} present SynMotion which synthesizes mmWave signals at high quality using widely available vision-based human motion datasets with the coordinates of body skeletal points and designs a few/zero-shot synthetic-to-real transfer learning framework for downstream human activity recognition.

\vspace{1mm}
\noindent \textbf{3D Human Mesh Construction.} 
mmwave signals can also be used for 3D human mesh construction by providing detailed information about the human body contours and structure.
\citet{xue2021mmmesh} present mmMesh, a DL-based real-time 3D human mesh construction framework to model the moving subject with commercial portable mmWave devices. mmMesh utilizes range and angle information to remove noisy reflections from static objects in the IF signals collected by commercial devices and generate the 3D point clouds as input to the DL model. 
\citet{kong2022m3track} propose $m^3Track$ to enable simultaneous tracking of the 3D postures of multiple users leveraging a single commercial mmWave device. $m^3Track$ obtains the Range-Doppler-Profile of the IF signals by range-FFT and doppler-FFT that contains information on the users and background objects. It distinguishes multiple users and backgrounds by sliding a convolutional kernel along the range bins of the Range-Doppler-Profile and performing convolution operations to detect the ranges that contain users. 
\citet{xue2022m4esh} develop $M^4esh$ for multi-subject 3D human mesh reconstruction. The tracking scheme of $M^4esh$ integrates techniques adopted by mmMesh and $m^3Track$, including subject detection, 3D point cloud generation for each subject, and per-subject mesh reconstruction. %
Similarly, \citet{xie2023mm3dface} propose mm3dFace to move towards the reconstruction of human face. It proposes to leverage commercial mmWave radar to reconstruct 3D human faces that continuously express facial expressions in a passive manner. mm3dFace captures human face information from the recorded IF signal. By applying range-FFT to the IF signal and AoA calculation, it obtains the range profile, azimuth profile, and elevation profile, which together form a Range-Angle-Profile in the three-dimensional space. The three-dimensional profile captures the side view and frontal view of human faces. 

\vspace{1.5mm}
\noindent \textbf{Voice Reconstruction.}
Voice Reconstruction refers to the process of capturing and reconstructing the human voice by detecting subtle vibrations with millimeter-wave signals.
\citet{xu2019waveear} propose WaveEar, which leverages mmWave devices to enable noise-resilient speech sensing for voice-user interface (VUI) in environments with audible and inaudible interference. The authors conducted an in-depth study of human voice generation to obtain insights into voice vibration caused by the integrated effort of three physiological organs, e.g., lungs, vocal cords, and articulators. WaveEar designs a low-cost mmWave probe that employs a phased directional array to locate the speaker by throat vibration and then transmits mmWave signals towards the near-throat region of the speaker and processes the reflected signal for voice reconstruction.

\vspace{1.5mm}
\noindent \textbf{Object Recognition.} 
The broadband nature of mmWave makes it also suitable for object recognition. 
\citet{he2023fusang} present Fusang, a system that adopts commercial off-the-shelf mmWave devices for accurate and robust 3D object recognition. Fusang leverages the large bandwidth of mmWave radars to capture a unique set of fine-grained responses reflected by objects with different shapes. It generates the High-Resolution Range Profile (HRRP) from the IF signal and constructs two novel graph-structured features, as the HRRP data of different objects in the spectrum is not always distinguishable. Fusang extracts the set of formants that denotes the peaks in the HRRP envelope and iteratively bisects the frequency bands to a point when there is no more than one formant falling in each subband to build a binary tree with subbands that contain formants as leaf nodes. 

\vspace{1.5mm}
\noindent \textbf{Indoor Mapping.} 
Indoor mapping using mmWave involves creating detailed maps or spatial representations of environments using the data obtained from mmWave radar sensors.
State-of-the-art mapping approaches are mainly based on optical sensors, such as lidar and cameras.
One of the advantages of mmWave over optical sensors is its ability to penetrate through certain materials and resilience to poor illumination.~\citet{lu2020see} present milliMap, which adopts a single-chip mmWave radar for dense indoor map generation and simple object annotation in low-visibility environments under emergency situations. milliMap adopts conditional GAN supervised by a co-located liar to generate dense patches similar to lidar ground truth from mmWave scans. In this way, milliMap overcomes the sparsity and multi-path noise of mmWave signals. It also identifies different objects from the spectral response of mmWave reflections by a CNN-based semantic recognizer. 

\vspace{1.5mm}
\noindent \textbf{Temperature Sensing.} 
Temperature sensing refers to the continuous or periodic process of measuring and recording temperature levels in a given environment, object, or individual.
While most wireless temperature monitoring solutions are not cost-effective and generate electronic wastes, ThermoWave ~\cite{chen2020thermowave} enables ecological, battery-less, and ultra-low-cost wireless temperature monitoring using mmWave signals. Specifically, ThermoWave is designed based on the principle of thermal scattering effect of mmWave. Specifically, it attaches ThermoTags made of cholesteryl material inked film or paper which aligns the molecular patterns at different temperatures and senses the temperature-induced pattern change by scattered mmWave signals. The ThermoTags are of low cost (less than 0.01 dollars per tag). ThermoWave adopts a mmWave-radar-based ThermoScanner to receive the temperature-modulated mmWave scattering and extract thermal features from it.

\subsubsection{LTE Sensing}
Long-Term Evolution (LTE) sensing leverages the capabilities of LTE wireless broadband communication technology for the task of sensing.
\revision{\citet{feng2021lte} explore the use of LTE signals for pervasive sensing applications both indoors and outdoors. It aims to address the limitations of existing wireless sensing technologies, such as Wi-Fi, which are constrained by coverage and performance issues. Specifically, the authors propose to leverage the widespread and diverse LTE infrastructure to achieve comprehensive and reliable sensing without affecting LTE data communication. Through advanced techniques to mitigate interference and noise, the authors demonstrate the effectiveness of LTE sensing in two key applications: indoor respiration monitoring and outdoor traffic monitoring.
In \cite{feng2022lte}, the authors leverage the infrastructure of LTE base stations to provide a cost-effective and energy-efficient solution for the application of soil moisture monitoring. By utilizing commercial off-the-shelf hardware, including software-defined radios and a Raspberry Pi, the proposed system achieves high accuracy comparable to high-end sensors but at a fraction of the cost. They have deployed their prototype system and examined its robustness across various soil types and conditions, demonstrating its potential for applications in precision agriculture and environmental monitoring.}

\subsubsection{LoRa Sensing}
\noindent
The long-range, low-power characteristics of LoRa networks make it popular among large-scale remote-area IoT applications. However, the use of LoRa for sensing tasks is yet to be explored due to challenges related to interference, sensing range, and many more.
To address these challenges, \revision{\citet{xie2020combating} introduce Sen-fence, which explores advanced signal processing techniques that maximize movement-induced signal variations, thereby increasing the sensing range. Additionally, the authors introduce a novel "virtual fence" method, which confines sensing activities to a specific area of interest, thereby reducing the impact of environmental noise and interference. Sen-fence achieves a 50-meter range for fine-grained human respiration detection while effectively managing interference for practical LoRa sensing applications.
Though the proposed method in Sen-fence is effective for detecting tiny movements like respiration but struggles with larger movements such as human walking. 
To address this issue, the authors in \cite{xie2023boosting} introduce ChirpSen, a system designed to enhance the sensing range of LoRa-based localization by fully exploiting the properties of chirp signals. ChirpSen employs a chirp concentration scheme that concentrates the power of all signal samples in a LoRa chirp at one timestamp, thus increasing the signal power as well as the sensing range.
Real-world experiments demonstrate that ChirpSen significantly enhances detection capabilities, extending the range for monitoring human respiration at a distance of 138 meters and tracking a walking human at up to 210 meters.}

\begin{figure}[]
\centering
\includegraphics[width=0.8\linewidth]{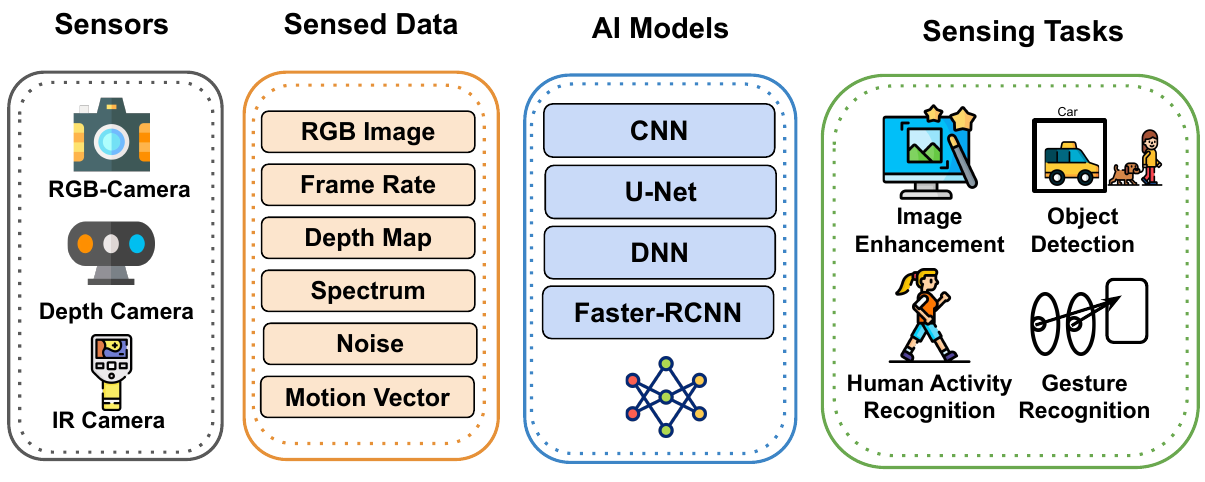}
\caption{Illustration of AI-empowered vision sensing pipeline.}
\label{fig:Pipeline of Vision Sensing}
\vspace{-4mm}
\end{figure}

\subsection{Vision Sensing} 
\label{subsec_vision_sensing}
Vision sensing involves the use of vision sensors such as \mz{RGB cameras, depth cameras, and near-infrared (NIR) image sensors} to capture and analyze visual information for various sensing tasks.
As summarized in Figure~\ref{fig:sensing-tree}, depending on the sensing tasks, existing works on AI-empowered vision sensing can be grouped into five categories:
\mz{human activity recognition, image enhancement, object detection, eye tracking, and pose estimation.}

\vspace{1.5mm}
\noindent
\textbf{Human Activity Recognition.}
DL-based models used in vision sensing for HAR can be computationally demanding, posing a significant challenge when it comes to execution on mobile and IoT devices. 
Moreover, vision systems that rely on RGB cameras are intrinsically susceptible to privacy leakage by hacking. To tackle this problem, \citet{shim2023mosaic} choose to use a Near-Infrared (NIR) image sensor to monitor human activities that inherently does not contain enough data to reveal personal identity. Although the NIR sensor loses a lot of spatial information, the authors have demonstrated that the temporal information and pixel-wise computation over DNN are enough to recognize the performed activities.

\vspace{1.5mm}
\noindent
\textbf{Image Enhancement.}
Image enhancement involves manipulating the image itself to improve its quality. A key technique within this area is super resolution, which aims to increase the resolution of the image. However, executing this task on-device poses significant challenges due to the immense computational complexity and substantial storage requirements. To mitigate these issues, \citet{lee2019mobisr} employ two distinct compressed DNNs and schedule their operations across CPU, GPU, and DSP. 
Captured images by vision sensors are often transmitted over low-power, unreliable IoT networks. However, traditional methods such as JPEG, designed for use on reliable networks, are still commonly employed for image transmission. To efficiently transmit and receive high-quality image data over this unstable network,  \citet{hu2020starfish} find the optimal encoder and decoder pair of DNN by employing neural architecture search methods. %
Motion blurs on IoT devices are a severe problem while capturing the image. Existing solutions to this problem often necessitate additional hardware or have high computational demands that are ill-suited to microcontrollers. To solve this problem, \citet{lee2023microdeblur} adopt depth-independent convolution operations on DNN to estimate the blur kernel. This predicted kernel is then applied to the blurred image to recover the original, clear image. Additionally, the algorithm employs a matrix transformation, converting it to a Toeplitz Matrix. This transformation yields computational advantages, making it particularly efficient for deployment in extremely resource-constrained microcontroller environments.

\vspace{1.5mm}
\noindent
\textbf{Object Detection.}
Object detection is one of the most fundamental and important tasks in vision sensing.
Recognizing faces in crowded environments is a critical challenge, particularly in applications like finding missing children. Existing DNN methods suffer from the low-resolution problem of the detected face. To solve the problem, \citet{yi2020eagleeye} design a three-step mult-DNN pipeline consisting of detection, clarification, and recognition. During the clarification phase, the system recovers missing elements of the low-resolution image by fine-tuning it with the target's face.
The research by \citet{sami2021lapd} leverage a Time of Flight (ToF) sensor embedded in mobile phones to locate and identify concealed spy cameras. Conventional methods typically necessitate manual interpretation to discern these hidden devices. However, the incorporation of a ToF sensor enables the system to detect distinctive reflections emitted by spy cameras. Following this, deep learning techniques are deployed to filter out false positives from the detected images and effectively pinpoint the hidden cameras in an automated manner.
\citet{sun2022detecting} have shown the use of a smartphone camera to detect counterfeit liquid products, eliminating the need for additional hardware. The method tracks the movement of bubbles in the liquid using Faster-RCNN and U-Net and verifies the product's authenticity using the AdaBoost algorithm.
Depth-contained images acquired from depth sensors can be employed in the detection and classification tasks of DNN~\cite{xie2023mozart,xie2021ultradepth,mithun2018odds}. These methods are effective compared to RGB cameras in low-light environments. 
Both~\citet{xie2023mozart,xie2021ultradepth} employ the indirect Time-of-Flight (iToF) depth camera to capture the high-resolution texture depth map while \citet{mithun2018odds} use Kinect for XBOX One to achieve it.
In particular, in the construction of the depth map, \citet{xie2023mozart} employ an autoencoder that exploits the phase components of the iToF camera. On the other hand, \citet{xie2021ultradepth} employs an additional distorting IR source and uses the energy difference of the signal depending on the texture.

\vspace{1.5mm}
\noindent
\textbf{Eye Tracking.}
Exploiting eyewear devices for eye tracking presents unique challenges due to their limited computational resources and the variability in eye characteristics across different individuals. To address this issue, \citet{wu2020emo} design EMO, a personalized DNN classifier that classifies emotions using images captured from a single eye by the eyewear. 
Likewise, \citet{lan2020gazegraph} employ eyewear devices for extracting gaze data and aims to use it for cognitive context sensing. However, this approach also suffers from the diversity of people. To address this issue, this research adopts the few-shot learning method. These allow for rapid adjustment to new environments when operating a spatial-temporal graph-based DNN system, which is used to classify activities from gaze information.
Tracking the gaze from the eye is highly demanding because of the small size of the iris and subtle hints concerning the directions. Existing commercial systems are expansive, while low-cost RGB camera approaches suffer from the insufficiency of datasets. 
To effectively track eye gaze, \citet{10.1145/3560905.3568544} have developed a geographical gaze model that maps the relationship between the smartphone screen and the iris boundary, which contains the gaze directions. To accurately extract the iris boundary over the eye, the authors employ U-Net and further refine the resulting pixels to enhance the accuracy of eye tracking. 

\vspace{1.5mm}
\noindent
\textbf{Pose Estimation.}
Pose estimation is the process of determining the position and orientation of the human body, in a 3D space using visual inputs. 
\citet{zhang2020mobipose} introduce MobiPose, a system designed to achieve efficient and accurate real-time multi-person pose estimation on mobile devices. MobiPose introduces a motion-vector-based approach that tracks human proposals across consecutive frames to eliminate the need for repeated human detection. It also introduces a mobile-friendly model employing lightweight, multi-stage feature extractions utilizing heterogeneous computing resources (CPU and GPU) to perform pose estimation in parallel, thereby minimizing latency.
Traditional 60Hz cameras have limited capabilities when it comes to tracking delicate finger movements due to their low sampling rate. Consequently, the performance of 3D hand pose reconstruction displays restricted accuracy. 
To address this issue, \citet{zhang2023rofin} have developed a 3D hand pose reconstruction method that utilizes the camera and wearable gloves embedded with LEDs on the fingertips and wrist. The camera captures the strip effect of the rolling shutter from the LEDs on the gloves, and a CNN identifies the location and bounding box of these strips. This information is then used to construct a 3D representation of the hand posture.

\begin{figure}[t]
\centering
\includegraphics[width=0.7\linewidth]{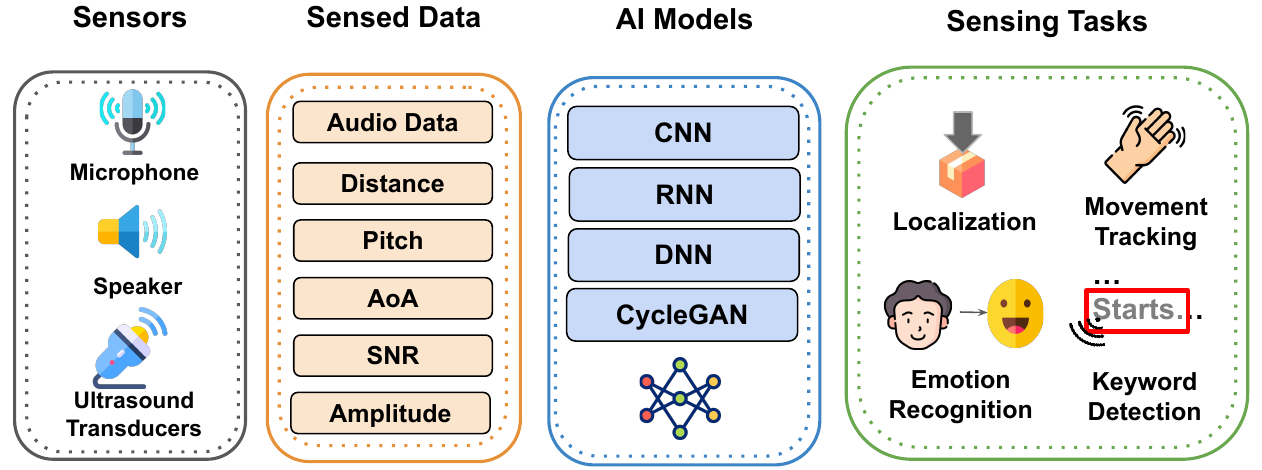}
\caption{Illustration of AI-empowered acoustic sensing pipeline.}
\label{fig:Pipeline of Acoustic Sensing}
\end{figure}

\subsection{Acoustic Sensing}
\label{subsec_acoustic_sensing}
Acoustic sensing involves utilizing acoustic sensors to capture, measure, and analyze acoustic signals for sensing purposes.
As summarized in Figure~\ref{fig:sensing-tree}, depending on the sensing tasks, existing works on AI-empowered acoustic sensing can be grouped into four categories: \mz{localization, movement tracking, emotion recognition, and keyword and event detection.}

\vspace{1.5mm}
\noindent
\textbf{Localization.}
Localization using acoustic sensing refers to the process of determining the position or location of objects or sources of sound using sound waves. 
\revision{\citet{DeepRange2020} introduce DeepRange, which investigates the limitations of traditional signal processing methods in localization tasks utilizing aquatic signals, particularly in scenarios with a low SNR environment. They pose the question of whether DNNs can automatically learn features from received acoustic signals to estimate distance, potentially surpassing the performance of conventional signal processing algorithms devised by domain experts. The study introduces a DNN-based ranging system, which directly employs raw acoustic signals without feature extraction and indicates superior performance compared to established signal processing approaches.}
Conventional methodologies for sound source localization require multiple microphone arrays, which is impractical for tiny devices. Addressing this, Owlet~\cite{garg2021owlet} place a microphone inside the stencil with sound holes. The incoming sound through these apertures indicates the direction based on hole patterns. Nevertheless, the approach remains susceptible to environmental factors, such as reflective wall signals. To mitigate this, the authors employed a CNN to estimate the Direction of Arrival (DoA), trained on a synthetic dataset representative of various environments.
\revision{\citet{DeepEar2024} introduce DeepEar, a DL-based framework to improve sound localization using only two microphones, particularly in scenarios with multiple sound sources. Drawing inspiration from the biological function of human ears, which shape sound waves to provide more spatial information, the authors design a neural network architecture that simulates human auditory processing. This includes a gammatone filterbank mimicking the cochlea's role by transforming audio into the time-frequency domain, followed by an autoencoder that extracts high-level sound representations. These features are then utilized by a deep neural network to pinpoint sound locations accurately. }

\vspace{1.5mm}
\noindent
\textbf{Movement Tracking.}
Movement tracking using acoustic sensing involves detecting and monitoring the movement of objects or individuals through the analysis of sound waves. 
Acoustic signals, as they propagate through the human body, undergo a range of transformations. By performing a comprehensive analysis of these signal changes, we can effectively track the movements.  %
\revision{\citet{sun2018vskin} introduce VSkin, a system that can detect finger movement on mobile devices using acoustic signals. VSkin utilizes both structure-borne and air-borne sounds to detect touch and measure finger movements on all surfaces of a device, not only limited to the touchscreen.}
The existing method of movement tracking frequently encounters challenges such as low SNR, interference, and mobility, which may affect the accuracy of the tracking. To overcome these issues, \citet{mao2019rnn} employ the 2D MUSIC algorithm~\cite{wax1984spatio} to produce joint of distance and Angle of Arrival (AoA) profiles derived from hand motion. Leveraging this profile, an RNN is utilized to precisely track both the distance and AoA of hand movements on a room-scale. %
\revision{\citet{yourtable2019} present Ipanel, a system that uses acoustic signals created by finger movements on a hardwood tabletop to extend mobile device interactions beyond the small screen and onto surrounding surfaces. Unlike traditional finger tracking systems that use a fixed frequency acoustic signal, Ipanel tracks the dynamically changing frequencies of acoustic signals produced when fingers slide on a surface. IPanel extracts distinctive features from both the spatio-temporal and frequency domain characteristics of the acoustic signals, converting them into images, which are then processed by a CNN for finger movement recognition. The system supports recognition of common gestures like clicks, flips, scrolls, and zooms, as well as handwriting recognition of numbers and alphabets with high accuracy.}
Acoustic signals can capture human breathing patterns, with a key advantage being the elimination of specialized wearable sensors. Leveraging this, \citet{xu2019breathlistener} designs a model to monitor the drivers through accurate breathing pattern extraction. The initial stage involves the isolation of environmental driving noise, which is then followed by the reconstruction of detailed breathing waveforms via the application of GAN. 
\revision{\citet{li2020fm} present FM-Track, a system for tracking multiple moving targets using acoustic signals without physical contact. The authors propose a chirp-based signal model that integrates range, velocity, and angle information from the reflected signals to accurately determine the position and movement of each target. FM-Track can track up to four targets simultaneously within a 3-meter range, demonstrating its efficacy through experiments on both smartphones and smart speakers.
}
\citet{fu2022svoice} employ ultrasound signals emitted from a smartphone to detect the articulatory movements of the mouth. Using the reflected signals from these movements, the study successfully reconstructs audible speech with a DNN named SiVoNet by training the network supervised way using paired audible speech.
\revision{They implemented a prototype for a comprehensive evaluation, using a Samsung Galaxy S8 to validate performance on a commercial smartphone platform. The evaluation results show that SiVoNet can reconstruct speech with a Character Error Rate (CER) as low as 7.62\%, outperforming state-of-the-art acoustic-based approaches.}
\revision{Experience~\cite{li2022experience} investigate the challenges and solutions related to the deployment of acoustic sensing system-based movement tracking in real-world scenarios. The authors identify several critical issues, such as audible sound leakage, high power consumption, and performance degradation due to device mobility. \citet{li2022experience} propose a power control mechanism by dynamically adjusting the transmission power and switching between idle and active states based on detected activity to reduce power consumption. They built a prototype of their proposed power control schemes for hand tracking on a Samsung S9+ smartphone, reducing average power consumption from 22\% to 10\% over two hours.
}

\vspace{1.5mm}
\noindent
\textbf{\mz{Emotion Recognition}.}
Emotion recognition through acoustic sensing involves analyzing voice and sound patterns to determine the emotional state of a speaker.
\citet{lane2015deepear} present DeepEar, a mobile audio sensing framework to perform audio inference tasks such as ambient scene analysis, emotion recognition, and stress detection. DeepEar is designed to address the challenge of diverse and noisy acoustic environments that mobile users encounter. The framework consists of multiple DNNs, each specialized in a specific audio sensing task, and employs advanced DL techniques for pre-training and fine-tuning.
\citet{georgiev2017low} address the challenge of performing multiple audio analysis tasks, including emotion recognition, on resource-constrained mobile and embedded devices. Existing solutions for audio sensing focus exclusively on the operation of a single DNN. However, \citet{georgiev2017low} have shown that by sharing layers among different audio task DNN models, it can reduce its computation cost while achieving comparable accuracy. 
Microphone variability, which refers to differences in audio data quality and characteristics recorded by different microphones, can significantly impact the robustness and accuracy of audio-sensing tasks. To address this challenge, \citet{mathur2019mic2mic} design Mic2Mic, which leverages Cycle-Consistent Generative Adversarial Networks (CycleGANs) to ensure that emotion recognition and other audio sensing tasks can be performed accurately across different devices. Mic2Mic learns a translation function between audio data recorded from different microphones, effectively reducing the domain shift caused by microphone variability.

\vspace{1.5mm}
\noindent
\textbf{\mz{Keyword and Event Detection}.}
Keyword detection in acoustic sensing involves the identification and recognition of specific words or phrases from audio signals.
Selecting the device with the best audio quality leads to clearer and more distinguishable audio features, which are critical for accurate keyword recognition. ~\citet{min2019closer} introduce a real-time assessment framework to determine the optimal audio input from various devices. This model routinely evaluates potential devices and selects the most suitable one for operation within the execution duty cycle. They introduced two models for this assessment: probability-based and data-driven DNN models. It demonstrates that it achieves higher accuracy while consuming less energy than its baseline counterparts in keyword detection tasks.
Event detection refers to the process of identifying and recognizing specific events or activities based on sound signals captured by acoustic sensors.
Traditional systems often miss parts of longer-duration events due to intermittent power, resulting in incomplete audio data. To mitigate these challenges, SoundSieve~\cite{monjur2023soundsieve} employ a regression neural network to predict the importance of upcoming audio segments and captures only the most relevant segments of an audio clip. With this predictive capability, the device can decide whether to enter a sleep cycle or remain awake to capture the signal.

\subsection{Multi-Modal Sensing}
\label{subsec_multimodal_sensing}
Multi-modal sensing involves the use of more than one sensing modality where the key advantage is its ability to combine distinct information provided by each of the included sensing modalities.
At the same time, determining which sensing modalities to include, and how to combine them effectively, are highly dependent on the specific application. 
As summarized in Figure~\ref{fig:sensing-tree}, depending on the sensing tasks, existing works on AI-empowered multi-modal sensing can be grouped into five categories: \mz{human activity recognition, human and object identification, tracking, localization, and speech enhancement.}

\begin{figure}[t]
\centering
\includegraphics[width=0.8\linewidth]{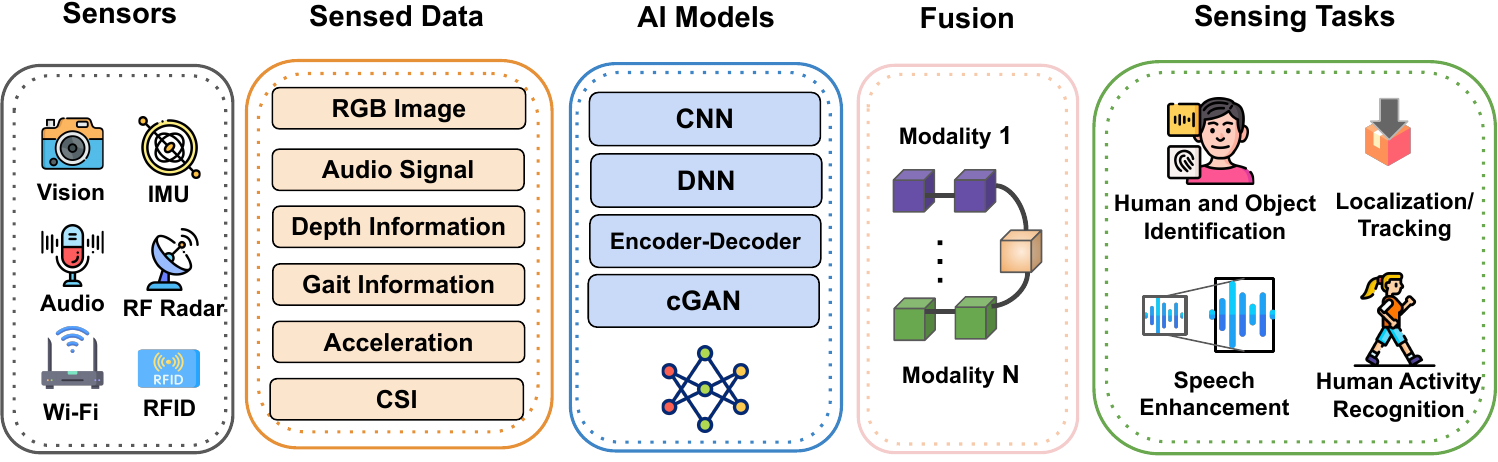}
\caption{Illustration of AI-empowered multi-modal sensing pipeline.}
\label{fig:Pipeline of Multi-Modal Sensing}
\vspace{-2mm}
\end{figure}

\vspace{1.5mm}
\noindent
\textbf{Human Activity Recognition.}
Human activity recognition (HAR) using multi-modal sensing integrates data from different sensory modalities to detect and identify human activities.
Traditional ML strategies typically employ one of two methods for sensor fusion: feature concatenation and Ensemble classifiers. Feature concatenation merges modalities but neglects inter-sensor correlation. Ensemble classifiers, on the other hand, uses separate classifiers but compromise intra-sensor correlation by fusing outputs later. \citet{radu2018multimodal} propose Modality-Specific Architecture that can learn both inter and intra-sensor correlation for the task of HAR. The network comprises multiple distinct branches, each dedicated to a specific modality. The outputs from these branches are then combined using fully connected layers. 
The task of HAR requires high accuracy with minimal inference latency. In multi-modal environments where sensors transmit data to a computing device, network fluctuations can cause asynchronous arrival of modalities. Straightforward approaches, such as waiting for delayed modalities or ignoring them, compromise both latency and accuracy. To address this challenge, \citet{li2021low} introduce speculative inference
Instead of waiting for delayed sensor data, it imputes the missing values and utilizes this generated data for subsequent inferences. If the accuracy falls below acceptable levels, the system executes a rollback of its results and re-initiates the inference process.
\citet{leite2021optimal} propose reducing the number of sensors used to lower computational demands, although this can potentially degrade accuracy. To mitigate this, the authors introduce a pipeline that prioritizes sensors based on their impact on accuracy. During the model training phase, sensors that have minimal or negative effects on accuracy are excluded. This approach significantly reduces memory usage and inference time while maintaining high accuracy in HAR.
In HAR, when a model trained in one domain is deployed in another domain, a degradation in performance occurs due to differences between the two domains. In multi-modal environments, these challenges are amplified due to the presence of additional variable factors. To tackle this issue, \citet{hu2022vma} propose  VMA, which transfers the DNN from one domain to another in the presence of multiple domains and modalities. The key idea is that changing one factor would have higher accuracy than changing multiple factors. Thus, VMA identifies pairs of domains wherein only one factor differs between them. Leveraging these pairs, it finds a path to effectively transition from one domain to the desired target domain by sequentially modifying only one factor at a time.
\citet{ouyang2022cosmo} propose Cosmo, a two-stage fusion learning system for enhancing HAR using multimodal data when labeled data are limited. In the initial stage, Cosmo leverages unlabeled data to discern consistent information, which denotes shared information that is uniformly present across different modalities. During the subsequent stage, Cosmo focuses on capturing complementary information, identifying the distinct and unique characteristics inherent to each modality, and leveraging the labeled data. As such, Cosmo achieve 26.73\% accuracy compared to the supervised fusion learning baseline.
Lastly, \citet{zhang2023cma} introduce CMA, a method for HAR by associating wearable IMU sensors with structural vibration signals. In CMA, all data is initially aligned and segmented according to the timestamp. Then, each data segment detects the activity in their data using a threshold. Lastly, the system utilizes a Temporal Convolutional Network to determine if the data segment sourced from distinct modalities points to an identical activity and individual.

\vspace{1.5mm}
\noindent
\textbf{Human and Object Identification.}
Human and object identification involves the ability to detect, recognize and categorize of individuals, objects or both.
One effective method for identifying individuals is analyzing their gait, as it constitutes a unique characteristic for each person.
While some studies utilize camera-based techniques for this purpose, they often struggle in low-light conditions and require the subject to be within the camera's field of view. 
Also, RF signals offer advantages like penetration through obstacles and not being affected by lighting conditions, but their accuracy may decrease when there is a significant difference between the training and testing environments.
To tackle this problem, \citet{korany2019xmodal} introduce XModal-ID, a gait-based identifying system using Wi-Fi signal and video footage. It determines whether a person within a Wi-Fi area is the same as the individual captured in the video footage. From the video, it creates the 3D mesh of a human and simulates how Wi-Fi signal would be after the signal is reflected from a 3D mesh human. 
This Wi-Fi signal implicitly contains gait information since it is reflected by the human body joint while moving. 
Thus, by comparing the simulated Wi-Fi data with the real-world Wi-Fi signal captured in the Wi-Fi area, it can identify whether the two sets of data correspond to the same individual or not. 
\citet{liu2021rfid} present an innovative system called RF-Camera, which combines RFID and computer vision techniques to recognize human interactions with physical objects in environments involving multiple subjects and objects. To achieve this goal, RF-Camera uses the Kinect DK system which is equipped with an RGB camera and depth camera to detect the human and its relevant hand trajectory. At the same time, an RFID system is used to identify the items and track their movement.
Current vision-based methods for video scene analysis excel at recognizing and identifying objects and people, i.e., extrinsic details. %
Nevertheless, they cannot be used when it comes to capturing intrinsic details, such as discerning the state of a washing machine. To bridge this gap, Capricorn~\cite{wang2022capricorn} integrates both RF and vision sensors, aiming to understand a scene's external and internal details comprehensively. Specifically, the camera provides data about types of objects and their respective bounding boxes. Concurrently, UWB radar detects object vibrations, leveraging this data to infer the internal states of these objects. 
In \cite{liu2022vi}, the authors propose Vi-Fi, which utilizes an RGB-D camera and smartphones to associate multiple individuals with their respective smartphone identifiers. It accomplishes this by capturing bounding box information and depth data from the RGB-D camera, as well as IMU sensor data and Wi-Fi Fine Timing Measurements (FTM) from smartphones. Subsequently, this diverse data is fed into LSTM models, and the output features are combined to generate an association score between the smartphones and their bounding boxes. Vi-Fi achieves an association accuracy of 81\% in real-time and 91\% in offline processing, demonstrating its effectiveness in identifying humans and objects in complex environments.
When it comes to object identification, capturing the material and shape information is of vital importance. mmWave signals can obtain rich information from the reflecting surfaces thanks to its broadband signals. However, the reflections from stationary objects contain less information than vibrating objects. To exploit its capability to the fullest, RFVibe~\cite{shanbhag2023contactless} fuses mmWave signals with acoustic signals for contactless material and object identification. Particularly, it plays an audio sound towards the object to generate micro-vibrations in the object and shines a millimeter wave radar signal on the object at the same time. By analyzing the physical properties of the reflected wireless signal, these micro-vibrations can be captured. RFVibe extracts several features, including frequency features, power features, and damping features. RFvibe adopts a CNN-based neural network to enable accurate identification of these features under different setups and locations. The neural network consists of three feature heads that transform features from different sources into a common latent space and a classification head that takes in the intermediate feature maps and outputs the probability distribution of possible classes. 

\vspace{1.5mm}
\noindent
\textbf{Tracking.}
Tracking humans or objects has been explored using various modalities. One method is utilizing the mmWave radar because it offers spatial information and the ability to construct data points in space. However, this sensor struggles in scenarios involving rapid movements. To address this limitation, \citet{lu2020milliego} introduce milliEgo, a robust egomotion estimation system that combines the capabilities of the IMU sensor and mmWave radar. To integrate the information from these two sensors, the authors proposed a two-stage intra- and inter-sensor cross-self attention mechanism, which interchangeably learns how to compensate for one another sensors during each step. Consequently, this approach outperforms its counterparts, which solely rely on the IMU sensor, combining RGB with IMU and integrating depth information with IMU.
Another combination of mmWave radar and IMU sensor has also been explored for tracking interpersonal distances by ImmTrack~\cite{dai2023interpersonal}. 
Since it requires tracking multiple individuals, IMU data from multiple individuals' smartphones and the corresponding mmWave data are generated. To associate them, cosine similarity metrics are employed. Once associated, the IMU data, initially in its local coordinate system, is translated to the mmWave's global coordinate system, making it suitable for monitoring interpersonal distances.

\vspace{1.5mm}
\noindent
\textbf{Localization.}
Multi-modal sensing can also enhance the performance of localization.
For example, \citet{boroushaki2021rfusion} introduce RFusion, a multi-modal localization system that utilizes both RF and vision sensing modalities. When estimating a location using a single RF antenna, there's a broad potential location area. Introducing an RGB-D camera can narrow down this area by leveraging depth information. Nevertheless, even with this refinement, multiple candidate locations remain, necessitating measurements from various positions. By optimizing this measurement trajectory through reinforcement learning, RFusion achieves centimeter-level accuracy, improving travel distance efficiency by twice as much compared to its baseline.
As another example, ELF-SLAM~\cite{luo2022indoor}
propose to combine both motion sensing and acoustic sensing for localization. IMU sensor inherently is susceptible to noise and biases that can accumulate over time. The authors propose to leverage the acoustic information emitted and captured by smartphones. As this acoustic data is reflected by surfaces, the captured echoes carry distinct spatial information based on their location. This enables precise indoor location alignment by compensating the inaccurate misaligned parts of the IMU sensor with spatial information in the acoustic data.

\vspace{1.5mm}
\noindent
\textbf{Speech Enhancement.}
Speech enhancement refers to the process of improving the quality and intelligibility of speech signals, typically in the presence of noise or other degrading factors.
Traditional research relying solely on audio data often requires multiple microphone arrays and is significantly influenced by the environment in which the data is captured. %
While multi-modal solutions exist that combine camera-captured lip movements with audio, their accuracy degrades in low-light conditions.
Consequently, \citet{sun2021ultrase} introduce UltraSE, a system that combines ultrasound signals with audible sound to enhance the user's speech. 
The user holds the phone near the mouth, and it emits ultrasound. Since this signal is reflected by the lip, it contains the articulation gestures that do not contain the noise of the audible sound. By fusing this noise-free ultrasound data with audio data in a cGan-based DNN, it produces de-noised audio output. 
However, this method has a limitation in terms of short working distances. Also, it has to hold the phone to capture the data. To address this issue, \citet{liu2021wavoice} design Wavoice, which aims to remove noise from the audio signal using mmWave that can operate long distances. They discovered a strong correlation between mmWave and audio signals, as both carry information about vocal fold vibrations, making them suitable for fusion. By integrating these two, the audio data offsets the motion interference inherent in mmWave signals, while the mmWave data counteracts the noise limitations of the audio signal. 
As a result, Wavoice surpasses its audio-only speech recognition baseline more than 20 times. 
Although Wavoice successfully separates the clear speech, it has the constraint that it needs the mmWave radar device. To address this issue, by narrowing the focus to scenarios involving head-mounted wearables like wireless earbuds or VR/AR headsets, VibVoice~\cite{he2023towards} leverage the IMU sensors that most of these devices are equipped with. IMU accelerometer attached to the head is capable of detecting vibrations generated by the speaker's voice via bone conduction through the skull, devoid of any external environmental noises. To integrate the IMU and audio modalities, they employ encoder-decoder architectures. The encoder extracts essential features from each modality and merges them while the decoder subsequently reconstructs human speech.

\vspace{-1mm}
\subsection{Earable Sensing}
\label{subsec_earables}
Earables are wearable devices attached to ears in the form of headphones or wireless earbuds.
As summarized in Figure~\ref{fig:sensing-tree}, depending on the sensing tasks, existing works on AI-empowered earable sensing can be grouped into three categories: \mz{facial expression sensing, user authentication, and sound localization}.

\vspace{1.5mm}
\noindent
\textbf{Facial Expression Sensing.}
Conventional methods for capturing facial expressions are primarily counted on video cameras. However, video cameras are limited in low-light environments and pose substantial risks of privacy infringement. In contrast, earables avoid such limitations and have demonstrated significant promise for a variety of facial expression sensing tasks.
For example, \citet{wu2021bioface} propose BioFace-3D, which leverages EMG and EOG signals captured by earables to detect the facial muscle activities, track 2D landmarks, and perform continuous 3D
facial reconstruction using a CNN. 
As another example, \citet{song2022facelistener} propose FaceListener, which uses the commodity headphone to recognize a user's facial expressions.
FaceListener emits ultrasound signals to detect face movements and uses this information to create a facial landmark model and recognize facial expressions based on an LSTM model.

\vspace{1.5mm}
\noindent
\textbf{User Authentication.}
Earables have also been utilized to identify unique individual characteristics, such as a person's gait for the purpose of user authentication. For gait-based user authentication, traditional methods often require special equipment, which is cost-prohibitive and limited in range. In contrast, \citet{ferlini2021eargate} propose EarGate, which employs an in-ear microphone to capture bone-conducted sounds induced by walking to detect the user’s gait for user identification. Furthermore, they demonstrate that classification performance can be notably improved through transfer learning.
\revision{\citet{MandiPass2021} introduce MandiPass, a biometric-based authentication system that utilizes intracorporal biometric called MandiblePrint, derived from the vibrations of human mandibles. It uses an Inertial Measurement Unit (IMU) embedded in an earphone to capture the MandiblePrint when a user voices some specific sound. This sound generates vibrations in the throat that propagate through the mandible to the ear, where they are sensed by the IMU. 
MandiPass validates the feasibility of MandiblePrint through theoretical modeling and experimental vibration propagation, demonstrating its potential as a user authentication method.}

\begin{figure}[t]
\centering
\includegraphics[width=0.8\linewidth]{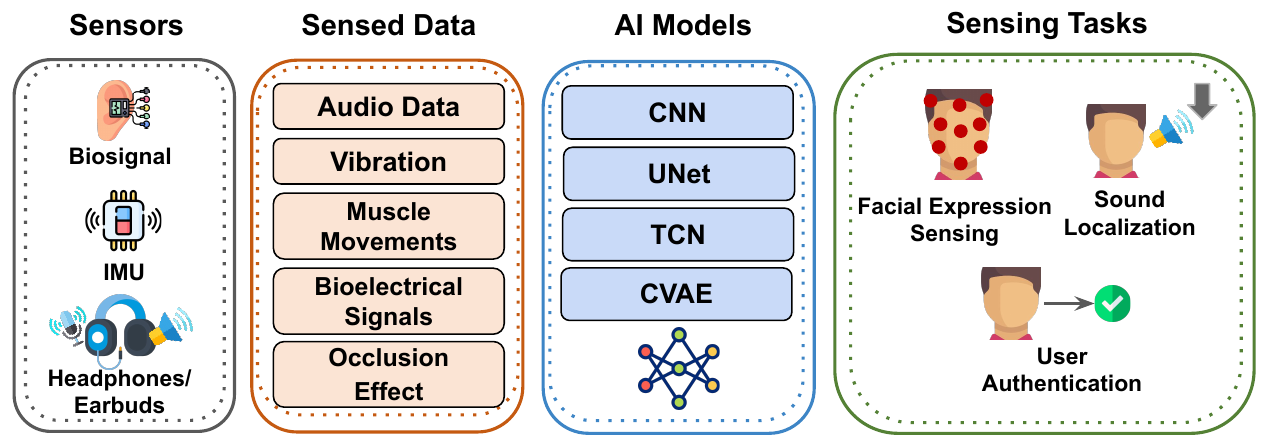}
\caption{Illustration of AI-empowered Earable sensing pipeline.}
\label{fig:Pipeline of Earable Sensing}
\vspace{-0mm}
\end{figure}

\vspace{1.5mm}
\noindent
\textbf{\mz{Sound Localization}.}
Sound localization in earable sensing refers to the ability of ear-worn devices to determine the direction of incoming sound sources. It is essential for enhancing spatial awareness and improving user experience in hearing aids, augmented reality, and personal assistants.
~\citet{chatterjee2022clearbuds} emphasize the importance of sound localization in enhancing user experience, particularly in distinguishing between the target speaker and background noise.
The authors use binaural wireless earbuds and dual-channel neural networks to separate the target voice from the noises. These networks consist of a time domain network called CB-Conv-TasNet and a frequency-based network called CB-UNet to exploit both spatial and acoustic information. As a result, it achieves a better scale-invariant signal-to-distortion ratio (SI-SDR) than AirPods Pro, which is based on beamforming.
Another critical task in sound localization is individualizing the Head-Related Transfer Function (HRTF). This individualization typically demands extensive and cost-intensive measurements in an anechoic chamber. To address this issue, \citet{zandi2022individualizing} introduce a simplified approach for conducting these measurements and propose to use a conditional variational autoencoder to achieve  HRTF individualization. 
Lastly, \revision{\citet{DeepEar2024} introduce DeepEar to address the issue of sound localization with two microphones. Unlike traditional methods that rely on extensive microphone arrays, DeepEar employs binaural microphones, which are more compact and thus more suitable for integration into devices like hearing aids. DeepEar leverages a multisector-based neural network that divides space into sectors for detecting multiple sound sources simultaneously.}

\vspace{-1mm}
\subsection{Generative AI for Sensing}
\label{subsec_genai_sensing}
Advancements in Generative AI have provided AIoT with opportunities to leverage state-of-the-art generative models such as Large Language Models (LLMs) to perceive, interpret, and present IoT sensor data in ways that are not attainable before \cite{wang2024iot}. 
Generative AI can correlate sensor readings with relevant contextual information, such as historical data, environmental conditions, and operational status so as to provide deeper insights into the sensor data and make decisions;
it can improve user experiences by allowing non-technical users to interact with sensor systems and perform data querying using natural language; it can also help translate raw sensor data into human-understandable reports and summaries, making it easier for users to understand key information contained inside sensor data.

Some efforts have been made to leverage such unique capabilities of Generative AI for sensing.
For example, \citet{ouyang2024llmsense} propose LLMSense, a prompting framework for LLMs to make sense of raw sensor data and low-level perception results. This framework can be implemented in an edge-cloud system, with small LLMs running on edge devices to summarize sensor data and high-level reasoning performed on the cloud to ensure data privacy. 
Two approaches are proposed to improve the performance of LLMSense: summarizing sensor data before reasoning and selectively including historical sensor data. Results show that LLMSense achieves high accuracy in tasks such as dementia diagnosis using behavior data and occupancy tracking with environmental sensor data.
In \cite{Xu2024Penetrative}, the authors propose Penetrative AI to explore how LLMs can be extended to interact with the physical world using IoT sensors and actuators. As a prompting framework, Penetrative AI shows how carefully constructed prompts can harness LLMs' embedded world knowledge for tasks such as user activity sensing and heartbeat detection. Specifically, Penetrative AI operates on two levels: textualized signal processing, where sensor data are converted into text for LLM analysis; and digitized signal processing, where LLMs directly interpret sensor data. Using heartbeat detection as an example, Penetrative AI demonstrates that LLMs can effectively analyze real-world sensor data with proper guidance, illustrating the potential of integrating LLMs into cyber-physical systems to enhance their intelligence and functionality.
Lastly, \citet{Wan2024Mar} go one step further beyond prompting and propose a multimodal LLM named MEIT that translates raw ECG sensor data into human-understandable reports.
For cardiologists, the task of interpreting ECG data and writing reports can be both intricate and time-consuming. 
MEIT aims to fill this gap by automating the ECG report generation task.
Specifically, MEIT involves instruction tuning a multimodal LLM to integrate raw ECG data with corresponding textual instructions, ensuring that the generated reports are clinically relevant and accurate. Experimental results demonstrate the superior performance of MEIT in generating accurate and professional ECG reports, underscoring its potential for real-world clinical applications.

\section{Computing} 
\label{computing}
\vspace{0mm}

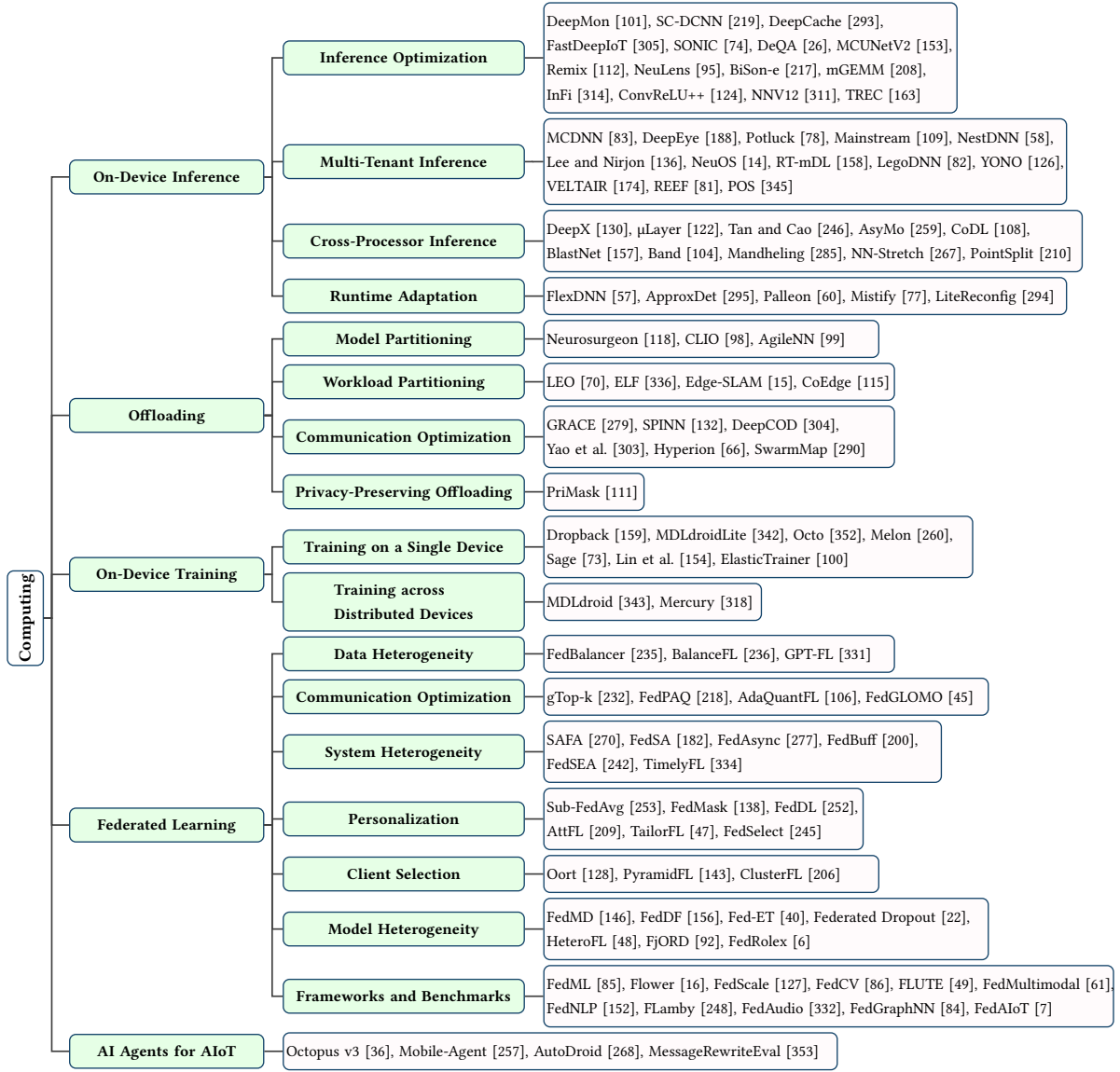
\begin{figure*}[t!]
    \centering
    \resizebox{\textwidth}{!}{
        \begin{forest}
            forked edges,
            for tree={
                grow=east,
                reversed=true,
                anchor=base west,
                parent anchor=east,
                child anchor=west,
                base=center,
                font=\large,
                rectangle,
                draw=hidden-draw,
                rounded corners,
                align=left,
                text centered,
                minimum width=4em,
                edge+={darkgray, line width=1pt},
                s sep=3pt,
                inner xsep=2pt,
                inner ysep=3pt,
                line width=0.8pt,
                ver/.style={rotate=90, child anchor=north, parent anchor=south, anchor=center},
            },
            where level=1{text width=12em,font=\normalsize,}{},
            where level=2{text width=15em,font=\normalsize,}{},
            where level=3{text width=17em,font=\normalsize,}{},
            [
                \textbf{Computing}, ver
                        [
                            \textbf{On-Device Inference}, fill=green!10
                            [
                                \textbf{Inference Optimization}, fill=green!10
                                [
                                DeepMon~\cite{huynh2017deepmon}{,}
                                SC-DCNN~\cite{ren2017sc}{,}
                        DeepCache~\cite{xu2018deepcache}{,} \\
                        FastDeepIoT~\cite{yao2018fastdeepiot}{,}
                        SONIC~\cite{gobieski2019intelligence}{,}
                        DeQA~\cite{cao2019deqa}{,}
                        MCUNetV2~\cite{lin2021mcunetv2}{,}\\
                        Remix~\cite{jiang2021flexible}{,}
                        NeuLens~\cite{hou2022neulens}{,}
                        BiSon-e~\cite{reggiani2022bison}{,}
                        mGEMM~\cite{park2022mgemm}{,}\\
                        InFi~\cite{yuan2022infi}{,}
                        ConvReLU++~\cite{kong2023convrelu++}{,}
                        NNV12~\cite{yi2023boosting}{,}
                        TREC~\cite{liu2023space}, leaf, text width=26em
                                ]
                                ]
                            [
                            \textbf{Multi-Tenant Inference} \\ , fill=green!10
                            [
                        MCDNN~\cite{han2016mcdnn}{,}
                        DeepEye~\cite{mathur2017deepeye}{,}
                        Potluck~\cite{guo2018potluck}{,}
                        Mainstream~\cite{jiang2018mainstream}{,}
                        NestDNN~\cite{fang2018nestdnn}{,}\\
                        \citet{lee2020fast}{,}
                        NeuOS~\cite{bateni2020neuos}{,}
                        RT-mDL~\cite{ling2021rt}{,}
                        LegoDNN~\cite{han2021legodnn}{,}
                        YONO~\cite{kwon2022yono}{,}\\
                        VELTAIR~\cite{liu2022veltair}{,}
                        REEF~\cite{han2022microsecond}{,}
                        POS~\cite{zhang2023pos}, leaf, text width=33em
                                ]
                            ]
                            [
                            \textbf{Cross-Processor Inference} \\ , fill=green!10
                            [
                        DeepX~\cite{lane2016deepx}{,}
                        \textmu\kern0pt Layer~\cite{kim2019mulayer}{,}
                        ~\citet{tan2021efficient}{,}
                        AsyMo~\cite{wang2021asymo}{,}
                        CoDL~\cite{jia2022codl}{,}\\
                        BlastNet~\cite{ling2022blastnet}{,}
                        Band~\cite{jeong2022band}{,} 
                        Mandheling~\cite{xu2022mandheling}{,} 
                        NN-Stretch~\cite{wei2023nn}{,}
                        PointSplit~\cite{park2023pointsplit},leaf, text width=34em
                                ]
                            ]
                            [
                            \textbf{Runtime Adaptation} \\ , fill=green!10
                            [
                        FlexDNN~\cite{fang2020flexdnn}{,}
                        ApproxDet~\cite{xu2020approxdet}{,}
                        Palleon~\cite{feng2021palleon}{,}
                        Mistify~\cite{guo2021mistify}{,}
                        LiteReconfig~\cite{xu2022litereconfig}, leaf, text width=33em
                            ]
                            ]
                        ]
                        [
                            \textbf{Offloading}, fill=green!10
                            [
                            \textbf{Model Partitioning}, fill=green!10
                            [
                            Neurosurgeon~\cite{kang2017neurosurgeon}{,}
                        CLIO~\cite{huang2020clio}{,}
                        AgileNN~\cite{huang2022real}, leaf, text width=21em
                            ]
                            ]
                            [
                            \textbf{Workload Partitioning}, fill=green!10
                            [
                            LEO~\cite{georgiev2016leo}{,}
                            ELF~\cite{zhang2021elf}{,}
                            Edge-SLAM~\cite{ben2022edge}{,}
                        CoEdge~\cite{jiang2023coedge}
                        , leaf, text width=22em
                            ]
                            ]
                            [
                            \textbf{Communication Optimization} 
                             , fill=green!10
                            [
                        GRACE~\cite{xie2019source}{,}
                        SPINN~\cite{laskaridis2020spinn}{,}
                        DeepCOD~\cite{yao2020deep}{,} \\
                        \citet{yao2021context}{,}
                        Hyperion~\cite{fu2022hyperion}{,}
                        SwarmMap~\cite{xu2022swarmmap}
                        , leaf, text width=22em
                            ]
                            ]
                            [
                            \textbf{Privacy-Preserving Offloading}, fill=green!10
                            [
                        PriMask~\cite{jiang_primask_2022}, leaf, text width=6em
                            ]
                            ]   
                        ]
                        [
                            \textbf{On-Device Training}, fill=green!10
                            [
                            \textbf{Training on a Single Device}, fill=green!10
                            [
                            Dropback~\cite{lis2019full}{,}
                        MDLdroidLite~\cite{zhang2020mdldroidlite}{,}
                        Octo~\cite{zhou2021octo}{,}
                        Melon~\cite{wang2022melon}{,}\\
                        Sage~\cite{gim2022memory}{,}
                        ~\citet{lin2022device}{,}
                    ElasticTrainer~\cite{huang2023elastictrainer}, leaf, text width=27em
                            ]
                            ]
                            [
                            \textbf{Training across } \\ \textbf{Distributed Devices} \\ , fill=green!10
                            [
                            MDLdroid~\cite{zhang2021mdldroid}{,}
                            Mercury~\cite{zeng2021mercury}, leaf, text width=13.5em
                            ]
                            ]   
                        ]
                        [
                            \textbf{Federated Learning}, fill=green!10
                            [
                            \textbf{Data Heterogeneity}, fill=green!10
                            [
                            FedBalancer~\cite{shin2022fedbalancer}{,}
                            BalanceFL~\cite{shuai2022balancefl}{,}
                            GPT-FL~\cite{zhang2023gpt}, leaf, text width=22em
                            ]
                            ]
                            [
                            \textbf{Communication Optimization} 
                             , fill=green!10
                            [
                            gTop-k~\cite{shi2019distributed}{,}
                            FedPAQ~\cite{reisizadeh2020fedpaq}{,}
                            AdaQuantFL~\cite{jhunjhunwala2021adaptive}{,}                            FedGLOMO~\cite{das2022faster}, leaf, text width=28em
                            ]
                            ]
                            [
                            \textbf{System Heterogeneity}, fill=green!10
                            [
                            SAFA~\cite{wu2020safa}{,}
                            FedSA~\cite{ma2021fedsa}{,}
                            FedAsync~\cite{jang2022asyncfl}{,}
                            FedBuff~\cite{nguyen2022federated}{,}\\
                            FedSEA~\cite{sun2022fedsea}{,}
                            TimelyFL~\cite{zhang2023timelyfl}, leaf, text width=25em
                            ]
                            ]
                            [
                            \textbf{Personalization}, fill=green!10
                            [
                             Sub-FedAvg~\cite{vahidian2021personalized}{,}
                             FedMask~\cite{li2021fedmask}{,} 
                             FedDL~\cite{tu2021feddl}{,}  \\ 
                             AttFL~\cite{park2023attfl}{,}
                        TailorFL~\cite{deng2022tailorfl}{,} FedSelect~\cite{tamirisa2023fedselect}                            , leaf, text width=20em
                            ]
                            ]
                            [
                            \textbf{Client Selection}, fill=green!10
                            [
                            Oort~\cite{lai2021oort}{,} 
                            PyramidFL~\cite{li2022pyramidfl}{,}
                            ClusterFL~\cite{ouyang2021clusterfl}, leaf, text width=21em
                            ]
                            ] 
                            [
                            \textbf{Model Heterogeneity}, fill=green!10
                            [
                            FedMD~\cite{li2019fedmd}{,}
                            FedDF~\cite{lin2020ensemble}{,}
                            Fed-ET~\cite{cho2022heterogeneous}{,}
                            Federated Dropout~\cite{caldas2018expanding}{,} \\
                        HeteroFL~\cite{diao2020heterofl}{,}  
                            FjORD~\cite{horvath2021fjord}{,}                            FedRolex~\cite{alam2022fedrolex}
                            , leaf, text width=28em
                            ]
                            ]
                            [
                            \textbf{Frameworks and Benchmarks}, fill=green!10
                            [
                            FedML~\cite{he2020fedml}{,} Flower~\cite{beutel2020flower}{,} 
FedScale~\cite{lai2022fedscale}{,}
FedCV~\cite{he2021fedcv}{,} 
FLUTE~\cite{dimitriadis2022flute}{,} FedMultimodal~\cite{feng2023fedmultimodal}{,}\\
FedNLP~\cite{lin2021fednlp}{,} FLamby~\cite{terrail2022flamby}{,} 
FedAudio~\cite{zhang2023fedaudio}{,} FedGraphNN~\cite{he2021fedgraphnn}{,} FedAIoT~\cite{alam2023fedaiot}, leaf, text width=36em
                            ]
                            ]
                        ]
                    [\textbf{AI Agents for AIoT}, fill=green!10 [ 
                    Octopus v3~\cite{chen2024octopusv3}{,}
                    Mobile-Agent~\cite{wang2024mobile}{,}
                    AutoDroid~\cite{AutoDroid2024Wen}{,}
                    MessageRewriteEval~\cite{zhu2023towards}
                    ,leaf, text width=35em]]
                    ]
        \end{forest}
}
    \caption{Summary of topics related to computing.}
    \label{fig:computing-tree}
\end{figure*}

\subsection{On-Device Inference}
\label{subsec_ondevice_inference}

One of the most fundamental and essential compute tasks of AIoT is to perform inferences on the device. On-device inference is particularly critical for latency-sensitive applications or scenarios where cloud connectivity is not available.
As summarized in Figure~\ref{fig:computing-tree}, existing works on on-device inference can be grouped into four categories: inference optimization, multi-tenant inference, cross-processor inference, and runtime adaptation.

\vspace{-1mm}
\subsubsection{Inference Optimization}
IoT devices are constrained in their onboard computing power, memory resources, and battery life. 
The objective of inference optimization is to enhance the computational and energy efficiency as well as to reduce memory demands and efficiently utilize memory resources during the inference process.
For example, \citet{huynh2017deepmon} propose DeepMon, an on-device inference framework that allows large DNNs to run on mobile devices at low latency for continuous vision applications. They propose a caching mechanism that exploits the similarities between consecutive images to cache intermediate processed data within CNN, which allows DeepMon to execute very deep models such as VGG-16 in near real-time.
\citet{ren2017sc} propose SC-DCNN, an optimization framework of stochastic computing (SC) for CNNs. They propose to apply SC to CNNs by designing function blocks and implementing hardware-oriented max-pooling in the SC domain. In addition, they propose to perform holistic optimizations for feature extraction blocks and weight storage schemes. By calculating multiplications and additions with AND gates and multiplexers in SC, SC-DCNN achieves a significant reduction in energy consumption.
\citet{xu2018deepcache} propose DeepCache, which adopts proven video compression techniques to systematically search for neighboring image blocks with similarities, rather than restricting matching solely to blocks in the same positions. They propose dividing video frames into regions, searching for similar regions in cached frames using a specialized matcher, and dynamically merging adjacent regions to maintain cache effectiveness.
In \cite{yao2018fastdeepiot}, the authors propose FastDeepIoT, 
which incorporates a profiling module and a compression steering module to optimize execution time and reduce energy consumption. The profiling module generates diverse training structures and builds an interpretable model for predicting the execution time, while the compression steering module enables existing DL compression algorithms to collaboratively minimize both execution time and energy consumption.
In SONIC \cite{gobieski2019intelligence}, the authors explore the opportunity of DNN inference intermittently on energy-harvesting systems. They propose loop continuation that significantly reduces the cost of ensuring accurate intermittent execution for DNN inference by modifying loop control variables within a loop nest, as opposed to dividing an extended loop into multiple tasks.
\citet{cao2019deqa} propose DeQA, a set of optimization techniques designed to enable Question Answering (QA) systems to run on mobile devices. DeQA reduces memory demands by loading partial indexes, dividing data into smaller units, and replacing in-memory lookups with a key-value database, altogether reducing the memory requirements of QA systems to just a few hundred megabytes.
\citet{lin2021mcunetv2} propose MCUNetV2, a scheduling technique in a patch-based manner to minimize memory usage for tiny DL. They propose initially executing the model on a limited spatial region, followed by the remainder of the network operating with a smaller peak memory consumption in the usual manner. Additionally, they propose to redistribute the receptive field to reduce the computation overhead caused by the patch-based initial stage.
\citet{jiang2021flexible} propose Remix, an adaptive image partitioning and selective execution strategy that involves the execution of existing DNNs on non-uniformly partitioned image blocks. They propose to leverage historical frames to learn the distribution of target objects and achieve higher detection accuracy with a given latency budget or higher inference speedup without accuracy deduction.
\citet{hou2022neulens} propose a dynamic inference mechanism known as the Assemble Region-Aware Convolution (ARAC) supernet, which removes redundant operations within CNN models by leveraging spatial redundancy and channel slicing. They propose to split the CNN inference flow into multiple micro-flows and load them into GPU as single models. In this way, NeuLens outperforms baseline methods in terms of latency reduction (up to 58\%) while achieving accuracy improvement (up to 67.9\%) within the same latency and memory constraints.
\citet{reggiani2022bison} propose BiSon-e, a RISC-V-based architecture that features a binary segmentation to enhance the CPU pipeline. They propose to perform Single Instruction Multiple Data (SIMD) operations on existing scalar Functional Units (FUs) to increase the performance of narrow integer applications on resource-constrained edge devices. In this way, BiSon-e achieves significant energy efficiency and execution time deduction.
To address the overload caused by the convolution layer, \citet{park2022mgemm} propose mGEMM, which expands the structure of the GEMM and eliminates the problems of memory overhead and low data reuse rate of the GEMM. They propose a reusable block of highly optimized computation on the inner computation kernel and partitioned the computation for the loops outside of the inner kernel.
In \cite{yuan2022infi}, the authors propose a learnable input filtering framework named InFi that unifies both approaches. They propose treating skip as a special case of reuse and designing a filter that supports both skip and reuse functions, requiring only maintaining an additional key-value table for reuse in the inference phase. In this way, InFi achieves lower energy consumption and latency.
\citet{kong2023convrelu++} propose a lossless acceleration method ConvReLU++, which achieves early negative result detection by employing reference-based upper-bound calculations. This approach guarantees that once intermediate results turn negative, the final results will be negative. When negative results are detected, the remaining computations can be skipped, leading to a significant latency reduction in ConvReLU++.
\citet{yi2023boosting} propose NNV12, an on-device framework that optimizes cold inference. They propose three optimization techniques encompassing kernel selection, weight transformation caching, and pipelined inference, to effectively reduce the latency of cold inference. In addition, they propose a heuristic-based kernel scheduling scheme, which fully harnessed three optimization techniques and led to substantial enhancements in the latency of cold inference.
Lastly, \citet{liu2023space} propose a set of optimization techniques for the Transient Redundancy Elimination-based Convolution (TREC), which recognizes and prevents redundant computations present in the form of identical tiles within input data or activation maps. They propose to repurpose parts of a matrix used in DNN computations as hashing vectors and embed a two-step stack for storing clustering IDs in TREC, aided by a reversed index for efficient entry location, which collaboratively eliminates significant memory overhead.

\vspace{-2mm}
\subsubsection{Multi-Tenant Inference}
Multi-tenant inference refers to the simultaneous execution of multiple distinct AI models, often originating from multiple concurrently running applications.
The key to multi-tenant inference is to efficiently manage and process inference requests from multiple tenants with limited resources on the device.
\noindent
\citet{han2016mcdnn} propose MCDNN, a framework for executing DNNs in video stream analytics using an approximation-based approach. They propose a heuristic scheduling algorithm designed to address approximate model scheduling, which allocates resources based on their usage frequency and utilizes a catalog to choose the most accurate model variant.
\citet{mathur2017deepeye} propose DeepEye, a small wearable camera running multiple models locally, enabling near real-time image analysis. They propose an inference pipeline that increased processor utilization by scheduling the execution of computation-heavy layers and the loading of memory-heavy layers across multiple models.
\revision{They also built prototype hardware powered by a quad-core Qualcomm Snapdragon 410 processor on a custom integrated carrier board to demonstrate the feasibility of their design.}
\citet{guo2018potluck} propose Potluck, which caches the previously computed results to provide cross-applications approximate deduplication. They propose a set of algorithms tuning the similarity threshold that regulates the degree to which various raw inputs are considered to be “the same”, which makes Potluck decreases processing latency for vision workloads.
\citet{jiang2018mainstream} propose Mainstream, a video processing system that addresses resource contention by sharing the same portion of DNN when inference is taken, which avoids redundant work. Additionally, they use an analytical model to estimate the effects of DNNs for an event and give the optimal model and sample rate option, resulting in significant overall event F1-score improvement.
In \cite{fang2018nestdnn}, the authors propose NestDNN, a framework that enables resource-aware on-device DL in multi-tenant settings. The key idea of NestDNN is to transform a DNN model into a multi-capacity model, where sub-models with smaller capacity are nested inside sub-models with larger capacity through shared parameters. At runtime, NestDNN incorporates a resource-aware scheduler which selects the optimal sub-model for each DNN model and allocates it the optimal amount of runtime resources so as to jointly maximize the overall performance of all the concurrently running applications.
\citet{lee2020fast} propose a concept of neural weight virtualization. Having each block of memory represent a block of weights for one or more DNNs makes it possible for multiple DNNs to be put into the main memory which has a smaller capacity than the total size of the DNNs. In this way, weight virtualization achieves significant improvement in execution time and energy efficiency.
\citet{bateni2020neuos} propose NeuOS, a latency-predictable framework for DNN-driven autonomous systems. They introduced the notion of a cohort, which represents a group of DNN instances capable of communication via a shared channel. They also propose a technique to predict the best system-level power configuration for each DNN of the cohort to meet the deadline for processing.
\citet{ling2021rt} propose RT-mDL, a framework that enables heterogeneous DL tasks to execute on edge devices by concurrently optimizing DNN model scaling and real-time scheduling. They propose a model scaling algorithm constrained by storage limitations that generates a range of model variants and overall optimizes the DL execution by identifying the optimal combination of task priorities and scaling levels of DL tasks.
\citet{han2021legodnn} explore a block-level scaling of DNNs, which only extracts and re-training descendant blocks from a complete DNN. Additionally, they employ a runtime scalar to determine the most effective combination of blocks to maximize accuracy. In this way, LegoDNN offers a wider range of model sizes without increasing time cost, resulting in significant improvement in accuracy and energy consumption reduction.
In \cite{kwon2022yono}, the authors propose YONO based on product quantization to compress heterogeneous models into two codebooks. Additionally, they enable in-memory model execution and support model switching for dissimilar multi-task learning on microcontrollers, achieving significant latency and energy consumption reduction.
\noindent
\citet{liu2022veltair} introduce VELTAIR, a scheduling approach that adapts its granularity to efficiently minimize scheduling conflicts. Additionally, they propose an adaptive compilation strategy that enables dynamic selection of programs with appropriate exclusive and shared resource usage patterns, aimed at mitigating overall performance degradation caused by interference.
In REEF~\cite{han2022microsecond}, the authors explore preemptive scheduling support for inference tasks on GPU. They propose a reset-based preemption mechanism that initiates a real-time kernel on the GPU through proactive termination and subsequent restoration of best-effort kernels.
\citet{zhang2023pos} propose POS, an operator-level scheduling framework combined with four operator-scheduling strategies. They propose abstracting the multi-model inference into a computation graph-based unified intermediate representation and finding optimal scheduling strategies for operators in the computation graph automatically with a learning-based operator-scheduling algorithm.

\vspace{-2mm}
\subsubsection{Cross-Processor Inference}
Cross-Processor Inference refers to the ability of a model to perform inference across different types of processors (i.e., CPUs, GPUs, TPUs) within a device. 
Modern IoT devices are often equipped with multiple heterogeneous processors, each of which is optimized for certain computing tasks. This provides a great opportunity to leverage these heterogeneous processing units to collaboratively perform inference in a cross-processor manner.
The realization of cross-processor inference involves a pivotal strategy: model partitioning. This technique capitalizes on the multiple processors to optimize inference tasks by partitioning the models and executing individual partitions on different processors.
For example, \citet{lane2016deepx} propose DeepX: a software accelerator for DL execution that allows any developer to use DL methods and automatically lowers resource usage.
They propose a deep architecture decomposition algorithm that can decompose models into unit blocks for heterogeneous local device processors, maximizing resource utilization.
In \cite{kim2019mulayer}, the authors propose \textmu\kern0pt Layer, a low latency on-device inference runtime that accelerates each layer by utilizing the onboard CPU and GPU simultaneously.
They propose channel-wise workload distribution to distribute the output channels of an NN layer to both CPU and GPU to fully utilize the resources, achieving a significant reduction in latency.
\citet{tan2021efficient} explore model partitioning between CPU and Neural Processing Units (NPUs). NPUs run DNN models faster but with less accuracy. Consequently, they propose heuristic-based algorithms and Machine Learning based Model Partition, which can explore a range of layer combinations to determine the part for CPU and NPU separately with optimal time-accuracy trade-off.
\citet{wang2021asymo} propose AsyMo, which focuses on partitioning the matrix multiplication blocks of DL models on asymmetric multiprocessors.
They propose cost-model-directed block partitioning and asymmetry-aware scheduling to balance the tasks. Additionally, they propose to set the frequency by offline profiling energy curves, which achieve more energy efficiency than baselines.
\citet{jia2022codl} propose CoDL, a concurrent DL inference framework that makes optimal use of diverse processors to expedite the execution at the operator level.
They propose to use hybrid-dimensional partitioning and operator chaining to reduce sharing-related overhead, and an accurate, lightweight method to predict latency by considering non-linearity and concurrency. In this way, CoDL achieves higher speedup and more energy saving compared with other methods.
\citet{ling2022blastnet} propose a model inference abstraction duo-block consisting of a CPU block and a GPU block. Such a duo-block is generated based on neural architecture search (NAS) techniques. They also propose a dynamic cross-processor scheduler that enhances the concurrent real-time DNN inference by optimizing CPU/GPU utilization.
Current mobile inference frameworks struggle to efficiently utilize diverse processors for multi-DNN workloads in applications due to a focus on a single DNN per processor, hampering performance and posing a challenge to serving multi-DNN tasks.
To address this issue, \citet{jeong2022band} propose Band, a mobile DNN runtime for scheduling multi-DNN requests based on a central component.
They propose using a model analyzer for model partitioning into subgraphs. A scheduler assigns subgraph-worker pairs, followed by execution of subgraphs on relevant processors by workers. In this way, Band outperforms TensorFlow Lite in terms of end-to-end performance.
\citet{xu2022mandheling} propose Mandheling, a system that leverages the benefits of Digital Signal Processors (DSP) in integer-based numerical computations during mixed precision training. They propose a co-scheduling technique between CPU and DSP to mitigate the overhead caused by DSP-unfriendly operators, which achieves latency improvement. In addition, they propose incorporating DSP compute subgraph reuse, self-adaptive rescaling, and batch splitting to collaboratively eliminate the preparation overhead on DSP.
\citet{wei2023nn} propose NN-Stretch, an automated model adaptation strategy that splits the DL model based on processor architecture traits.
They propose structure-preserved meeting point identification and capacity-guaranteed depth-width scaling. They also propose a sub-graph-based spatial scheduler for parallel inference across heterogeneous processors.
Another crucial component of cross-processor inference is distributing the workload across various processing units to minimize idle time.
\citet{park2023pointsplit} propose PointSplit, a 3D object detection framework for multi-accelerator edge devices. They propose a 2D semantics-aware biased sampling method to sample two complementary point sets and schedule them to be processed on GPU and NPU separately.

\begin{figure}[t]
\centering
\includegraphics[width=0.58\linewidth]{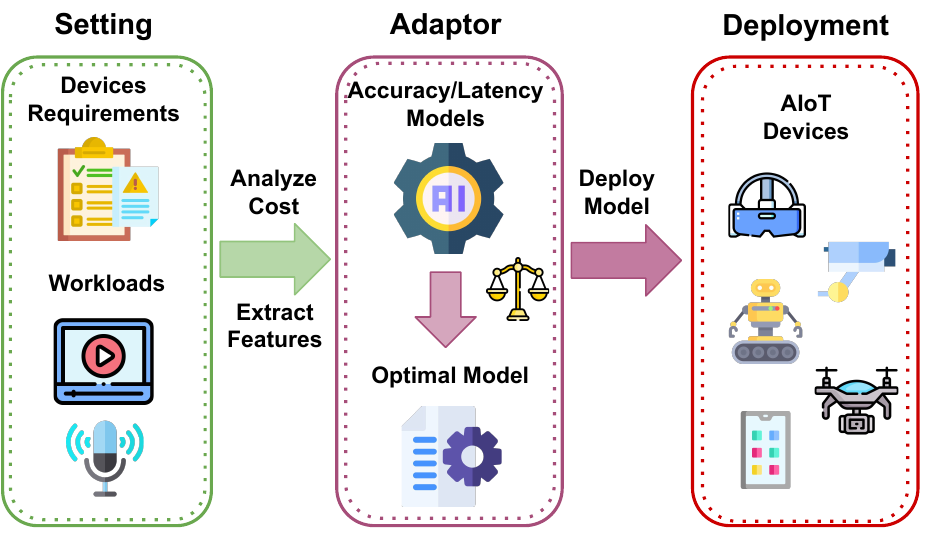}
\caption{Illustration of runtime adaptation pipeline.}
\label{fig:adaptation}
\vspace{-3mm}
\end{figure}

\vspace{-1mm}
\subsubsection{Runtime Adaptation}
Runtime adaptation in on-device inference refers to the ability of AI models to adjust and tailor their runtime behaviors in response to the changing available resources of the devices and evolving data inputs over time to deliver optimized system performance. 
For example, input images with contents that are easy to recognize do not need a large DNN model to process. Given that, in \cite{fang2020flexdnn}, the authors propose FlexDNN, an input-adaptive framework which leverages the early exit mechanism to construct
a single DNN model but dynamically adapts its model capacity to matching the difficulty levels of the input images at runtime.
In this way, FlexDNN is able to achieve a significant reduction in frame drop rate and energy consumption while maintaining accuracy.
\citet{xu2020approxdet} propose ApproxDet, a multi-branch framework employed to identify the optimal configuration branch for adaptive video object detection based on the characteristics of video content and available resources at runtime. They propose an accuracy and latency-driven scheduler to select the optimal execution branch for the specific user requirement, which achieves 52.9\% latency reduction with higher accuracy over YOLOv3 and lower switching overhead compared to other baselines.
\citet{feng2021palleon} propose Palleon, which dynamically selects an optimal DNN model by automatically detecting class skews. They propose a class-skew detector to generate precise class skew profiles and catch class skew switches. In addition, they propose Bayesian filter and separability-aware model selection techniques to improve accuracy and overall energy consumption.
\citet{guo2021mistify} propose Mistify, an intermediate layer that automates the process of porting a cloud-based model to a range of models optimized for edge devices across different points in the design space. They propose an architecture adaptor and a parameter-tuning coordinator, which collaboratively selects the optimal model that adapts to users' hardware profiles and performance targets.
Lastly, LiteReconfig proposed in \cite{xu2022litereconfig} consists of two components that collaborate as a scheduler to determine the execution branch to activate at runtime. The first component analyzes the cost and benefits associated with all potential features, and the scheduler selects which features to utilize for selecting the execution branch. The second component chooses the optimal execution branch within the execution kernel to adapt to different video contents and available resources.

\subsection{Offloading}
\label{subsec_offloading}
\mz{Given the limited memory and computing capacities of IoT devices, some of them may not be able to run the most efficient AI models by just using their own onboard resources. In such scenarios, it is necessary to offload the execution of part or even the whole model to nearby resourceful edges or the cloud.
As summarized in Figure~\ref{fig:computing-tree}, existing works on offloading can be grouped into four categories: model partitioning, workload partitioning, communication optimization, and privacy-preserving offloading.}

\vspace{1.5mm}

\begin{wrapfigure}{r}{0.33\textwidth}
\vspace{-4mm}
\includegraphics[width=0.32\textwidth]{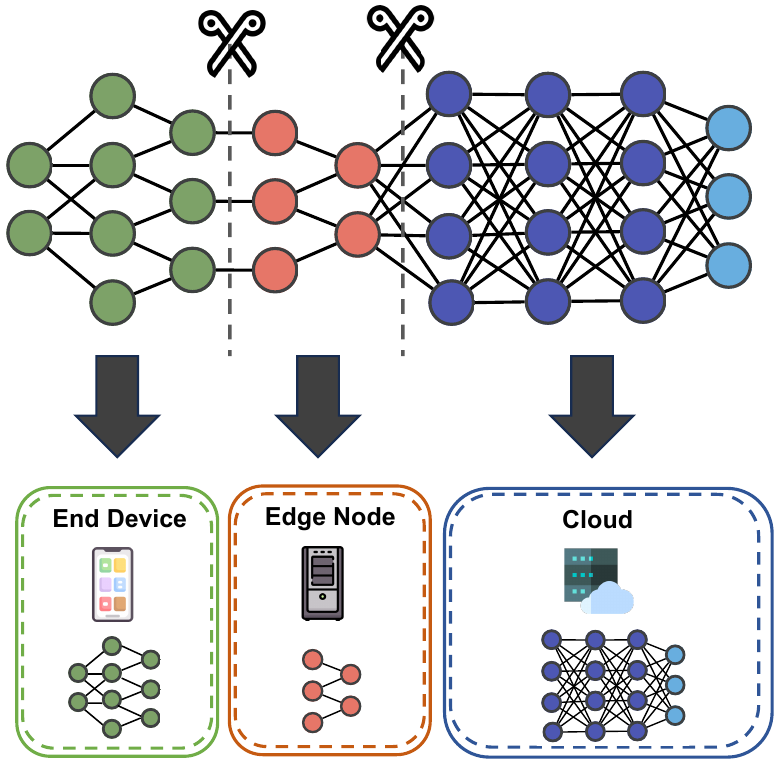}
\caption{Illustration of model partitioning.}
\vspace{-4mm}
\label{fig:model-partitioning}
\end{wrapfigure}

\noindent
\textbf{Model Partitioning.}
Model partitioning refers to the task of partitioning the AI model between the IoT devices and the nearby resourceful edge or cloud server such that different parts of the AI model are executed in a distributed manner. 
For example, \citet{kang2017neurosurgeon} propose Neurosurgeon, a framework that automatically partitions the DNN computation at the layer level. Neurosurgeon partitions the DNN into two parts for computation on mobile devices and the cloud, respectively, and trains a predictive model during the deployment phase to identify the optimal partition point of the model. In this way, Neurosurgeon achieves significant end-to-end inference latency and energy consumption reduction over cloud-only methods.
\citet{huang2020clio} propose CLIO, a framework enabling model compilation for extremely resource-constrained devices. They propose a novel technique for progressively partitioning models between the cloud and an end device, offering a variety of accuracy-bandwidth tradeoffs. This technique can be integrated with existing model compression and adaptive model partitioning techniques to achieve enhanced performance.
In \cite{huang2022real}, the authors propose AgileNN, an offloading technique that minimizes online computation and communication costs by putting a few valuable features computed locally and thus reducing the size of the local model.
They propose using eXplainable AI to estimate the most important features in the top k and retained by the local network to make a part prediction combined with the prediction by the remote network from other less important features for the final result.

\vspace{1.5mm}
\noindent
\textbf{Workload Partitioning.}
Workload partitioning refers to the distribution of workloads such as input data (e.g., images, SLAM map) and different DL models within the same processing pipeline across various edge devices and cloud servers to optimize performance, reduce latency, and improve resource utilization.
In \cite{georgiev2016leo}, the authors propose a sensing algorithm scheduler LEO that specializes in offloading workloads generated by sensor applications to heterogeneous processors. They propose to bring together critical ideas scattered in existing offloading solutions to maximize the performance without changing accuracy, and LEO runs as a service on LPU to perform both frequent and joint schedule optimization for concurrent pipelines, which also makes LEO more energy efficient compared with other baseline methods.
Current offloading solutions always assume the presence of a dedicated and robust server to which all inferences can be offloaded. However, it’s possible not to be able to find such a server in reality. To address this issue, \citet{zhang2021elf} propose ELF, a framework that accelerates mobile deep vision applications through parallel offloading, without being restricted to specific server provisioning. They propose a recurrent region proposal algorithm by predicting a new video frame’s region proposals based on the ones detected in previous frames, which achieve less latency compared with other baseline methods. Then, these predicted RPs are partitioned into “RP boxes” and offloaded to multiple servers, both partitioning and parallel processing make ELF achieve less resource demands.
\citet{ben2022edge} propose Edge-SLAM, a system that leverages edge computational resources to offload parts of Visual-SLAM. They propose to run the tracking module of VIsual-SLAM on mobile devices and move the left to nearby edge devices, which makes Edge-SLAM achieve significantly reduced latency. Additionally, they propose adding a partial global map as a fixed-size local map on the mobile device to achieve constant memory usage with minimal loss of accuracy in the final map.
Another line of research in workload partitioning involves dividing different DL models within the same processing pipeline across edge devices and cloud server.
In \cite{jiang2023coedge}, the authors propose CoEdge, a cooperative edge system for distributed real-time deep learning tasks.
They propose a hierarchical DL task scheduling framework integrated with global task dispatching and local batched real-time DL execution to maximize the utilization of edge resources. Additionally, a GPU-aware concurrent DL containerization method is proposed to furnish an isolated execution environment for every task. These techniques make CoEdge achieve less deadline missing rate and less end-to-end latency compared with other baseline methods.

\vspace{1.5mm}
\noindent
\textbf{Communication Optimization.}
Communication between IoT devices and the cloud is often conducted through wireless channels in which the bandwidth can be quite limited. To ensure a timely exchange of migrated workloads between IoT devices and the cloud while minimizing bandwidth usage and power consumption, efficient communication is crucial.
\citet{xie2019source} explore a DNN-aware compression algorithm measuring the perception model of a DNN to compress the input while maintaining inference accuracy.
They propose to use the gradient concerning the input to characterize the DNN’s perception. Using this estimated perceptual model, GRACE addresses a series of optimization challenges to ascertain the optimal codec parameters within the existing codec framework. In this way, GRACE achieves considerable compression ratio gains with little loss of accuracy.
In \cite{laskaridis2020spinn}, the authors propose SPINN, a synergistic progressive inference system that simultaneously employs an early-exit policy both in the cloud and locally. They propose an early-exit-aware cancellation mechanism that allows the interruption of the inference when having a confident early prediction evaluated by the wrapper of an intermediate classifier to provide robust operation under uncertain connectivity. Additionally, they propose a CNN-specific packing mechanism and an SLA- and condition-aware scheduler that make SPINN achieve higher throughput, higher accuracy, and less energy cost compared with other baseline methods.
\citet{yao2020deep} propose Deep Compressive Offloading, an asymmetric encoder/decoder framework that uses an efficient encoder on a local device while utilizing a relatively complex decoder on a server. In this way, most of the processing burden can be put on the server side and achieve a significant latency improvement. Additionally, they propose an effective system DeepCOD which incorporates a performance predictor and a runtime partition decision maker, which achieves higher speedup for inference.
In \cite{yao2021context}, the authors explored an edge-cloud training pipeline by harnessing parallel processing capabilities spanning both edge and cloud environments. They propose to apply scheduled feature replay and error-feedback compression, which fully utilize the computing capabilities available at the edge. Additionally, they offered a context-aware decision engine to adaptively organize parallel execution and compression, which keeps the overall latency low.
\citet{fu2022hyperion} propose Hyperion, a distributed mobile offloading framework that supports various applications and heterogeneous hardware. They propose a regularity-aware kernel analyzer to break down the tasks into smaller parts while ensuring that only the necessary data is transmitted, which makes Hyperion more efficient. Before scheduling, they propose a context-aware computing time predictor to predict the runtime duration of a given slice and a pipeline-enabled and network-adaptive scheduler to determine the optimal number of slices to be offloaded for each computational unit, both achieve superior speedup compared with the baseline.
However, as the number of agents increases, the operational overhead, which relies on a central node, also increases.
To address this issue, \citet{xu2022swarmmap} propose SwarmMap, a framework that scales up collaborative edge-based Visual-SLAM service.
They propose a change log-based map information tracker to achieve the minimum bandwidth consumption for map synchronization. Additionally, they propose a SLAM-specific task-aware scheduler that makes decisions based on the status of agents to minimize the procession time. Further, they propose a map backbone profiling technique to mitigate storage overhead without reducing accuracy.

\vspace{1.5mm}
\noindent
\textbf{Privacy-Preserving Offloading.}
IoT devices often collect personal data that may contain privacy-sensitive information.
In scenarios where data are also needed to offload along with the workloads to edge or cloud servers, it is imperative to ensure that this data is handled in a way that preserves the privacy of users.
PriMask \cite{jiang_primask_2022} introduce a small-scale neural network -- named MaskNet -- to mask the data before its transmission to the cloud. The data masked by MaskNet cannot be recovered by the cloud, thus preserving the privacy after offloading.
Moreover, each mobile device has its own unique MaskNet, which ensures that a privacy breach affecting the MaskNet of one device does not compromise the privacy of data on other devices.

\subsection{On-Device Training}
\label{subsec_ondevice_training}
\mz{Besides on-device inference, another fundamental and essential compute task of AIoT is on-device training.}
As summarized in Figure~\ref{fig:computing-tree}, existing works on on-device training can be grouped into two categories: training on a single device, and training across distributed devices.

\vspace{1.5mm}
\noindent
\textbf{Training on a Single Device.}
In the case of single-device training, 
the entire training process takes place on a single device.
To achieve effective training on a single device, existing efforts have mainly focused on the exploration of memory optimization.
For example, \citet{lis2019full} propose Dropback, which only trains a fraction of the weights who have the highest accumulated gradients while keeping the remaining weights not stored in memory, which significantly reduces the memory access cost.
\citet{zhang2020mdldroidlite} propose MDLdroidLite, a learning framework that transforms regular DNNs into resource-efficient models for on-device learning. They propose a Release-and-Inhibit Control (RIC) technique to wisely grow each layer independently from tiny to backbone, which avoids redundant resource overhead. In addition, they propose a RIC-adaption pipeline that transfers existing parameters to new-born parameters during growth. In this way, MDLdroidLite achieves 28X to 50X fewer model parameters compared with other baselines.
In \cite{zhou2021octo}, the authors propose Octo, a cross-platform system designed for lightweight on-device learning that leverages the fixed-point computational capabilities of embedded processors. They propose an INT8 training technique with loss-aware compensation and parameterized range clipping methods to efficiently apply quantization in forward pass and backward pass, respectively. In this way, Octo achieves higher training efficiency compared with other baselines.
\citet{wang2022melon} propose Melon, a memory-optimized on-device training framework that retrofits established recomputation and micro-batch techniques to fit into resource-constrained devices. They further propose a lifetime-aware memory pool to optimize memory utilization based on the characteristics of DNN training. In addition, they propose an on-the-fly memory adapting technique to quickly adjust to changes in the memory budget and resume execution using the partial results. In this way, Melon achieves higher training throughput with the same batch size.
In \cite{gim2022memory}, the authors propose Sage, an on-device training framework that incorporates memory-optimized techniques. They propose to separate differentiable operations from computable operations by employing a two-layer abstraction to represent a node in the computational graph, then Sage applies operator fusion and subgraph reduction to minimize the graph size. Additionally, they propose to dynamically adapt to the memory budgets by using gradient accumulation and checkpointing.
\citet{lin2022device} propose an on-device training framework with algorithm-system co-design. They propose a quantization-aware scaling technique to align the accuracy with the floating-point counterpart by automatically scaling the gradient with varying bit-precision. To save memory during backward computation, they propose a sparse update technique to skip the computation of less important layers and sub-tensors.
In \cite{huang2023elastictrainer}, the authors propose ElasticTrainer, a technique that can dynamically select the optimal trainable network portion at training time. They propose a tensor importance evaluator by leveraging the XAI technique to define the importance of a tensor in a specific epoch. On the other hand, they propose a tensor timing profiler to compute the backward pass timing of each tensor. Based on importance and time, they propose a tensor selector to select the optimal trainable network portion, which makes ElasticTrainer achieve higher training speedup with less energy consumption compared with baselines.

\vspace{1.5mm}
\noindent
\textbf{Training across Distributed Devices.}
In the case of training across distributed devices, DL models are trained collaboratively across a network of IoT devices where data on each device can be exchanged with other devices. In doing so, the collective computational power and data across the multiple devices can be leveraged to jointly train and update the DL models.
For example, \citet{zhang2021mdldroid} propose MDLdroid, a decentralized mobile DL training framework for mobile sensing applications. They propose a chain-directed synchronous stochastic gradient descent algorithm that dynamically aggregates and manages the model with one of the neighbors based on runtime resource status. Additionally, they propose a chain-scheduler, an agent-based multi-goal reinforcement learning technique, incorporating an accelerated reward function to effectively and equitably manage and allocate resources. In this way, MDLdroid achieves high training accuracy with low  overhead.
As another example, in \cite{zeng2021mercury}, the authors propose Mercury, an importance sampling-based on-device distributed training framework. The key principle behind the design of Mercury is that not all the data samples contribute equally to model training. Given that, in each training iteration, Mercury identifies and selects data samples that provide more important information. By focusing on those more important data samples, Mercury considerably enhances the training efficiency of each iteration. As a result, the total number of iterations and total training time is reduced.

\subsection{Federated Learning}
\label{subsec_federated_learning}
As data collected by IoT devices often contain privacy-sensitive information, federated learning (FL) emerges as a privacy-preserving approach that can train models across decentralized devices while keeping data on each device to preserve data privacy \cite{kairouz2021advances, wang2021field, fliotvision2022ieeeiotm}. Unlike fully on-device training, FL has the advantage of allowing information to be shared among devices, making it suitable for more complex applications that require more data volume.
Instead of gathering data from different devices into a central server for training, the model is disseminated to the participating devices in FL. These devices then conduct local training for a number of rounds and communicate only their model updates or gradients back to the central server for aggregation. The updated global model is subsequently broadcasted to the next set of participating devices for further training rounds \cite{mcmahan2017communication}. 
As summarized in Figure~\ref{fig:computing-tree}, existing works on FL for IoT can be grouped into seven categories: data heterogeneity, communication optimization, system heterogeneity, personalization, client Selection, model heterogeneity, as well as frameworks and benchmarks.

\vspace{1.5mm}
\noindent
\textbf{Data Heterogeneity.}
Unlike centralized training, data distributed across the devices participating in the FL process is generally non-IID (non-independent and identically distributed). Such data heterogeneity could make the local models overfit to local data, and aggregating these models could lead to convergence issues. 
\citet{shuai2022balancefl} propose BalanceFL, which scales the model weights making it behave as if it was trained on uniform distributed data. As such, it allows the global model to effectively learn both common and rare classes from a long-tailed real-world dataset, and thus mitigates the bias caused by data heterogeneity.
\citet{shin2022fedbalancer} propose FedBalancer, which uses a data selection strategy to select informative samples with adaptive deadline control. In doing so, the global model avoids overfitting caused due to data heterogeneity and makes convergence more stable. 
Lastly, \citet{zhang2023gpt} propose GPT-FL, which pre-trains the global model using synthetic data generated by generative models before fine-tuning with federated training. This makes the global model start from a more stable point instead of starting from scratch such that data heterogeneity does not strongly affect convergence.

\vspace{1.5mm}
\noindent
\textbf{\mz{Communication Optimization}.}
Communication between client devices and the central server in FL is often conducted through bandwidth-limited wireless networks. Therefore, reducing bandwidth usage between client devices and the central server can significantly enhance FL efficiency.
\citet{shi2019distributed} introduce gTop-k. Instead of accumulating the local top-k gradients from all the clients to update the model in each iteration, gTop-k chooses the global top-k gradients from a subset of clients, which considerably reduces the amount of gradients to communicate.
\citet{reisizadeh2020fedpaq} propose FedPAQ, which quantizes model updates to reduce their sizes before uploading to the server while the server only periodically averages the updates. The quantized updates and the periodic averaging on the server lead to lower communication costs.
Similarly, \citet{jhunjhunwala2021adaptive} propose an adaptive quantization scheme called AdaQuantFL, which achieves communication efficiency through quantization while maintaining a low error floor by changing the number of quantization levels during training.
Lastly, \citet{das2022faster} propose FedGLOMO to reduce the variance of local updates by global aggregation with momentum. This results in faster convergence and an overall lower number of communication rounds.

\vspace{1.5mm}
\noindent
\textbf{System Heterogeneity.}
\label{sec:comm_ov}
The participating devices in FL can be heterogeneous in their available on-device computing resources and network bandwidths. Such system heterogeneity would inevitably cause different participating devices to complete their local training at different times. Consequently, the slowest clients become the bottlenecks in the FL process.
One key technique to address system heterogeneity is the design of semi-asynchronous or asynchronous communication protocols. 
For example, 
\citet{wu2020safa} introduce SAFA, which uses a lag-tolerant model distribution algorithm and version-aware aggregation method based on a cache system. This decouples the global model broadcast and gradient upload process, making the system more tolerant of lagging clients.
\citet{ma2021fedsa} propose FedSA, which is a semi-asynchronous mechanism where the server aggregates a subset of local models by their arrival order in each round. The authors show that this approach improves convergence both theoretically and experimentally.
\citet{jang2022asyncfl} propose FedAsync, where the updates to the server and the broadcast to the clients are done asynchronously with a buffer. The updates from clients that are far behind the server schedule are deprioritized or excluded entirely. This avoids the destabilizing effects of stragglers and increases the number of communication rounds the system can complete within a time frame. 
\citet{nguyen2022federated} propose FedBuff, which also uses a buffered asynchronous aggregation scheme sending updates asynchronously but aggregating and broadcasting updates synchronously. This not only makes the system lag-tolerant, but also makes it compatible with Secure Aggregation and Differential Privacy.
\citet{sun2022fedsea} introduce FedSEA in which the authors design a scheduler that can efficiently predict the arriving time of local updates from devices and adjust the synchronization time point according to the devices' predicted arriving time. In doing so, it reduces the total number of straggling clients.
\citet{zhang2023timelyfl} propose an asynchronous FL framework named TimelyFL. The key idea of  TimelyFL is adaptive partial training, which allows each client to train part of the model based on the available resources of each client at runtime. In doing so, more clients are able to join in the global update without staleness.

\vspace{1.5mm}
\noindent
\textbf{Personalization.}
Besides training a global model, another use case of FL is to personalize the global model for participating clients such that the personalized model can better fit the needs of the end user.
For example, Sub-FedAvg~\cite{vahidian2021personalized} creates a personalized sub-network for each client from the global model by applying structured pruning on convolutional filters and unstructured pruning on fully connected layers.
\citet{li2021fedmask} propose FedMask where each device learns a sparse binary mask and applies the learned sparse binary mask to local models to  create personalized and sparse local models for each client. 
Instead of creating a personalized model for each user, \citet{tu2021feddl} propose FedDL, a clustering approach in which the client pool is grouped into several clusters, and one personalized model is assigned to each cluster.
Similarly, AttFL~\cite{park2023attfl}, designed for time series mobile and embedded sensor data, groups clients with similar contextual goals using cosine similarity, and redistributes updated personalized model parameters for improved inference performance at each local device.
\citet{deng2022tailorfl} propose TailorFL, a resource-aware and data-directed pruning strategy that makes each device's sub-model structure match its available resource and correlate with its local data distribution. 
Lastly, FedSelect~\cite{tamirisa2023fedselect} incrementally expands sub-networks to personalize client parameters, concurrently conducting global aggregations on the remaining parameters. This enables the personalization of both client parameters and sub-network structure during the training process.

\vspace{1.5mm}
\noindent
\textbf{Client Selection.}
In each round of FL, the central server selects a subset of clients to participate in the federated training process. The client selection strategy to determine which subset of clients to be included in each round plays a significant role in FL.
For example, \citet{lai2021oort} introduce Oort, a utility-based client selection scheme that takes both data and system utilities into account, where data utility is measured by the importance of model update and system utility is measured by the local training speed and the available network bandwidth for communication. By selecting clients with the highest utilities, Oort enhances both data and system efficiency and outperforms random client selection in terms of time-to-accuracy performance.
PyramidFL \cite{li2022pyramidfl} moves one step further and proposes to exploit data and system utilities within the selected clients to further enhance the time-to-accuracy performance of federated training. 
Lastly, \citet{ouyang2021clusterfl} introduce ClusterFL that minimizes the empirical training loss of multiple learned models while automatically capturing the intrinsic clustering relationship among the clients. This helps select and drop the clients with little correlation with others in each cluster, which speeds up the federated training process.

\vspace{1.5mm}
\noindent
\textbf{Model Heterogeneity.}
In standard FL, the participating clients and the central server collaboratively train the same model. However, imposing the same model on all devices would exclude low-end devices that do not have the enough memory. Moreover, state-of-the-art AI is increasingly reliant on large models, such as LLMs. Requiring the server and client models to be identical makes it impossible for standard FL to train such large models due to the resource limitations of client devices. Given that, model-heterogeneous FL was introduced to address this issue, allowing for the training of models with varying capacities across the server and clients.
One primary approach for model-heterogeneous FL is based on knowledge distillation (KD).
For example, \citet{li2019fedmd} propose FedMD, where clients train their own local models on a public dataset and upload their logit vectors to the server for KD. Since only logits are sent, clients' local models can have different architecture and sizes.
\citet{lin2020ensemble} propose FedDF to train the global model through ensemble distillation in which client models with different sizes and architectures are used as teachers. An unlabeled dataset is used and the predictions of the teacher models on that dataset are used to distill the global model.
Similarly, \citet{cho2022heterogeneous} propose Fed-ET in which models of different architectures and sizes are trained on clients' private data and then used to train a larger model at the server. However, Fed-ET uses weighted consensus distillation where the client updates are weighed based on a consensus function. This deprioritizes underperforming clients, resulting in higher accuracy.
The other primary approach for model-heterogeneous FL is based on partial training where different parts of the global model are extracted and disseminated to different clients for local training.
For instance, Federated Dropout~\cite{caldas2018expanding} propose to randomly extract sub-models of different sizes from the global model. Given the random nature, the sub-models extracted from the global model in each round can be different. During the update step, the server aggregates the sampled client updates with weighted averaging based on how many updates each part of the global model receives. 
Different from Federated Dropout~\cite{caldas2018expanding}, HeteroFL~\cite{diao2020heterofl} and FjORD~\cite{horvath2021fjord} propose static sub-model extraction schemes where the sub-models extracted from the global model in each round are always the same.
However, the key issue of static sub-model extraction schemes is that part of the global model cannot be trained on data across all the clients. This inevitably biases the global model training, especially data heterogeneity across the clients is high.
To address this key issue, \citet{alam2022fedrolex} propose FedRolex, which is a rolling sub-model extraction scheme that allocates sub-models of different sizes to clients and progressively rolls the sub-model extraction window across the entire global model. In doing so, all parts of the global model are evenly trained on the entire client data.

\vspace{1.5mm}
\noindent
\textbf{\mz{Frameworks and Benchmarks}.}
Frameworks and benchmarks play important roles in enabling FL on IoT devices. 
Popular FL frameworks include FedML~\cite{he2020fedml}, which implements a wide range of FL algorithms and datasets to facilitate developing and evaluating FL algorithms for a wide range of applications.
Flower~\cite{beutel2020flower} is another FL framework that is built on top of Ray~\cite{moritz2018ray} and is heavily customizable to different FL algorithms.
FedScale~\cite{lai2022fedscale} and FLUTE~\cite{dimitriadis2022flute} provide high-level APIs to implement, deploy, and evaluate FL algorithms at scale. 
These frameworks are, however, not specifically geared towards IoT devices.
In terms of benchmarks, existing FL benchmarks are predominantly conducted on datasets in domains of computer vision (FedCV \cite{he2021fedcv}), natural language processing (FedNLP \cite{lin2021fednlp}), medical imaging (FLamby \cite{terrail2022flamby}), speech and audio (FedAudio \cite{zhang2023fedaudio}), multimodal (FedMultimodal~\cite{feng2023fedmultimodal}), and graph neural networks (FedGraphNN \cite{he2021fedgraphnn}). 
These datasets, however, do not come from genuine IoT devices and therefore do not accurately reflect the distinctive characteristics of IoT data.
In contrast, in \cite{alam2023fedaiot}, the authors propose FedAIoT, an FL benchmark designed for IoT devices. FedAIoT includes eight datasets collected from IoT devices such as smartphones, smartwatches, Wi-Fi routers, drones, and smart home sensors. It also includes an FL framework customized for IoT, which supports IoT-friendly models and facilitates non-IID data partitioning, IoT-specific data preprocessing, quantized training, and noisy label emulation.

\vspace{-1mm}
\subsection{AI Agents for AIoT}
\label{subsec_genai_computing}
Traditional machine learning approaches focus on low-level basic recognition tasks. However, real-world applications can be complicated and require not only basic perception but also performing more complicated tasks such as making higher-level plans and decisions based on reasoning.
AI agents, powered by advanced generative AI models such as LLMs, can autonomously perform such complicated tasks, thereby significantly enhancing the capabilities of AIoT. 
Some efforts have been made to build AI agents for AIoT.
For example, %
multimodal input is crucial for developing AI agents for AIoT as IoT devices in general collect data from multiple sensing modalities such as language, vision, and audio.
\citet{chen2024octopusv3} introduce Octopus v3, a multimodal model with functional tokens tailored for AI agents, which supports both English and Chinese and operates efficiently on various edge devices such as Raspberry Pi. 
In \cite{wang2024mobile}, the authors introduce Mobile-Agent, an AI agent designed for mobile devices. Mobile-Agent can interpret user instructions to identify and locate elements on the mobile app's interface. It then autonomously plans and executes tasks, navigating apps step-by-step without requiring system-specific customization. 
\citet{AutoDroid2024Wen} introduce AutoDroid, a mobile task automation framework designed to handle arbitrary tasks on an Android application without manual intervention. AutoDroid combines the capabilities of LLMs with dynamic app analysis to manage unseen tasks. During the offline stage, it gathers app-specific knowledge by exploring UI relationships and creating simulated tasks. In the online stage, AutoDroid uses memory-augmented LLMs to guide the next actions and complete tasks based on these suggestions. Experimental results demonstrate that AutoDroid effectively automates tasks and outperforms existing training-based and LLM-based methods.
Lastly, as another line of research, text rewriting is a crucial feature of AI agents, as it can enhance communication by transforming informal or incorrect text into well-structured content. Despite advancements in LLMs for text summarization and rewriting, their large size and computation time make them challenging to use on mobile devices. Developing a smaller model with similar capabilities is also challenging due to the need to balance size and performance and the requirement for expensive data labeling. \citet{zhu2023towards} present MessageRewriteEval, a compact yet powerful language model for text rewriting tasks that operate efficiently on mobile devices. They present an innovative method for fine-tuning instructions for a mobile-centric text rewriting model, enabling high-quality training data generation without human labeling. 

\section{Networking \& Communication}
\label{4_Networking_and_Communication}

\begin{figure*}[t!]
    \centering
    \resizebox{0.95\textwidth}{!}{
        \begin{forest}
            forked edges,
            for tree={
                grow=east,
                reversed=true,
                anchor=base west,
                parent anchor=east,
                child anchor=west,
                base=center,
                font=\large,
                rectangle,
                draw=hidden-draw,
                rounded corners,
                align=left,
                text centered,
                minimum width=2em,
                edge+={darkgray, line width=1pt},
                s sep=3pt,
                inner xsep=20pt,
                inner ysep=1pt,
                line width=0.8pt,
                ver/.style={rotate=90, child anchor=north, parent anchor=south, anchor=center},
            },
            where level=1{text width=13em,font=\normalsize,}{},
            where level=2{text width=18em,font=\normalsize,}{},
            where level=3{text width=17em,font=\normalsize,}{},
            [
                \textbf{Networking \& Communication}, ver
                        [
                            \textbf{Cellular/Mobile Networks}, fill=blue!10
                            [
                            \textbf{Network Configuration}, fill=blue!10
                            [
                    Auric~\cite{mahimkar2021auric}{,}
                        FIRE~\cite{liu2021fire},leaf, text width=10em
                            ]
                            ]
                            [
                            \textbf{Resource Allocation} \\ , fill=blue!10
                            [
                            \citet{xiao2019depth}{,}
                            Microscope~\cite{zhang2020microscope}
                            , leaf, text width=15em
                            ]
                            ]
                            [
                            \textbf{Traffic Analysis} \\ , fill=blue!10
                            [
                            DMM~\cite{shen2020dmm} ,leaf, text width=5em
                            ]
                            ]
                            [
                            \textbf{Signal Generation} \\ , fill=blue!10
                            [
                            NeRF$^2$~\cite{zhao2023nerf2}{,}~RF-Diffusion~\cite{RFDiffusion2024Chi} ,leaf, text width=13em
                            ]
                            ]
                        ]
                        [
                            \textbf{Wi-Fi Networks}, fill=blue!10
                            [
                            \textbf{Coverage Estimation}, fill=blue!10
                            [
                            Supreme~\cite{li2020supreme},leaf, text width=6em
                            ]
                            ]
                            [
                            \textbf{Inference Cancellation}, fill=blue!10
                            [AiFi~\cite{chen2022aifi}, leaf, text width=3.5em
                            ]
                            ]   
                        ]
                        [
                            \textbf{Visible Light Communication}, fill=blue!10
                            [
                            \textbf{Optical Camera Communication}, fill=blue!10
                            [
                            CORE-Lens~\cite{liu2022core}{,}
                            WinkLink~\cite{xiao2024practical}
                            ,leaf, text width=15em
                            ]
                            ]
                            [
                            \textbf{Screen Camera Communication} 
                             , fill=blue!10
                            [
                            DeepLight~\cite{tran2021deeplight}
                            , leaf, text width=7em
                            ]
                            ]
                            [
                            \textbf{Passive-VLC}, fill=blue!10
                            [
                            U-star~\cite{zhang2022u}{,}
                            SpectraLux~\cite{ghiasi2023spectralux},leaf, text width=12.5em
                            ]
                            ]
                        ]
                        [
                            \textbf{LoRa/LoRaWAN}, fill=blue!10
                            [
                            \textbf{Link Estimation}, fill=blue!10
                            [
                            DeepLoRa~\cite{liu2021deeplora}{, }~\citet{ren2022lorawan}
                            ,leaf, text width=14em
                            ]
                            ]
                            [
                            \textbf{Modulation/Demodulation Enhancement}
                             , fill=blue!10
                            [
                            NELoRa~\cite{li2021nelora}{,}
                            LLDPC~\cite{yang2022lldpc}{,}\\
                            SRLoRa~\cite{du2023srlora}{,}
                            ChirpTransformer~\cite{ren2024chirp}, leaf, text width=16em
                            ]
                            ]
                        ]
                        [
                            \textbf{Other Networks}, fill=blue!10
                            [
                            ZiSense~\cite{zheng2014zisense}{,} 
                            \citet{shi2021adapting}{,}
                            DeepRadar~\cite{sarkar2021deepradar}{,}\\ 
                            DeepGANTT~\cite{perez2023deepgantt}{,} 
                            Sirius~\cite{garg2023sirius},leaf, text width=21em
                            ]
                            ]
                    ]
        \end{forest}
}
    \caption{Summary of topics related to networking \& communication.}
    \label{fig:networking-tree}
\end{figure*}
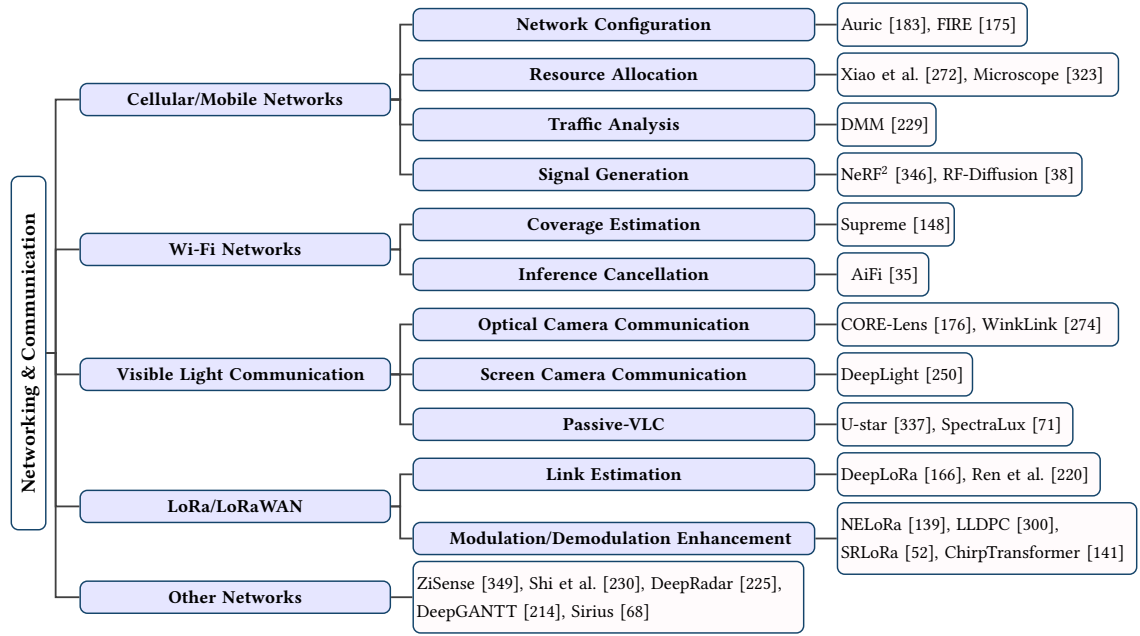

\subsection{Cellular/Mobile Networks}
\label{communication-sub-cellular}

\noindent As cellular networks evolve over many generations, they play an increasingly important role in providing mobile, reliable, and evolving communication. %
As summarized in Figure~\ref{fig:networking-tree}, existing works on AI-empowered cellular/mobile networks can be grouped into four categories: network configuration, resource allocation, traffic analysis, and signal generation.

\vspace{1.5mm}
\noindent
\textbf{Network Configuration.}
Cellular/mobile network parameters are typically manually configured based on rulebooks. Unfortunately, this process is time-consuming, error-prone, and difficult to maintain. AI-guided network configuration has been explored to provide a data-driven approach to improve network performance and service robustness.
For example, adding new carriers to accommodate increasing voice and data traffic can make cellular network configuration tasks very challenging.
To address this issue, \citet{mahimkar2021auric} propose Auric, which uses a series of carrier attributes as inputs to train a DL model and outputs network configuration based on geographical proximity. Experimental results show that Auric leads to 96\% accuracy across a large number of carriers and configuration parameters when evaluated on real-world LTE network data. 
As another line of research, \citet{liu2021fire} introduce FIRE, a system that employs a variant of variational autoencoders (VAE) for downlink channel estimation. In doing so, it eliminates the overhead of requesting feedback from client devices and improves the quality of FDD (Frequency Domain Duplexing) MIMO systems. Moreover, FIRE effectively supports MIMO transmissions in real-world settings, achieving an SNR enhancement of over $10$ dB compared to the state of the arts.

\vspace{1.5mm}
\noindent
\textbf{Resource Allocation.}
AI techniques can also enhance the performance of cellular/mobile networks by managing and distributing critical network resources based on various factors such as demand, network conditions, and service requirements in a data-driven manner.
For example, \citet{xiao2019depth} conduct an extensive measurement study on the ecosystem of mobile virtual network operators (MVNO). 
Based on the findings, the authors propose to leverage big data analytics and ML-based techniques to optimize an MVNO’s service such as predicting monthly data usage to optimize data plan reselling, customer churn profiling, and mitigation. 
As another example, \citet{zhang2020microscope} propose Microscope, a DL-based framework that decomposes per-service level resource demand based on spatio-temporal features hidden in traffic aggregates. In doing so, Microscope reduces relative demand estimation error to below 1.2\%, allowing cellular operators to allocate network resources more accurately. 

\vspace{1.5mm}
\noindent
\textbf{Traffic Analysis.}
Traffic analysis refers to the techniques to monitor, analyze, and optimize the flow of data across the network. AI-based traffic analysis can help in forecasting future traffic demands and making adjustments to enhance the overall network efficiency.
\citet{shen2020dmm} propose a fast map-matching system named DMM for cellular data. DMM utilizes a recurrent neural network (RNN) to determine the most probable road trajectory given a series of cell tower locations. To make DMM practically useful in real-world scenarios, DMM also incorporates a number of techniques such as spatial-aware representation of cell tower sequences, an encoder-decoder structure for variable input and output lengths, and a reinforcement learning-based model to optimize the matched results. 

\vspace{1.5mm}
\noindent
\textbf{Signal Generation.}
Lastly, 
the success of Generative AI in natural language processing and computer vision has sparked interests in \mz{using Generative AI in the domain of cellular/mobile networks}.
For example, NeRF$^2$~\cite{zhao2023nerf2} introduces a radio-frequency (RF) radiance field that uses a Neural Radiance Network to model a continuous volumetric scene function, which captures the propagation of RF signals in complex environments. The model trained with signal measurements and a physical model of ray tracing can generate synthetic RF datasets that can be adopted to boost the training of application-layer artificial neural networks (ANNs). Experimental results demonstrate the effectiveness of NeRF$^2$ in the fields of indoor localization and 5G MIMO. 
As another example, \citet{RFDiffusion2024Chi} present RF-Diffusion, a novel approach for generating high-quality time-series RF signals using a generative model. The method involves using time-frequency diffusion theory and a hierarchical diffusion transformer to generate high-quality synthetic RF signals by leveraging the unique characteristics of RF signals in both time and frequency domains.
\revision{RF-Diffusion demonstrates superior performance compared to other generative models including DDPM, DCGAN, and CVAE, achieving higher structural similarity and better SNRs.}

\subsection{Wi-Fi Networks}
\label{subsec_wifi_networks}

As summarized in Figure~\ref{fig:networking-tree},
existing works on AI-empowered Wi-Fi networks can be grouped into two categories: coverage estimation and interference cancellation.

\vspace{1.5mm}
\noindent
\textbf{Coverage Estimation.}
\noindent 
AI-empowered Wi-Fi coverage estimation aims to leverage AI algorithms to obtain the distribution and strength of Wi-Fi signals in a specific area with higher resolution.
For example, inspired by advancements in image super-resolution, \citet{li2020supreme} propose Supreme, which constructs fine-grained radio maps based on coarse-grained radio maps crowd-sourced across sites with a deep spatial-temporal reconstruction network consisting of 3D convolutions, spatial-temporal residual blocks, and reconstruction subnets. The authors have conducted experiments on a dataset consisting of six months of data collected from two university campuses. Experimental results show that Supreme outperforms state-of-the-art baselines based on coarse-grained radio maps and achieves lower localization error in a Wi-Fi fingerprint-based localization case study.%

\vspace{1.5mm}
\noindent
\textbf{Interference Cancellation.}
As the number of wireless devices increases, multiple devices may simultaneously transmit data within the same unlicensed Wi-Fi band. This can cause severe performance degradation. To ensure reliable communication, advanced interference cancellation techniques are needed.
\citet{chen2022aifi} introduce AiFi, an AI-empowered interference cancellation method for commodity Wi-Fi devices to estimate interference using knowledge gathered from the Wi-Fi receiver's physical layer without extra RF hardware. AiFi leverages the domain knowledge of Wi-Fi physical layer information including pilot information (PI) and channel state information (CSI) to guide the DL model design. Specifically, AiFi first extracts the interference features from Wi-Fi physical layer, estimates interference via an attention network using these features, and finally removes those interference from the received signal using a fully-connected network and an LSTM.
Experiments show that AiFi effectively boosts the MAC frame reception rate by $18 \times$ with a cancellation delay under 1ms per frame.

\subsection{Visible Light Communication}
\label{subsec_visible_light_communication}

\noindent Visible light communication (VLC) uses visible light as a data transmission medium to connect devices and communicate. VLC requires bit encoding using visible light sources, and light-sensitive sensors as receivers. 
As summarized in Figure~\ref{fig:networking-tree}, in VLC, AI has been used to improve the performance of optical-camera communication, screen-camera communication, and passive-VLC.

\vspace{1.5mm}
\noindent
\textbf{Optical-Camera Communication.}
Optical-camera communication (OCC) relies on LED lighting infrastructures as transmitters and cameras as receivers. The coded information is either transmitted directly from LED lights or reflected from the illuminated objects and is received by the camera. 
\citet{liu2022core} introduce CORE-Lens, which addresses the challenges posed by the mutual interference between OCC and object recognition (OR) in indoor environments. Traditional OCC systems often suffer from the entanglement of light patterns used for communication with the background, which degrades both OR accuracy and OCC decoding performance. CORE-Lens addresses these challenges by employing a disentangled representation learning (DRL) approach combined with GAN-based image reconstruction. Experimental results show that CORE-Lens achieves superiority in both visible light sensing and communications compared to conventional approaches. 
\citet{xiao2024practical} propose WinkLink, an OCC system that enables robust transmission behind complex backgrounds even under low SNR conditions. They design a two-stage DNN and a context-aware demodulation protocol to extract subtle signals in the lossy OCC channel. WinkLink is trained solely on a synthesized dataset yet generalizes well to unseen real-world backgrounds.%

\vspace{1.5mm}
\noindent
\textbf{Screen-Camera Communication.}
Screen-camera communication (SCC) encodes video content in a human-imperceptible manner on a screen as the light source, and uses cameras capturing images of such screen content work as receivers. 
Existing techniques on SCC often suffer from high decoding errors due to screen extraction inaccuracies and perceptible flickers on common refresh rate screens.
To address this issue, \citet{tran2021deeplight} present DeepLight, an innovative approach for SCC that addresses the challenges of decoding inaccuracies and perceptible screen flickers. For the bit encoder, DeepLight %
applies a Manchester coding strategy. For the decoder, Deeplight adopts the state-of-the-art deep object detection pipeline to extract the screen from a camera frame and then adopts a DNN-based model to decode spatially encoded bits in the frame simultaneously. %
Experimental results show that DeepLight is able to achieve high decoding accuracy (frame error rate < 0.2) and moderately high data throughput ($\geq 0.95$Kbps) at extended distances.%

\vspace{1.5mm}
\noindent
\textbf{Passive-VLC.}
Instead of relying on active light sources for data transmission, passive-VLC uses ambient light which can be modulated and then detected by a receiver to decode the transmitted information. Essentially, passive-VLC systems leverage changes in light intensity or other properties of ambient light to convey information.
\citet{zhang2022u} design U-star, a system consisting of passive Underwater Optical Identification (UOID) tags and DL-enabled camera-based tag readers, providing objects/human identification and location-based services as underwater navigation assistance in scenarios such as dive and rescue. %
U-star employs a three-dimensional multi-color cube-shaped design for the UOID tags %
and adopts the CycleGAN-based underwater denoising model that converts underwater UOID images into clear ones.
Experiments under different underwater scenarios show that U-star achieves a bit error rate of 0.003 at 1m and less than 0.05 at up to 3m, which is sufficient for guiding underwater navigation.
\citet{ghiasi2023spectralux} present SpectraLux, an  approach to transmit and decode data using low-power liquid crystal (LC) cells.
It utilizes the physical characteristics of LC shutters toggling between being translucent and opaque when switching the voltage from 0V to 5V, emitting different spectrums of the incident light. 
SpectraLux adopts a spectrometer that captures 256 bands of incoming light %
and achieves multi-symbol decoding by feeding PCA-reduced spectrum features to CNNs for classification. SpectraLux shows the potential of utilizing the wide spectrum of ambient light in passive-VLC.

\subsection{LoRa/LoRaWAN}
\label{subsec_lora}
\noindent LoRa (Long Range) is a rising low-power wide-area communication technology. LoRa's physical layer adopts the chirp spread spectrum (CSS) modulation which is known for its resistance to interference and capacity to travel long distances, making it particularly suitable for various IoT applications. LoRaWAN (Long Range Wide Area Network) refers to the protocol and system architecture for networks of LoRa nodes which is an open standard that ensures interoperability among different manufacturers and developers. 
As summarized in Figure~\ref{fig:networking-tree}, existing works on AI-empowered LoRa/LoRaWAN can be grouped into two categories: link estimation and modulation/demodulation enhancement.

\vspace{1.5mm}
\noindent
\textbf{Link Estimation.}
To study LoRa link coverage in the wild in supporting smarter LoRa deployments,
\citet{liu2021deeplora} propose DeepLoRa, a DL-based framework for LoRa path loss estimation of long-distance links in real-world environments. To do so, DeepLoRa extracts land-cover types along a LoRa link from multi-spectral remote sensing images, %
and exploits the order dependency of the land-cover sequence by utilizing Bi-LSTM (Bidirectional Long Short Term Memory) for path loss estimation. Experimental results on a real LoRaWAN dataset show that DeepLoRa is able to achieve less than 4dBm estimation error, which is $2\times$ smaller than state-of-the-art approaches.
\revision{Moreover, the study conducted in \cite{ren2022lorawan} further corroborates that DeepLoRa outperforms other link estimation approaches in terms of LoRa localization accuracy.}

\vspace{1.5mm}
\noindent
\textbf{Modulation/Demodulation Enhancement.}
Enhancements in LoRa modulation and demodulation are essential for improving the performance, efficiency, and reliability of data transmission in LoRa systems.
For example, \citet{li2021nelora} present NELoRa, a neural-enhanced LoRa demodulation framework that takes advantage of the powerful feature learning capability of DL to enable LoRa communication under ultra-low SNR. %
The key idea of NELoRa is the dual-DNN design: the first DNN is used as a noise filter to extract clean chirp symbols from the noisy LoRa packets, and the second DNN is used as a decoder that decodes the extracted clean chirp symbols.
Experimental results show that NELoRa outperforms the standard LoRa demodulation method under a wide range of LoRa configurations in both indoor and outdoor deployments.
\citet{yang2022lldpc} propose LLDPC, %
which enables low-density parity-check (LDPC) coding in LoRa networks under the inspiration of the wide usage of LDPC coding in other wireless networks. %
LDPC requires the Log-likelihood Ratio (LLR) for decoding which %
is not applicable to the CSS modulation adopted by LoRa. Moreover, the mainstream decoding algorithms for LDPC need multiple iterations to achieve effective error correction, resulting in long decoding latency that exceeds the maximum ACK time of the LoRa gateway. To tackle these challenges, LLDPC extracts LLR %
by treating CSS demodulation as a classification task and outputs the probability of all possible decoding results. %
It further utilizes a Graph Neural Networks (GNN) for fast belief propagation to achieve efficient LDPC decoding. %
\citet{du2023srlora} propose SRLoRa, which decodes LoRa signals by leveraging spatial diversity from multiple gateways. Specifically, SRLoRa employs CNN-based interleaving denoising layers to extract features under ultra-low SNR and consolidates features from different gateways in the merging layers. The merged signals with accumulated energy are then fed to a CNN decoder for decoding.
Lastly, \citet{ren2024chirp} further establish an encoding framework, providing four features including on-air time, selective initial frequency, chirp repeating, and symbol hopping, to combat various challenges of weak signals, signal collisions, and environment dynamics. On the decoder side, the neural-enhanced decoder is adopted and optimized for decoding the symbols with symbol hopping based encoding in terms of input and parameter sizes.

\subsection{\mz{Other Networks}}
\label{subsec_other_networks}
\noindent 
Besides the wireless networks mentioned above, AI has also been applied to various other types of networks for diverse objectives.
For instance, ZiSense~\cite{zheng2014zisense} is proposed to enhance the energy efficiency of sensor nodes in co-existence environments by using a sequence of received signal strength (RSS) values to predict the presence of ZigBee signals through a decision tree model.
\citet{shi2021adapting} propose to improve the configuration of wireless mesh networks (WMN) by DL-based domain adaption that adapts models for network configuration prediction trained on simulation to its corresponding physical network. %
In particular, the authors develop a teacher-student neural network that learns robust configuration prediction models from large-scale inexpensive simulation data with minor physical measurements to close the simulation-to-reality gap.  
\citet{perez2023deepgantt} present DeepGANTT, a DL-based scheduler that leverages GNN to provide a near-optimal solution for the NP-hard carrier scheduling problem in RFID backscatter networks. In those networks, battery-free RFID tags harvest energy from excitation in the environment, and IoT devices equipped with RFID readers provide them with the carrier for communication. 
To avoid collisions, DeepGANTT trains a carrier scheduler based on GNN to handle and capture the interdependence of nodes in the irregular network topology graphs. 
DeepGANTT breaks the scalability constraints of the optimal scheduler used for training and can generalize to networks $6\times$ larger in the number of nodes and $10\times$ larger in the number of tags. 
\citet{sarkar2021deepradar} propose DeepRadar that utilizes DL to detect radar signals and estimate their spectral occupancy for incumbent protection and efficient spectrum sharing. This approach involves spectrogram image learning (SIL) based on YOLO (You Only Look Once) model that learns an object detection model using spectrograms, including both radar and non-radar data. 
Lastly, \citet{garg2023sirius} design Sirius, a self-localization system, where the node computes its own location onboard, using a single receiver for low-power IoT nodes to close the gap between the needs for accurate and robust localization and the lack of efficient solutions in the low-power scenario. Instead of relying on strictly synchronized antenna arrays to estimate angle-of-arrival (AoA) and time-of-flight (ToF) which requires resources low-power nodes do not possess, 
Sirius uses antennas whose gain pattern can be reconfigured by the on/off of controllable switches in real-time to embed direction specific encoding to the received signal. The gain patterns are passed to AI models to estimate the angle in degrees. Experimental results show that Sirius is able to obtain competitive performance compared to state-of-the-art antenna array-based systems, achieving 7-degree median error in AoA estimation and 2.5-meter median error in localization.

\begin{figure*}[t!]
    \centering
    \resizebox{\textwidth}{!}{
        \begin{forest}
            forked edges,
            for tree={
                grow=east,
                reversed=true,
                anchor=base west,
                parent anchor=east,
                child anchor=west,
                base=center,
                font=\large,
                rectangle,
                draw=hidden-draw,
                rounded corners,
                align=left,
                text centered,
                minimum width=7em,
                edge+={darkgray, line width=1pt},
                s sep=3pt,
                inner xsep=5pt, %
                inner ysep=1pt, %
                line width=0.8pt,
                ver/.style={rotate=90, child anchor=north, parent anchor=south, anchor=center},
            },
            where level=1{text width=13em,font=\normalsize,}{},
            where level=2{text width=17em,font=\normalsize,}{},
            where level=3{text width=17em,font=\normalsize,}{},
            [
                \textbf{Domain-specific AIoT Systems}, ver
                        [
                            \textbf{Healthcare and Well-being}, fill=orange!10
                            [
                            \textbf{Vital Sign Monitoring}, fill=orange!10
                            [
                            MoVi-Fi~\cite{chen2021movi}{,}~\citet{zhang2023passive}{,}
                            VitaMon~\cite{huynh2019vitamon}{,}\\
                            RF-SCG~\cite{ha2020contactless}{,}~VocalHR~\cite{xu2022hearing}{,}~Crisp-BP~\cite{cao2021crisp},leaf, text width=21em
                            ]
                            ]
                            [
                            \textbf{In-Situ Illness Detection}\\ \textbf{and Monitoring} \\ , fill=orange!10
                            [
                            StudentLife~\cite{wang2014studentlife}{,}
                            ~\citet{saeb2015mobile}{,}
                            \citet{adler2020predicting}{,}\\
                            PDVocal~\cite{zhang2019pdvocal}{,}
                            PTEase~\cite{yin2023ptease}{,}
                            SpiroSonic~\cite{song2020spirosonic}{,}\\
                            EarHealth~\cite{jin2022earhealth}{,}
                            ~\citet{chan2022off}{,}~OAEbuds~\cite{chan2023wireless},leaf, text width=22em
                            ]
                            ]
                            [
                            \textbf{Assistive Technology} \\ , fill=orange!10
                            [
                            MobileDeepPill~\cite{zeng2017mobiledeeppill}{,}
                            iBlink~\cite{xiong2017iblink}{,}\\
                            DeepASL~\cite{fang2017deepasl}{,}
                            SignSpeaker~\cite{hou2019signspeaker},leaf, text width=16em
                            ]
                            ]
                            [
                            \textbf{Personal Health Insight Generation}, fill=orange!10
                            [
                            PH-LLM~\cite{Cosentino2024Jun}{, }PHIA~\cite{Merrill2024Jun}{, }~\citet{Englhardt2024Jun},leaf, text width=20em
                            ]
                            ]    
                        ]
                        [
                            \textbf{Video Streaming}, fill=orange!10
                            [
                            \textbf{Adaptive Video Streaming}, fill=orange!10
                            [
                            Pensieve~\cite{Pensieve2017}{,}
                            Concerto~\cite{Concerto2019}{,}
                            OnRL~\cite{OnRL2020}{,}
                            Fugu~\cite{Fugu2020}{,}\\
                            PERCEIVE~\cite{lee2020perceive}{,}
                            Loki~\cite{Loki2021}{,}
                            SENSEI~\cite{Sensei2021}{,}
                            Swift~\cite{swift2022}
                           ,leaf, text width=24em
                            ]
                            ]
                            [
                            \textbf{Video Enhancement} \\ , fill=orange!10
                            [
                            NAS~\cite{NAS2018}{,}
                            NEMO~\cite{Nemo2020}{,}
                            LiveNAS~\cite{LiveNas2020}{,}\\
                            NeuroScaler~\cite{NeuroScaler2022}{,}
                            YuZu~\cite{YuZu2022}{,}
                            OmniLive~\cite{OmnniLive2023}
                           ,leaf, text width=20em
                            ]
                            ]
                            [
                            \textbf{Efficiency Optimization} \\ , fill=orange!10
                            [
                            CHESS~\cite{chess2017}{,}
                            Pano~\cite{Pano2019}{,}
                            SalientVR~\cite{wang_salientvr_2022}{,}
                            Tambur~\cite{Tambur2023}
                           ,leaf, text width=23.5em
                            ]
                            ]    
                        ]
                        [
                            \textbf{Video Analytics}, fill=orange!10
                            [
                            \textbf{Continuous Learning}, fill=orange!10
                            [
                            Ekya~\cite{Ekya}{,}
                            RECL~\cite{RECL}
                           ,leaf, text width=10em
                            ]
                            ]
                            [
                            \textbf{Runtime Adaptation} 
                             , fill=orange!10
                            [
                             Camtuner~\cite{camtuner}{,}
                            Chameleon~\cite{Chameleon2018}{,}
                            AWStream~\cite{awstream2018}{,}\\
                            Distream~\cite{distream2020}{,}
                            Turbo~\cite{Turbo2023}
                           ,leaf, text width=22em
                            ]
                            ]
                            [
                            \textbf{Efficiency Optimization}, fill=orange!10
                            [
                            FilterForward~\cite{filterforward2019scaling}{,}
                            Reducto~\cite{reducto2020}{,}
                            AccMPEG~\cite{AccMpeg2022Apr}{,}\\
                            CoVA~\cite{cova2022}{,}
                            Gemel~\cite{Gemel2023}
                           ,leaf, text width=22em
                            ]
                            ]
                            [
                            \textbf{Query Optimization}, fill=orange!10
                            [
                            ELF~\cite{elf2020}{,}
                            DIVA~\cite{Diva2021}{,}
                            Privid~\cite{privid2022}{,}
                            Tutti~\cite{Tutti2022}{,}
                            Boggart~\cite{Boggart2023}
                           ,leaf, text width=25em
                            ]
                            ]
                        ]
                        [
                            \textbf{Autonomous Driving}, fill=orange!10
                            [
                            \textbf{Perception Enhancement}, fill=orange!10
                            [
                            Pointillism~\cite{bansal_pointillism_2020}{,}
                            EMP~\cite{zhang_emp_2021}{,}
                            Vi-Eye~\cite{he_vi-eye_2021}{,}\\
                            VIPS~\cite{shi_vips_2022}{,}
                           Automatch~\cite{he_automatch_2022}
                           ,leaf, text width=17em
                            ]
                            ]
                            [
                            \textbf{Localization, Tracking, and Mapping} 
                             , fill=orange!10
                            [
                            MVP~\cite{wang_mvp_2021}{,}
                            VeTrac~\cite{tong_large-scale_2021}{,}
                            \citet{lin_architectural_2018}{,}
                            CarMap~\cite{ahmad_carmap_2020}
                           ,leaf, text width=24em
                            ]
                            ]
                            [
                            \textbf{Automatic Testing} 
                             , fill=orange!10
                            [
                            BigRoad~\cite{liu_bigroad_2017}{,}
                            \citet{li_automatic_2018}
                           ,leaf, text width=13em
                            ]
                            ]
                            [
                            \textbf{Control and Actuation} 
                             , fill=orange!10
                            [
                            EAGLE~\cite{Sandha2023May}
                           ,leaf, text width=6em
                            ]
                            ]
                        ]
                        [
                            \textbf{AR/VR/MR}, fill=orange!10
                            [
                            \textbf{Object Detection and Tracking}, fill=orange!10
                            [
                            \citet{liu_edge_2019}{,}
                            MARLIN~\cite{apicharttrisorn_frugal_2019}{,}
                            DeepMix~\cite{guan_deepmix_2022}
                           ,leaf, text width=19em
                            ]
                            ]
                            [
                            \textbf{User Inputs} 
                             , fill=orange!10
                            [
                            HandSense~\cite{nguyen_handsense_2019}{,}
                            ExGSense~\cite{chen_exgsense_2021}
                           ,leaf, text width=14.5em
                            ]
                            ]
                            [
                            \textbf{Performance Enhancement}, fill=orange!10
                            [
                            NEAR~\cite{trinelli_transparent_2019}{,}
                            Heimdall~\cite{yi_heimdall_2020}{,}
                            CollabAR~\cite{liu_collabar_2020}{,}
                            FreeAR~\cite{apicharttrisorn_breaking_2022}
                           ,leaf, text width=25em
                            ]
                            ]
                            [
                            \textbf{Omnidirectional AR}, fill=orange!10
                            [
                            Xihe~\cite{zhao_xihe_2021}
                            ,leaf, text width=6em
                            ]
                            ]
                        ] 
                    ]
        \end{forest}
}
    \caption{Summary of topics related to domain-specific AIoT systems.}
    \label{fig:system-tree}
\end{figure*}
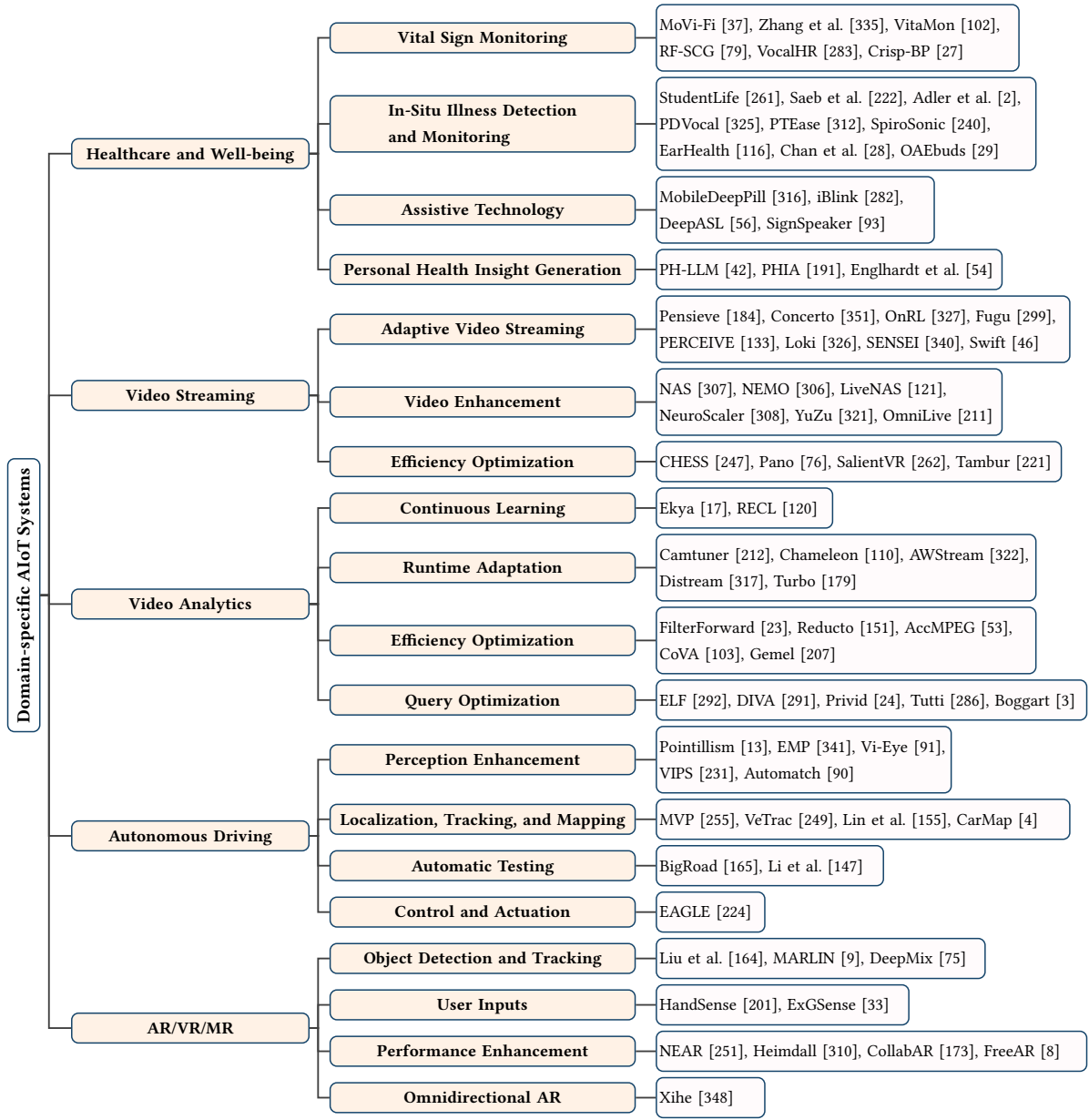
\section{Domain-specific AIoT Systems} 
\label{systems}

\subsection{Healthcare and Well-being}
\label{subsec_health}
One important application domain of AIoT systems is healthcare and well-being.
As summarized in Figure~\ref{fig:system-tree}, existing works on AIoT systems for healthcare and well-being can be grouped into four categories: vital sign monitoring, in-situ illness detection and monitoring, assistive technology, and personal health insight generation.

\vspace{1.5mm}
\noindent
\textbf{Vital Sign Monitoring.}
One of the primary use cases of AIoT systems developed for healthcare and well-being is monitoring an individual's vital signs such as cardiac signals, breathing, and blood pressure.
For instance, one of the key challenges of vital sign monitoring is motion artifacts caused by body movements.
In \cite{chen2021movi}, the authors introduce MoVi-Fi to monitor breathing and heartbeats in a contactless way using RF signals under the existence of body movements. MoVi-Fi utilizes deep contrastive learning to separate vital signs from the body movements and further uses an encoder-decoder model to refine and recreate the vital sign waveforms. 
In \cite{zhang2023passive}, the authors observe that vital signs including breathing and heartbeats cause subtle facial vibrations. They propose to use the motion sensors inside the commodity AR/VR headsets to capture those subtle facial vibrations and employ an LSTM-based model to reconstruct the vital sign waveforms.
VitaMon \cite{huynh2019vitamon}, RF-SCG \cite{ha2020contactless}, and VocalHR \cite{xu2022hearing} focus on monitoring cardiac signals. 
Specifically, VitaMon \cite{huynh2019vitamon} proposes to use video to measure the inter-heartbeat interval (IBI). Since blood absorbs more light than other tissues, video can effectively detect the changes in blood vessel volume that occur with each heartbeat. Based on this principle, VitaMon employs a CNN to identify and reconstruct the peak of each heartbeat across consecutive facial image frames. %
RF-SCG \cite{ha2020contactless}, on the other hand, uses mmWave to reconstruct the seismocardiogram (SCG) waveforms that detect fine-grained cardiovascular events. RF-SCG emits mmWave radar signals and captures the reflections from the human body, and proposes a CNN-based model to translate the
mmWave reflections to SCG waveforms.
VocalHR \cite{xu2022hearing} proposes to infer cardiac activities from human voice production. It extracts phonation and articulatory features from human voice that are related to cardiac activities, and transforms these vocal features into cardiac activities through an LSTM-based model. Lastly, 
\citet{cao2021crisp} introduce Crisp-BP, a blood pressure monitoring system that leverages wrist-worn devices equipped with PPG sensors. The sensors emit green and infrared light, which measures volume changes in blood vessels, which are processed by a BLSTM-based model to estimate both diastolic and systolic blood pressure. 

\begin{figure}[t]
\centering
\includegraphics[width=0.65\linewidth]{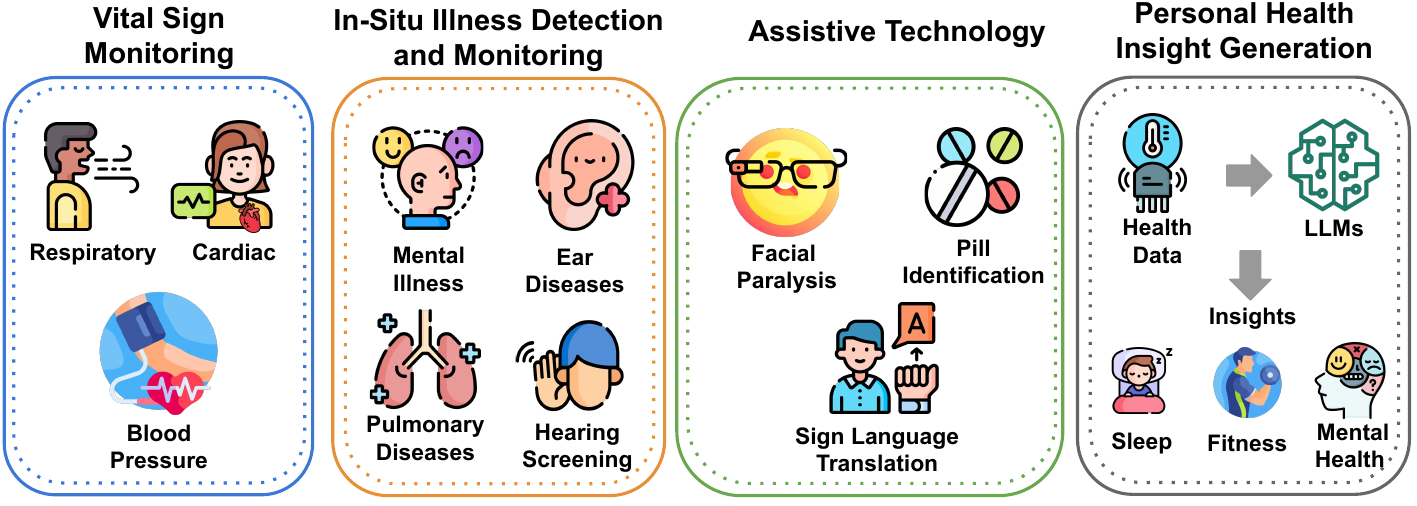}
\caption{Summary of AIoT systems for healthcare and well-being.}
\label{fig:overview of healthcare}
\vspace{-3mm}
\end{figure}

\vspace{1.5mm}
\noindent
\textbf{In-Situ Illness Detection and Monitoring.}
Another important use case of AIoT systems developed for healthcare and
well-being is to detect or monitor the progress of illnesses such as mental illnesses, \mz{lung and ear diseases} in non-clinical settings.
Mental illnesses are a leading cause of disability worldwide \cite{mohr2017arcp}.
One pioneering work for mental illnesses is StudentLife \cite{wang2014studentlife}, where the study identifies relationships between smartphone sensor data and students' mental health and academic performance. 
As another pioneering work in this domain, \citet{saeb2015mobile} propose to use smartphone  GPS data and phone usage data to capture and detect various daily-life behavioral markers from individuals with depression and utilize AI models to analyze the collected sensor data to infer depressive symptom severity.
\citet{adler2020predicting} focus on leveraging smartphone sensor data to predict early warning signs of psychotic relapse in patients with schizophrenia spectrum disorders. They develop encoder-decoder neural network models that could identify behavioral anomalies occurring within 30 days before a relapse.
Parkinson's disease is another use case. It is observed that non-speech body sounds~\cite{rahman2014bodybeat}, such as breathing and throat-clearing sounds, are highly correlated to Parkinson's disease. \citet{zhang2019pdvocal} propose PDVocal, which leverages everyday smartphone voice activities such as calls and chats to capture these sounds and employs a ResNet-based DL model to assess the presence probability of Parkinson's disease.
In terms of lung diseases, \citet{yin2023ptease} introduce PTEase, a 3D-printed mouthpiece that attaches to a smartphone for pulmonary disease detection. The smartphone emits acoustic waves via its speaker. These waves travel through the airway and are captured by the smartphone's microphone, providing detailed information about the user's airway conditions which is crucial for pulmonary disease detection and lung function assessment. 
Similarly, \citet{song2020spirosonic} introduce SpiroSonic, a smartphone-based system for conducting spirometry tests by monitoring the motion of the chest wall during breathing.
SpiroSonic emits an ultrasound wave from the smartphone speaker and captures the reflected wave from the chest wall. It extracts specific features such as the maximum velocity of chest wall motion, the chest wall displacement during the first second of exhalation, and the peak chest wall displacement. These features are then used as inputs to a regression neural network, which provides an assessment of the user's lung function.
In terms of ear diseases,  \citet{jin2022earhealth} propose EarHealth, an earphone-based system that detects three ear diseases: otitis media, ruptured eardrums, and earwax blockages. By emitting sound waves into the ear and capturing the echoes using its integrated microphone, EarHealth analyzes the captured data that contains crucial information about the ear through a multi-view DL model to detect and monitor these ear diseases.
\citet{chan2022off} present the development and clinical evaluation of a low-cost otoacoustic emissions (OAE) probe designed to facilitate early hearing screening. Conventional OAE tests require highly sensitive and expensive acoustic hardware, making it inaccessible to low and middle-income countries. To fill this gap, the authors propose to develop a low-cost probe using off-the-shelf microphones and earphones connected to a smartphone. The probe functions by emitting two pure tones through the earphones, prompting the cochlea to generate distortion-product OAEs, which are then captured by a microphone.
In \cite{chan2023wireless}, they further improve the design of the hearing screening probe using wireless earbuds, and propose OAEbuds, which employs a two-step protocol combining frequency-modulated continuous wave (FMCW) signals and wideband pulses to separate OAEs from in-ear reflections. The clinical study shows that OAEbuds achieves sensitivity and specificity comparable to commercial medical devices, demonstrating its potential to make hearing screening more affordable and accessible.

\vspace{1.5mm}
\noindent
\textbf{Assistive Technology.}
AIoT systems for healthcare and
well-being has also been developed as assistive technologies, which help individuals with disabilities perform tasks that might otherwise be difficult or impossible. 
For instance, \citet{zeng2017mobiledeeppill} introduce MobileDeepPill, a mobile assistive technology that automatically identifies prescription pills in real-world settings using smartphone cameras.
MobileDeepPill identifies pills by employing a multi-CNN model to extract a pill's three distinctive characteristics including color, shape, and imprints. It also adopts knowledge distillation to reduce the size of the multi-CNN model for on-device inference.
\citet{xiong2017iblink} propose iBlink, a smart glasses-based assistive technology for individuals with facial paralysis. Most individuals with facial paralysis are not able to blink on one side of the face, which could lead to blindness.
iBlink aids individuals with facial paralysis to blink by detecting the non-paralyzed side's blinking using a camera and CNN and applying electrical stimulation to trigger blinking on the paralyzed side.
As another line of research, DeepASL \cite{fang2017deepasl} and SignSpeaker \cite{hou2019signspeaker} focus on developing sign language translation systems that bridge the communication gap between deaf people and people with normal hearing ability. Specifically, DeepASL uses infrared light-based sensing to capture and extract skeleton joint information of fingers, palms, and forearms when the user performs sign language. %
On the other hand, SignSpeaker derives sign-related information using motion sensors from a smartwatch. Both systems utilize the Connectionist Temporal Classification (CTC) technique to construct the sentence-level translation from the word-level translation.

\vspace{1.5mm}
\noindent
\textbf{Personal Health Insight Generation.}
The emergence of LLMs opens up a wide range of possibilities in the application domain of healthcare and well-being. One of the most promising capabilities is to generate personal health insights based on data collected from health-related sensors inside an individual's mobile and wearable devices.
For example, \citet{Cosentino2024Jun} introduce PH-LLM, a Personal Health Large Language Model based on a fine-tuned version of Gemini designed to generate insights and recommendations for improving sleep and fitness behaviors. PH-LLM collects data from multiple sources, including medical records, wearable sensor data, and self-reported health data from each individual. By integrating this information, it seeks to understand each individual's unique health profile and to provide tailored health recommendations and predictions. %
Similarly, \citet{Merrill2024Jun} present PHIA, a Personal Health Insights Agent to analyze behavioral health data from wearable sensors using LLMs. %
PHIA can address both factual and open-ended health queries, and generate personalized, actionable health insights with high accuracy. %
Lastly, \citet{Englhardt2024Jun} explore the potential of using LLMs to derive clinically relevant insights from multi-sensor data collected from mobile and wearable devices. The authors develop chain-of-thought prompting methods to facilitate LLMs in reasoning about activity, sleep, and social interaction data, and their relation to mental health conditions such as depression and anxiety. While the authors initially focused on using LLMs for diagnostic task, they found greater potential in generating detailed, natural language summaries that integrate multiple data streams, offering a more comprehensive understanding of a patient's health condition. %
\subsection{Video Streaming}
\label{subsec_video_streaming}
Video streaming involves the continuous and seamless transmission of video and audio content from a server to a client over a network. 
It has become one of the most widely used technologies that enables services such as live streaming and video conferencing, which are integral to people's daily lives.
As summarized in Figure~\ref{fig:system-tree}, existing works on AIoT systems for video streaming can be grouped into three categories: adaptive video streaming, video enhancement, and \mz{efficiency optimization.}

\begin{figure}[t]
\centering
\includegraphics[width=0.67\linewidth]{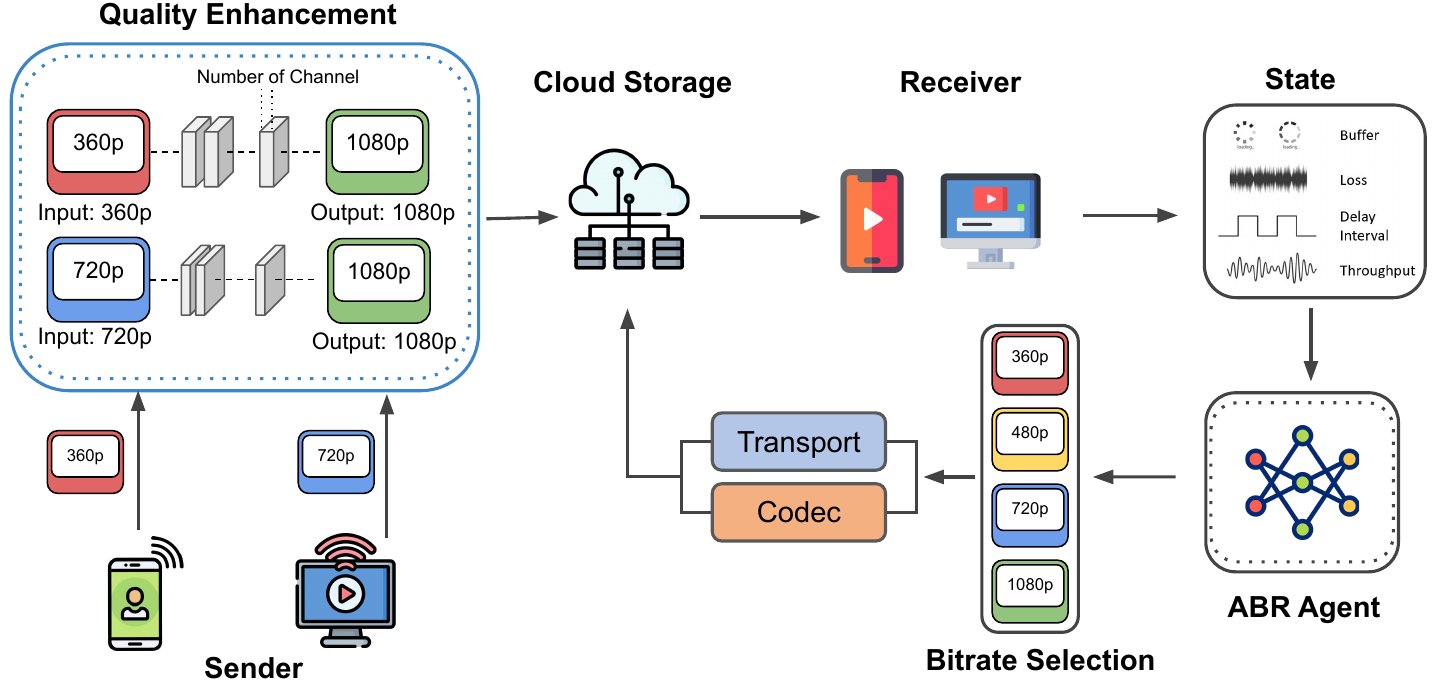}
\vspace{-1mm}
\caption{Illustration of the architecture of AIoT systems for video streaming.}
\label{fig: Components of a Video Streaming pipeline}
\vspace{-3mm}
\end{figure}

\vspace{1.5mm}
\noindent
\textbf{Adaptive Video Streaming.}
One key challenge of video streaming is to maintain a consistent high-quality viewing experience and uninterrupted playback when network bandwidth fluctuates due to factors such as congestion, interference, and user mobility. Adaptive video streaming addresses this challenge by dynamically adjusting the video quality in real-time based on the available network bandwidth, ensuring a smoother viewing experience.
For example, \citet{Pensieve2017} propose Pensieve, an adaptive video streaming framework that employs reinforcement learning (RL) to autonomously learn adaptive bitrate (ABR) algorithms to eliminate the need for pre-programmed control rules. %
\citet{Concerto2019} present Concerto, which identifies an important factor of poor quality of experience (QoE): the lack of coordination between application-layer video codecs and the transport-layer protocols. To address this issue, Concerto introduces a video bitrate adaptation strategy based on deep imitation learning, which is abel to identify the most suitable bitrate for codec and transport layers and successfully boosts the QoE.
While both Pensieve and Concerto exhibit their potential, a notable challenge arises from the fact that the learning models are commonly trained within simulators or emulators. Unfortunately, this can result in poor performance when applied in real-world scenarios. \citet{OnRL2020} present OnRL, which effectively bridges the gap between simulation and real-world scenarios by introducing an online RL framework designed for real-time mobile video telephony applications. 
One challenge with RL is that the algorithm might make incorrect exploitation decisions.
OnRL addresses this issue by introducing a hybrid learning approach: if the RL model performance deviates from the expected, the system switches to a rule-based ABR algorithm; otherwise, it continues to follow the RL strategy. 
Another challenge with RL-based approaches is that acquiring suitable training data and creating a suitable environment is not trivial. \citet{Fugu2020} address this challenge by developing an ABR algorithm and training it directly within the real deployment environment using in-situ data. 
As another line of research, \citet{lee2020perceive} introduce PERCEIVE, which utilizes a 2-stage LSTM model for cellular uplink channel throughput prediction and adapts the video encoding bitrate based on the prediction results to improve user experience in mobile live streaming applications.
\citet{Loki2021} show that attempts to combine such hybrid approaches do not effectively utilize the combined strengths of both methods, often resulting in suboptimal performance. Therefore, they propose Loki, which strives for a more profound collaboration of rule-based methods with learning-based methods. This is achieved by converting a "white-box" rule-based approach into a similar "black-box" neural network model using a customized imitation learning model. 
\mz{\citet{Sensei2021} introduce SENSEI, a streaming optimization scheme that capitalizes on users' varying sensitivity levels to different segments of a video. This approach is rooted in the understanding that users are more attuned to crucial moments (e.g., goal-scoring moments in a sports video) and are more displeased by buffering interruptions during these instances compared to less critical parts. Given that, SENSEI reduces the current video quality to conserve bandwidth, which can later be allocated to deliver higher quality during moments of heightened user sensitivity. This strategy enhances the QoE within the same bandwidth constraints by efficiently adapting video quality to users' sensitivity patterns.}
Lastly, \citet{swift2022} introduce Swift, an adaptive video streaming system featuring a layered encoder.
Instead of encoding video segments separately in various qualities, Swift encodes the video segments into layers. In doing so, it significantly reduces bandwidth usage and achieves a quicker response time to fluctuations in network conditions.

\vspace{1.5mm}
\noindent
\textbf{Video Enhancement.}
Another key challenge of video streaming is the inherent limitation in the resolution of the original source video, which can affect the viewing experience on high-definition displays. 
Video enhancement techniques address this challenge by enhancing the quality of videos by upgrading their resolution beyond the resolution of the original source video, thereby providing a better viewing experience for users with high-definition displays.
\mz{\citet{NAS2018} introduce NAS, a super resolution-based video delivery framework that leverages client-side computation and DNNs to enhance user QoE. Their approach involves combining scalable DNNs with adaptive predictions that can adjust their processing requirements dynamically in response to the available resources. Reinforcement learning is used to determine the best time to download a DNN model and choose the appropriate video bitrate for each video segment.}
However, one key limitation of NAS \cite{NAS2018} is its high computational demand and power consumption, making it less competitive to be deployed on mobile devices. To make video enhancement feasible for mobile devices, \citet{Nemo2020} propose NEMO, which capitalizes on the inherent temporal redundancy in videos by applying super-resolution to only some specific frames while reusing the super-resolution results to enhance the entire video. 
However, due to the involvement of resource-intensive offline computation, NEMO is not ideal for live video streaming. In contrast, \citet{LiveNas2020} design LiveNAS specifically for live video streaming scenarios.
LiveNAS utilizes real-time online training and incorporates recently trained outcomes for super-resolution within the context of live video.
Similarly, \citet{NeuroScaler2022} present NeuroScaler, a streamlined and scalable neural-enhancing framework for live video streaming. NeuroScaler focuses on reducing the costs of live video streaming and includes cost-reducing algorithms for video super-resolution and a specialized hybrid video codec that drastically cuts encoding expenses for selective super-resolution outputs.
\citet{YuZu2022} move one step further and propose YuZu, a super-resolution-based video streaming system for 3D video streaming. This system addresses key limitations of existing 3D video streaming methods, such as high bandwidth consumption and the ineffectiveness of viewport-based streaming when the entire scene is within the view.
Lastly, \citet{OmnniLive2023} explore omnidirectional video (i.e., 360° video) streaming, and develop OmniLive, a super-resolution-based omnidirectional video streaming system that utilizes GPU to sustain a high super-resolution quality at 30 frames per second across a range of mobile devices.

\vspace{1.5mm}
\noindent
\textbf{Efficiency Optimization.}
Enhancing the efficiency of video streaming services is also important.
\citet{chess2017} present CHESS, a video popularity prediction scheme designed to forecast the future popularity of videos. Since only a small fraction of videos gain significant popularity and contribute to the majority of watch time, by prioritizing these popular videos rather than processing all videos uniformly, CHESS effectively allocates processing resources to optimize the user experience.
Omnidirectional video streaming consumes more bandwidth compared to standard video streaming. One potential solution for bandwidth optimization is the viewport-driven approach, which focuses on streaming only the region that the viewer is watching (viewport) in high quality. However, this approach comes with constraints as it requires predicting the viewer's future gaze direction, and any prediction errors can lead to rebuffering or drops in quality. To address this challenge, \citet{Pano2019} develop Pano, a method that leverages the sensitivity of users to variations in quality distortion, which effectively balances the trade-off between quality and bandwidth allocation. This approach allows for increasing quality to the highest noticeable extent when there is surplus bandwidth and decreasing quality to an almost unnoticeable degree when bandwidth is limited. 
Improving QoE for mobile omnidirectional video streaming is crucial, particularly in bandwidth-limited wireless networks. Previous research on omnidirectional video streaming has attempted optimization based on head movement trajectory (HMT) but often falls short in achieving precise HMT predictions. To address this challenge, \citet{wang_salientvr_2022} introduce SalientVR, a framework that integrates gaze information into a saliency-driven mobile 360° video streaming system. SalientVR holds the potential to elevate the QoE by utilizing user gaze patterns to deliver content more accurately and effectively for mobile VR devices.
Lastly, \citet{Tambur2023} introduce Tambur, a scheme designed to address bandwidth-efficient loss recovery for video conferencing. Existing streaming codes fall short of meeting the specific demands of video conferencing due to the frequent loss of packets, often occurring in bursts, which can impede the rendering of video frames. Tambur introduce a learning-based predictive model for effectively configuring bandwidth overhead and achieves a noteworthy reduction in both the frequency and cumulative duration of freezes.

\subsection{Video Analytics}
\label{subsec_video_analytics}
Video cameras have been deployed at scale at places such as streets and intersections, stores and shopping malls, as well as homes and office buildings. Analyzing video streams collected from these distributed cameras enables many applications such as security and surveillance, traffic management, and customer behavior analysis.
As summarized in Figure~\ref{fig:system-tree}, existing works on AIoT systems for video analytics can be grouped into four categories: \mz{continuous learning, runtime adaptation, efficiency optimization, and query optimization.}

\begin{figure}[t]
\centering
\includegraphics[width=0.67\linewidth]{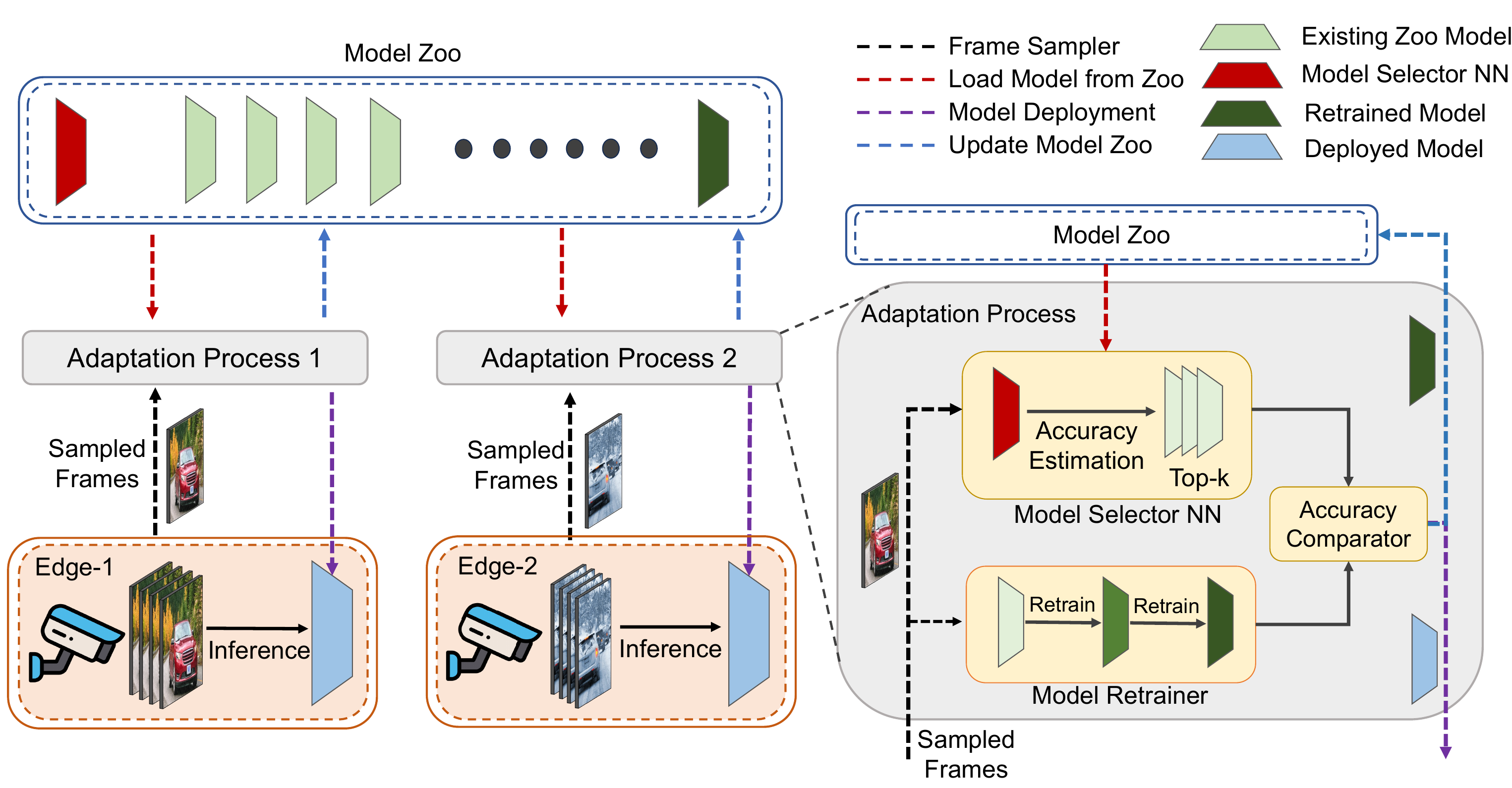}
\vspace{-1mm}
\caption{Illustration of continuous learning for video analytics.}
\label{fig: Continuous Learning for video analytics}
\vspace{-3mm}
\end{figure}

\vspace{1.5mm}
\noindent
\textbf{Continuous Learning.}
In video analytics, it is inevitable that new video data emerge. Therefore, it is critical for video analytics systems to adapt to such data drift.
Although continuous learning can effectively tackle data drift by periodically retraining models on new data, supporting continuous learning on video analytics systems is not trivial.
\citet{Ekya} propose Ekya, a video analytics system that addresses the challenge of jointly supporting inference and continuous learning on edge servers.
The key idea of Ekya is to identify models that need retraining the most while balancing the resources for joint retraining and inference.
Ekya can enhance the performance of video analytics, particularly in dynamically changing environments where data drift is a significant factor in performance. However, since the retraining process consumes the majority of the time, relying solely on model retraining may not be resource-efficient for real-time video analytics tasks.
\citet{RECL} propose RECL, an end-to-end system that integrates model reusing with model retraining to overcome this problem. RECL performs continuous model retraining as well as leverages historical specialized DNNs and shares this model zoo across various edge devices. Additionally, RECL efficiently allocates GPU resources by utilizing an iterative training scheduler, which prioritizes retraining jobs based on their progression rate. 
\revision{RECL shows remarkable improvement in both accuracy and mAP for object detection and image classification tasks, outperforming all baseline models, including Ekya.}

\vspace{1.5mm}
\noindent
\textbf{Runtime Adaptation.}
Another critical capability of video analytics systems is runtime adaptation.
Adapting camera parameters and settings has a significant impact on video analytics performance, particularly due to weather and lighting conditions. To maintain high accuracy, it becomes essential to adapt camera parameters in response to these conditions. However, the task of identifying the optimal camera settings for specific scenes is challenging. To mitigate the impact of environmental condition changes on video analytics performance, \citet{camtuner} propose Camtuner, a reinforcement learning-based approach to dynamically adapt non-automated camera parameters. 
Apart from camera parameters, various other factors within a video analytics pipeline can impact its performance, including frame resolution, frame sampling rate, and the choice of DNN models. 
Collectively, these components can be referred to as the overall configuration. Choosing a suitable configuration can impact both the resource utilization and accuracy of a video analytics application. 
Although adapting model configurations frequently can optimize resource usage, it incurs high costs due to the large number of possible configurations. 
To address this issue, \citet{Chameleon2018} introduce Chameleon, a technique for achieving a balance between resource allocation and accuracy by choosing the appropriate neural network configuration. 
\citet{awstream2018} propose AWStream, an adaptive stream processing system with low latency and high accuracy. AWStream's main contribution is its runtime system that consistently monitors and adjusts to network conditions. It optimizes streaming data rate based on available bandwidth and employs learned Pareto-optimal configurations to maintain high accuracy.
\citet{distream2020} propose Distream, which focuses on runtime adaptation to the dynamic workloads generated by distributed video cameras. Depending on the deployment location, the number of objects captured by each camera and its corresponding workload is different and varies throughout the day. The key idea of Distream is to adaptively balance the workloads across the cameras and also partition the workloads between cameras and the edge server. As such, Distream fully utilizes the compute resources at cameras and the edge server to enhance system performance.
Lastly, \citet{Turbo2023} propose Turbo, which capitalizes on managing latent computing resources to enhance overall performance, particularly in object detection tasks. The proposed approach revolves around a multi-exit GAN structure, which is paired with an adaptive scheduler that dynamically determines the optimal enhancement level for each incoming frame, thereby maximizing object detection accuracy in real-time. The adaptive scheduler makes on-the-fly decisions about the most appropriate enhancement levels based on the current resource availability.
In terms of results, Turbo presents remarkable improvements in absolute mAP.

\vspace{1.5mm}
\noindent
\textbf{Efficiency Optimization.}
Enhancing the efficiency of video analytics systems is also important.
The widespread deployment of video cameras, numbering in the thousands and operating continuously, leads to a massive amount of data that needs to be transmitted and processed. Transmitting and processing all video frames from the edge to the server can be extremely expensive due to the bandwidth constraints and computational resources required. 
\citet{filterforward2019scaling} propose FilterForward, which only selects and transmits the relevant video frames to save bandwidth. 
Similarly, \citet{reducto2020} introduce Reducto, another filtering-based technique which implements on-camera filtering and dynamically adjusts filtering decisions to cater to live video analytics requirements. Experimental results show that Reducto outperforms FilterForward by 93\% in terms of frame filtering efficiency. %
As another line of research, \citet{AccMpeg2022Apr} introduce AccMPEG.
The proposed key techniques involve the design of a cheap camera-side model to efficiently decide which regions of the frames should be encoded high-quality and which regions should be subjected to lower-quality encoding. Additionally, AccMPEG allows for quick customization to different DNNs, with training times reduced to mere minutes, further demonstrating its efficiency. 
\citet{cova2022} introduce CoVA, a cascade architecture that reduces the need for full video decoding. By leveraging compressed-domain analysis, CoVA efficiently detects and tracks objects across frames, only decoding a minimal subset of frames necessary for DNN processing. CoVA's design not only optimizes computational efficiency but also supports both temporal and spatial queries, broadening its applicability in video analytics.
Lastly, video analytics systems often host multiple tasks like object detection, face recognition, and semantic segmentation, where different models can be used for different tasks. Given the limited GPU resources of edge devices, attempting to load all models can exceed GPU memory limit. \citet{Gemel2023} introduce Gemel, a model merging technique that can efficiently merge and share layers from models with the same architectures.
In doing so, Gemel effectively reduces the number of swaps required and the amount of data loaded into GPU memory, resulting in fewer frame drops and improved accuracy.

\vspace{1.5mm}
\noindent
\textbf{Query Optimization.}
A video analytics query is an inquiry submitted to a video analytics system to retrieve useful information and insights from video data. 
Summarizing a video scene with object count is a common query type to get insight from a video stream. Due to the energy constraint of edge devices, continuously transmitting video streams is a challenging task. ELF, presented by ~\citet{elf2020}, is a framework designed to continually summarize video scenes through the aggregation of object counts, all while operating within the confines of limited energy resources. Rather than transmitting raw video data, the approach involves sending only numerical data, such as count numbers or other relevant query-related values.
In many camera setups, a significant portion of cameras often remain inactive and unqueried. This scenario can be referred to as "zero-streaming" where the inactive cameras store video data in their local storage and communicate with the server only when a specific query is requested. \citet{Diva2021} propose DIVA, an approach to effectively query video analytics on zero streaming cameras. When it comes to tasks like retrieval, tagging, and counting, DIVA consistently demonstrates superior performance compared to other baseline methods. As zero streaming cameras primarily store data on their local storage, a drawback of DIVA is its susceptibility to video data loss in the event of camera storage failures.
Video analytics queries can sometimes raise concerns about privacy violations, as users may request sensitive information about others, potentially infringing on their privacy. \citet{privid2022} introduce Privid, a method aimed at extracting valuable information from video data without compromising privacy. Privid's approach involves breaking the video into smaller segments and executing processing code on each segment individually rather than processing the entire video at once. A Privid query comprises a set of statements in a PrividQL language, similar to SQL, along with executable video processing components. They run an experiment on video data collected from three cameras and apply Detectronv2 for object detection and DeepSORT for object tracking.
The rise of 5G technology has propelled the expansion of ultra-fast video analytics, largely due to the growing need for low-latency processing capabilities. Tutti, developed by ~\citet{Tutti2022}, combines the 5G radio access network and edge computing at the user level to ensure optimal performance for video analytics tasks with low latency. 
Tutti achieves a remarkable reduction in response latency and demonstrates substantial progress in enhancing QoE for video analytics applications.
In response to diverse applications, video analytics platforms have gradually moved away from providing pre-defined video processing results. Instead, they now enable users to utilize their customized models, all while ensuring a consistent commitment to specified accuracy standards. Recent optimization efforts involve preprocessing video data in advance to construct indices that can expedite subsequent queries. However, these optimizations were tailored for scenarios where models were predefined and not user-provided. \citet{Boggart2023} introduce Boggart, a comprehensive pipeline for a video analytics platform that can function as a versatile accelerator using the model provided by the users.
\subsection{Autonomous Driving}
\label{subsec_autonomous_driving}
Autonomous driving enables a vehicle to navigate and operate partially with or fully without human intervention. 
By fusing real-time data collected through IoT sensors with AI-driven perception and decision-making algorithms, AIoT systems are contributing to making autonomous vehicles safer, more efficient, and adaptable to changing road conditions.
As summarized in Figure~\ref{fig:system-tree}, existing works on AIoT systems for autonomous driving can be grouped into four categories: \mz{perception enhancement, localization, tracking, and mapping, automatic testing, and control and actuation.}

\begin{figure}[t]
\centering
\includegraphics[width=0.8\linewidth]{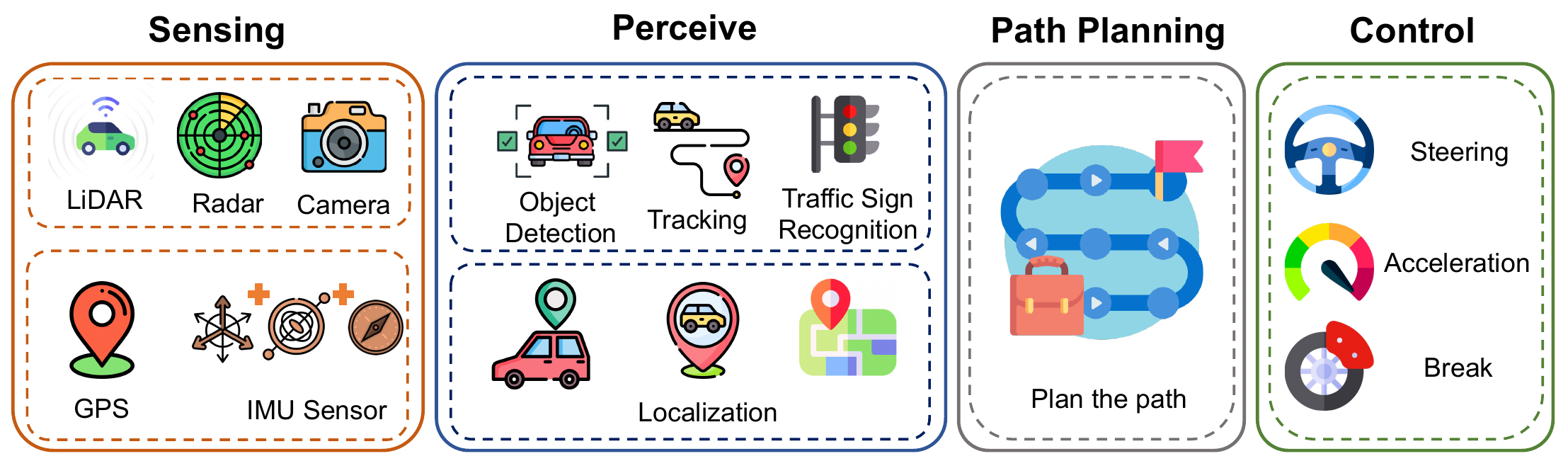}
\vspace{-1mm}
\caption{Illustration of the architecture of AIoT systems for autonomous driving.}
\vspace{-3mm}
\label{fig: Components of Autonomous Driving Pipeline}
\end{figure}

\vspace{1.5mm}
\noindent
\textbf{Perception Enhancement.}
Perception enhancement involves the use of AI to process data from various sensors such as cameras, LiDAR, and radar more accurately, allowing an autonomous vehicle to have a more comprehensive understanding of its surroundings. 
While LiDAR-based systems offer detailed spatial mapping, they fail in adverse weather conditions because LiDAR beams struggle to penetrate through elements like fog. To address this limitation, \citet{bansal_pointillism_2020} introduce Pointillism, an innovative concept called cross-potential point clouds, which leverages the spatial diversity generated by utilizing multiple radar systems to effectively address the issues related to noise and sparsity in radar-based point clouds.
Single-vehicle 3D sensors have two primary limitations: susceptibility to occlusion by non-transparent objects and reduced detail perception at greater distances. \citet{zhang_emp_2021} introduce EMP, a collaborative approach where all nearby connected autonomous vehicles (CAVs) share sensor data with each other. This sharing allows each vehicle to create a more comprehensive and higher-resolution perception compared to relying solely on its own sensors. 
The emerging paradigm of infrastructure-assisted autonomous driving leverages infrastructure elements like smart lampposts to assist autonomous vehicles. However, a challenge arises when the vehicle and infrastructure point clouds do not hold a significant overlap or similarity, resulting in a drop in accuracy and delays. Vi-Eye, presented by \citet{he_vi-eye_2021} is a pioneering system capable of aligning vehicle-infrastructure point clouds with centimeter-level accuracy in real time. 
VIPS, developed by \citet{shi_vips_2022}, takes this capability a step further, achieving decimeter-level accuracy while still maintaining real-time performance. VIPS distinguishes itself from Vi-Eye by adopting an alternative strategy. While Vi-Eye relies on highly accurate point cloud transmission between infrastructure and vehicles, VIPS focuses on aligning two graphs generated from simplified and diverse representations of objects detected by the vehicle and infrastructure. 
\citet{he_automatch_2022} present Automatch, an innovative solution by utilizing traffic cameras to enhance the perception and localization capabilities of autonomous vehicles, particularly at intersections. The pioneering aspect of the system is that it enables vehicles to expand their range of perception by correlating images taken by both traffic cameras and on-vehicle cameras. 

\vspace{1.5mm}
\noindent
\textbf{Localization, Tracking, and Mapping.}
Localization is the process of determining the precise position of a vehicle within a known environment by comparing sensor data to pre-existing maps or reference points. Accurate localization in environments like tunnels and underpasses, where Global Navigation Satellite Systems (GNSS) signals are unavailable, can be a challenging task in autonomous driving. MVP~\cite{wang_mvp_2021} address this challenge by extracting magnetic fingerprints from anomalies in the geomagnetic field. These magnetic fingerprints are then compared to a magnetic map, allowing for precise positioning of vehicles without relying on GNSS signals.
In the context of autonomous driving, tracking refers to the continuous monitoring and prediction of the movements and trajectories of vehicles on a roadway. VeTrac~\cite{tong_large-scale_2021} employs traffic cameras as a sensing network to reconstruct large-scale vehicle trajectories, addressing the limitations of GPS-dependent solutions. It achieves this through a vision-based vehicle detection and tracking algorithm applied to video frames collected from the traffic cameras.
\citet{lin_architectural_2018} identifies three primary computational bottlenecks in autonomous driving systems: object detection, object tracking, and localization, which collectively consume over 94\% of computational resources in the system. In response to these challenges, the authors have developed an end-to-end autonomous driving system that draws from the most cutting-edge system designs found in both academic research and industry practices.
Mapping is a continuous process that involves creating and continually updating a detailed map of the surroundings of a vehicle through the use of various sensors such as LiDAR, cameras, and radar. Maps used in autonomous driving systems require continuous updates to account for significant changes in the environment, which can affect the features visible to a vehicle. CarMap, developed by \citet{ahmad_carmap_2020} offer an innovative solution by collecting 3D maps from vehicles equipped with LiDAR and advanced cameras, ensuring near real-time map updates. As each vehicle travels through a road segment, it uploads map updates to a cloud service over a cellular network, making these updates accessible to other vehicles. %

\vspace{1.5mm}
\noindent
\textbf{Automatic Testing.}
Automatic testing involves identifying and analyzing various events or scenarios that autonomous vehicles may encounter on the road and testing the vehicle's AI-driven systems to ensure they respond appropriately to these events.
BigRoad~\cite{liu_bigroad_2017} provides a cost-effective and dependable solution for collecting extensive driving data by utilizing a smartphone and an Inertial Measurement Unit (IMU) installed within the vehicle. This system extracts internal driver inputs, such as steering wheel angles, driving speed, and acceleration, and also discerns external perceptions of road conditions, including the distinction between wet and dry surfaces. This information can be highly valuable for various purposes, including autonomous vehicle testing and evaluation.
Automatic testing of autonomous driving technology is a complicated process due to the necessity of addressing unusual events and corner cases like road obstacles, pedestrians on highways, or wildlife encounters. To address this challenge, \citet{li_automatic_2018} introduce an automatic system that utilizes an algorithm to identify and respond to unusual driving events effectively. The results of detecting unusual events can be valuable for retraining and enhancing a self-steering algorithm, particularly in more complex driving scenarios.

\vspace{1.5mm}
\noindent
\textbf{Control and Actuation.}
Autonomous control systems manage components that interact with their environments while making decisions independently, without human intervention. Prior works in autonomous AIoT control systems involve multiple stages, including data acquisition from sensors, processing with deep neural networks, and control of configuration parameters to interact with the external environment. The multiple stages suffer from performance bottlenecks due to the difficulty in tuning each step. For instance, even lightweight deep neural networks for object detection have millions of parameters and are too complex for embedded platforms. This complexity makes it infeasible to run multi-stage AIoT control algorithms in real-time on platforms with memory and computation constraints. \citet{Sandha2023May} present EAGLE, an end-to-end deep reinforcement learning (RL) solution that trains a neural network policy to directly use images as input for controlling the PTZ camera. The proposed system bypasses the conventional multi-stage process of object identification, tracking, and control by directly mapping raw photos to control actions using a neural network policy. The paper demonstrates Eagle's effectiveness in various scenarios and its successful transfer from simulation to real-world applications, making significant contributions to the fields of edge AI and autonomous camera control.

\subsection{Augmented, Virtual, and Mixed Reality}
\label{subsec_ar}
Augmented Reality (AR), Virtual Reality (VR), and Mixed Reality (MR) redefine our perception of the world. Specifically, AR enriches reality by overlaying digital content on our surroundings; VR immerses us in entirely digital environments; and MR provides an interactive experience between the virtual and real worlds. 
As summarized in Figure~\ref{fig:system-tree}, existing works on AIoT systems for AR/VR/MR can be grouped into four categories: 
\mz{object detection and tracking, user inputs, performance enhancement, and omnidirectional AR.}

\vspace{1.5mm}
\noindent
\textbf{Object Detection and Tracking.}
Object detection and tracking is one of the most fundamental tasks in AR/MR.
\citet{liu_edge_2019} present an efficient offloading-based object detection and tracking system for AR/MR, which offloads the object detection task to the cloud while conducting tracking on AR devices. The key technique incorporated in the system is a dynamic region of interest (RoI) encoding technique that encodes regions where objects are not likely to be detected in lower quality. As such, the proposed system reduces both offloading latency and bandwidth consumption while maintaining object detection accuracy. 
\citet{apicharttrisorn_frugal_2019} propose MARLIN, a lightweight object detection and tracking framework for AR. Instead of running computationally expensive DNN on each frame, it initiates the DNN execution on the initial frame and then assesses if there are significant scene changes using a change detector specifically designed to identify alterations in the background. If such frame changes are not detected, MARLIN opts for a more lightweight tracking scheme, conserving computational resources while maintaining tracking accuracy.
\citet{guan_deepmix_2022} move one step further and propose DeepMix that focuses on 3D object detection for AR/MR, aiming to provide an efficient solution in this computationally demanding domain. Instead of relying on computationally intensive DNN-based 3D object detection models for bounding box inference, DeepMix offloads 2D RGB images to the edge for 2D object detection and then utilizes the returned 2D bounding boxes in conjunction with depth data captured by headsets to estimate 3D bounding boxes. 
\revision{DeepMix was prototyped on a Microsoft HoloLens 2.
Evaluation results show that compared to existing methods based on 3D object detection, DeepMix not only enhances detection accuracy but also considerably decreases end-to-end latency.}

\begin{figure}[]
\centering
\includegraphics[width=0.8\linewidth]{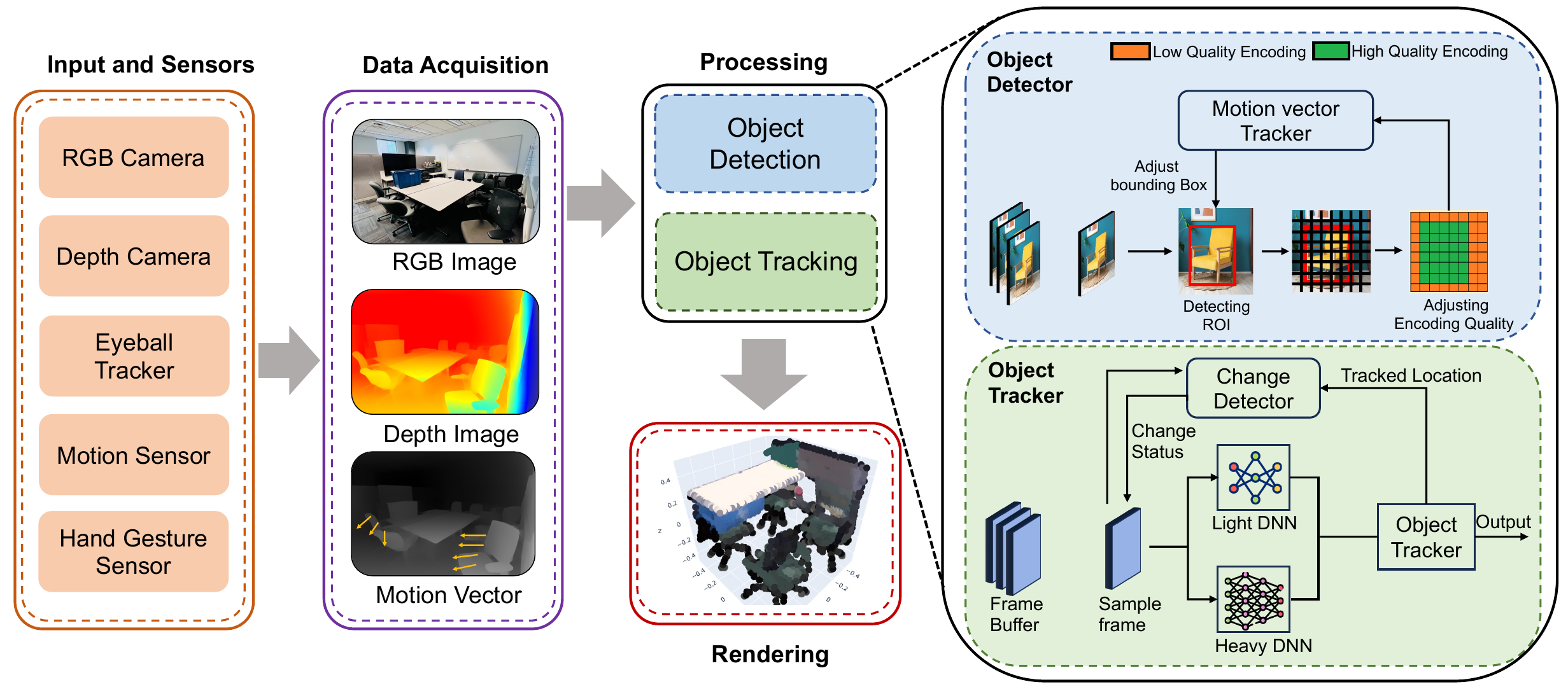}
\vspace{-1mm}
\caption{Illustration of the architecture of AIoT systems for AR/VR/MR.}
\vspace{-3mm}
\label{fig: Overview of AR/VR/MR pipeline}
\end{figure}

\vspace{1.5mm}
\noindent
\textbf{User Inputs.}
Capturing user inputs in an accurate, intuitive and user-friendly manner is another important task in AR/VR/MR.
Existing systems face challenges in capturing user-friendly inputs, particularly in detecting subtle and low-effort finger gestures, which are more suitable for head-mounted devices (HMD) controllers. 
\citet{nguyen_handsense_2019} introduce HandSense, a system using capacitively coupled electrodes to precisely capture and recognize micro-finger gestures for interaction with HMD. They develop an electrode placement configuration on fingertips that minimizes the need for extensive hand movements and utilize several DNN-based methods to recognize the gestures. \revision{Experimental results show that HandSense is able to achieve a 97\% accuracy in recognizing 14 gestures performed by 10 subjects.}
As another line of research, in interactive VR applications, conventional techniques have limitations as they cannot capture the upper face of users, which is mostly occluded by the head-mounted display. To address this limitation, \citet{chen_exgsense_2021} propose ExGSense, which detects and recognizes eye and mouth gestures as VR inputs. This capability is made possible through the utilization of sparse near-eye biopotential signal measurements combined with a DNN-based classifier.
\revision{They evaluated their prototype with 42 facial gestures, achieving 93\% accuracy for user-specific and 77\% for user-independent evaluation.}

\vspace{1.5mm}
\noindent
\textbf{Performance Enhancement.}
Performance enhancement involves optimizing software and hardware to reduce latency, increase processing speed, and improve resource management.
\citet{trinelli_transparent_2019} present NEAR, a transparent AR processing system designed to reduce latency and enhance performance when integrating AR features into streaming videos from lightweight IoT devices. NEAR introduces a simplified SOCKS 5 proxy, a video decoder, and an encoder for the extraction and re-injection of video streams into network flows. This setup enables offloading heavy computational tasks, like object detection, to edge devices, reducing the processing load on both source and consumer devices. NEAR operates without requiring modifications to the IoT streaming devices or client-side applications, ensuring a seamless integration of AR and other computationally intensive functions directly within the network.
Mobile DL frameworks often encounter limitations, particularly related to multi-DNN GPU contention, which can significantly increase inference latency. Unlike desktop GPUs, mobile GPUs cannot effectively implement multi-tasking approaches due to their constraints. Heimdall, introduced by \cite{yi_heimdall_2020}, can efficiently manage the demands of multiple DNN rendering tasks on mobile devices, ensuring minimal latency and optimal performance in emerging AR applications. Heimdall introduces an innovative GPU coordinator that effectively handles multiple DNN rendering tasks on both GPU and CPU by decomposing complex DNNs into smaller units and adopting flexible scheduling techniques. The approach significantly reduces the contention between DNNs and rendering tasks, which typically degrades performance on mobile devices, thus enhancing overall system performance.
\revision{Heimdall was prototyped on various mobile GPUs and AR applications, showing it boosts frame rates from 12 to 30 fps and reduces worst-case DNN inference latency by up to 15 times compared to the baseline multi-threading approach.}
\citet{liu_collabar_2020} introduce CollabAR, the concept of collaborative image recognition into its design, capitalizing on users' temporally and spatially correlated images to enhance image recognition accuracy. The edge-assisted design of the system significantly reduces end-to-end latency, ensuring seamless and efficient performance on commodity mobile devices.
CollabAR attains a recognition accuracy rate exceeding 96\% even for images with substantial distortions. 
FreeAR, presented by \cite{apicharttrisorn_breaking_2022}, enhances the performance of mobile AR by introducing infrastructure-free AR experiences through collaborative time slicing and efficiently distributing compute-intensive tasks across multiple user devices. In FreeAR, all devices unite under a common coordinate system. The chosen primary device takes charge by executing DNNs, enabling it to update the device pose, physical object locations, and 3D virtual overlays, much like traditional AR systems. Meanwhile, secondary devices shift into a low-power mode, where they track their locations within the converged coordinate system using lightweight methods. With this approach, FreeAR can establish a low-power framework, enabling users to seamlessly engage in AR experiences without relying on infrastructure support.

\vspace{1.5mm}
\noindent
\textbf{Omnidirectional AR.}
Lastly, omnidirectional AR refers to AR experiences that provide a 360-degree view of the environment, allowing users to interact with and view augmented content from any direction. 
In mobile AR applications, achieving accurate omnidirectional lighting is crucial to avoid undesirable visual effects. However, accurately estimating omnidirectional lighting in practical scenarios can be challenging, primarily due to the influence of environmental lighting conditions and the dynamic nature of mobile users. \citet{zhao_xihe_2021} introduce Xihe, a mobile AR application capable of real-time and precise omnidirectional lighting estimation by employing a sphere-based point cloud sampling technique. Combined with 3D vision-based lighting estimation pipeline, this sampling technique delivers significantly improved results over farthest point sampling techniques.

\section{Discussions}
\label{sec.discussion}

Lastly, in the field of AIoT, addressing issues such as bias and fairness, security and privacy, as well as legal and ethical concerns is as crucial as tackling the technical challenges in sensing, computing, and networking \& communication, and domain-specific AIoT systems as we have covered in previous sections. 
In this section, we provide a brief discussion on these issues.

\vspace{1.5mm}
\noindent
\textbf{Bias and Fairness in AIoT.}
The integration of IoT and AI significantly broadens the functional capabilities of AIoT. As a result, it raises the need for fairness considerations of AIoT since its extended capabilities allow it to be widely deployed in daily life. 
Bias is referred to the systematic deviation in data or algorithms used by  AIoT that leads to unfair or discriminatory outcomes. 
\citet{Balasingam2021Jun} address the challenge of balancing throughput and fairness in mobility platforms that allocate tasks to vehicles for applications such as food delivery and ridesharing. They show that current ridesharing platforms often fail to ensure that riders from different neighborhoods receive equitable service. This issue arises when the algorithm prioritizes ride requests from certain neighborhoods over others, typically favoring areas with higher demand or shorter and more profitable trips. Given that, they introduce Mobius, a system engineered to balance high throughput and fairness among customers by effectively managing the inherent trade-offs in shared mobility, which enhances the overall performance and fairness of mobility platforms.
Bias in AIoT may arise through federated learning (FL), where models are trained across multiple edge devices, influenced by the heterogeneous resources and data distributions of these devices. 
\citet{Selialia2022Nov} observe that sample feature heterogeneity, resulting from different feature representations at various devices, is a major contributor to bias in FL. Their results show that existing bias mitigation techniques, such as normalization do not fully eliminate bias, with bias levels being proportional to the degree of heterogeneity in sensor sampling features. 
Lastly, \citet{Bae2022Nov} focus on biases in pedestrian trajectory prediction models used in autonomous vehicles. They highlight that many DL models trained on pedestrian data are biased, particularly against vulnerable demographics like children and the elderly, who exhibit different walking patterns compared to the general adult population. This bias can lead to higher prediction errors for these groups and increasing their risk of involvement in vehicle crashes.

\vspace{1.5mm}
\noindent
\textbf{Security in AIoT.}
The vulnerabilities inherent in AIoT pose critical security concerns. 
One of the root causes is the limited resources of AIoT devices, which makes it challenging to implement robust security measures. 
For example, to make an effective balance between security needs with resource limitations, \citet{Luo2018Nov} propose ShieldScatter, a lightweight solution to enhance IoT security by utilizing battery-free backscatter tags. These tags create fine-grained multi-path propagation signatures, allowing for the identification of legitimate users and the detection of attackers.
ShieldScatter provides a cost-effective method that does not require expensive hardware modifications, offering a practical security solution for resource-constrained IoT devices.
As another example, in contact-free smart sensing devices, limited storage resources necessitate the use of cloud storage. However, data stored in the cloud is particularly vulnerable due to the open nature of cloud environments, making it susceptible to potential third-party attacks. To mitigate these risks, \citet{Yaxin_Mei_privacy_2023]} introduce a novel Cloud-Edge-End cooperative storage scheme that leverages the distinct characteristics of the cloud, edge, and endpoint layers. This scheme employs a strategically designed data partitioning strategy, which involves storing sensory data across the three layers separately. By doing so, it increases the difficulty of potential security breaches while offering robust protection against both internal and external attacks.
To protect from malicious attacks in IoT environment, several DL-based detection mechanisms are proposed~\cite{Khan2022Security, Li2024Practical, Farrukh2020FaceRevelio}.
\citet{Khan2022Security} investigate the robustness of SplitFed Learning -- a hybrid of split learning and federated learning (FL) -- against model poisoning attacks, where attackers deliberately inject fake data into the network. SplitFed combines the parallel computation efficiency of FL with the resource efficiency and improved privacy of split learning. The study shows that SplitFed, due to its smaller client-side model portions, is inherently more robust to model poisoning attacks compared to FL. 
\citet{Li2024Practical} focus on physical adversarial attacks on DL-based Wi-Fi sensing systems. This attack manipulates Wi-Fi packet preambles to subtly alter the Channel State Information, thereby influencing the DL models that rely on this data,  without interrupting normal communication. Demonstrating high success rates of attack in activity recognition and user authentication, this study exposes significant security vulnerabilities in current Wi-Fi sensing systems. 
Lastly, \citet{Dong2024RefreshChannels} explore a critical security vulnerability in modern mobile devices that utilize dynamic refresh rate switching to optimize power consumption. The authors present an innovative attack vector named RefreshChannels, where two colluding apps modulate the display’s refresh rate to covertly transmit sensitive information, bypassing the operating system's sandboxing and isolation measures. They also propose countermeasures to mitigate the RefreshChannels attack such as restricting refresh rate API access, limiting refresh rate change frequency, introducing delays and randomization, and detecting abnormal refresh rate patterns.

\vspace{1.5mm}
\noindent
\textbf{Privacy in AIoT.}
Since AIoT could gather a diverse array of data such as an individual's location, personal healthcare record, behavior patterns, and biometric information that is rich in personal information, the collection and processing of such personal data can raise significant privacy concerns.
To protect the privacy of individuals, various regulations have been implemented. The European Union (EU)'s General Data Protection Regulation (GDPR) offers comprehensive data protection rules for handling EU citizens' personal data~\cite{GDPR2016}. In the U.S., the California Consumer Privacy Act (CCPA) outlines consumer rights regarding personal information collected by businesses, while the Health Insurance Portability and Accountability Act (HIPAA) stringently controls the handling of personal healthcare data~\cite{CCPA2018,HIPAA1996}. 
Alongside these legal frameworks, numerous research efforts are underway to tackle privacy-related challenges.
\citet{abadi2016deep} introduce Differential Privacy (DP), a technique that injects noise into data to preserve sensitive personal information. They introduce DP into DL model training with their proposed DP-SGD method, which has proven to maintain high accuracy while effectively preserving privacy.
Fully Homomorphic Encryption (FHE) is another privacy-preserving mechanism which enables computation to be performed over encrypted data. FHE ensures that original data remains hidden and is not decrypted during processing. However, due to its significant computational demands, AIoT is exploring alternatives like Partially Homomorphic Encryption (PHE) and Somewhat Homomorphic Encryption (SHE) to reduce computational overhead. 
~\citet{shafagh2017secure} propose Pilatus, a PHE scheme for IoT while sharing the data with the cloud. Pilatus  protects data privacy by ensuring that the cloud stores only encrypted data while still enabling operations like summation. 
\citet{Mo2021PPFL} introduce PPFL, a framework that leverages Trusted Execution Environments (TEEs) to prevent private information leakage in federated learning scenarios. Though federated learning enables decentralized training across multiple devices without aggregating user data, model updates can still leak sensitive information, posing significant privacy risks. To address this, PPFL employs TEEs to securely process model updates, ensuring that both local training on clients and secure aggregation on servers are protected from potential adversaries.
\citet{Akash2021IAlways} introduce SnoopDog, a framework designed to address the privacy issues arising from hidden wireless sensors, such as secret cameras and microphones. SnoopDog identifies Wi-Fi-based sensors monitoring users by detecting causal patterns between trusted sensor data like IMU readings and Wi-Fi traffic. Although the current implementation of SnoopDog is limited to Wi-Fi-connected devices, future enhancements could extend its capabilities to other wireless communication standards like Zigbee or Bluetooth.
Conventional privacy-preserving machine learning (PPML) methods often face significant latency issues due to  computation overhead of encryption processes. To address this issue, \citet{Chien2023Enc2} introduce Enc$^2$, a hybrid method that combines encoding and homomorphic encryption to enhance PPML for resource-constrained IoT devices. The proposed method performs most of the computations on plaintext, thus reducing latency and shifting the encoding burden from the IoT device to the cloud.
Lastly, \citet{Corbett2023BystandAR} introduce BystandAR, which addresses the privacy concerns posed by Augmented Reality (AR) devices that unintentionally capture the visual data of bystanders. BystandAR leverages eye gaze and voice indicators to differentiate between subjects and bystanders, protecting the bystander's privacy in real-time without offloading data to external servers.

\vspace{1.5mm}
\noindent
\textbf{Legal and Ethical Concerns in AIoT.}
Finally, AIoT must adhere to ethical norms and legal obligations.
\citet{Mittelstadt2017Sep} discusses the intersection of ethical issues and the deployment of health-related IoT technologies, emphasizing the importance of designing these technologies in ways that are both ethically responsible and legally compliant. 
It also underscores the need for responsible design and deployment of IoT technologies, ensuring they are trustworthy, respect user rights, and enhance healthcare delivery without compromising ethical standards.
\citet{Gill2021Dec} highlights the importance of addressing ethical dilemmas in the adoption of autonomous vehicles. The study focuses on the ethical dilemma of programming autonomous vehicles to make decisions in situations where harm is unavoidable, such as whether to protect passengers or pedestrians. Despite industry and policymakers' tendencies to downplay these ethical issues, the findings underscore the necessity of addressing these dilemmas to ensure the successful deployment and acceptance of autonomous vehicles.
\citet{Bouderhem2024Mar} proposes a comprehensive ethical framework to govern the use of AI in healthcare. This framework is based on values such as human dignity, fairness, transparency, accountability, and inclusivity. \citet{Bouderhem2024Mar} also discusses the role of the European Union's General Data Protection Regulation (GDPR) and the AI act as models for creating robust regulatory frameworks. 

\vspace{-2mm}

\section{Concluding Remarks}
\label{sec.conclusion}

In this survey, we present a comprehensive review of AIoT research. 
We organize the AIoT literature into a taxonomy that includes four categories: sensing, computing, networking \& communication, and domain-specific AIoT systems, and review key topics within each category. 
We hope our survey serves as a foundational reference, enabling researchers and practitioners to gain a comprehensive understanding of AIoT and inspiring further contributions to this exciting and important field. 
\vspace{-2mm}

\section{Acknowledgement}
\label{sec.ack}

We would like to thank the editorial board and the anonymous reviewers of ACM Transactions on Sensor Networks (TOSN) for their helpful and constructive comments. We would also like to thank Christopher Ellis, Vishnu Chhabra, Imran Kibria, and Anwesha Roy for their help.
Ness Shroff has been supported in part by National Science Foundation (NSF) grants NSF AI Institute (AI-EDGE) CNS-2112471, CNS-2312836, CNS-2106933, CNS-2106932, CNS-2312836, CNS-1955535, and CNS-1901057, by Army Research Office under Grant W911NF-21-1-0244 and was sponsored by the Army Research Laboratory (ARL) and was accomplished under Cooperative Agreement Number W911NF-23-2-0225. 
Bhaskar Krishnamachari is supported in part by Defense Advanced Research Projects Agency (DARPA) under Contract Number HR001120C0160 and in part by ARL under Cooperative Agreement W911NF-17-2-0196. 
Mani Srivastava is supported in part by Air Force Office of Scientific Research (AFOSR) under Cooperative Agreement FA95502210193, DEVCOM ARL under Cooperative Agreement W911NF-17-2-0196, and National Institutes of Health (NIH) mDOT Center under Award 1P41EB028242.
Zhichao Cao and Mi Zhang are supported in part by NSF under award NeTS-2312675.
Icons inside Figure~\ref{fig: Motion Sensing Pipeline},
~\ref{fig:Pipeline of Wi-Fi Sensing},
~\ref{fig:Pipeline of Vision Sensing},
~\ref{fig:Pipeline of Acoustic Sensing},
~\ref{fig:Pipeline of Multi-Modal Sensing},
~\ref{fig:Pipeline of Earable Sensing},
~\ref{fig:adaptation},
~\ref{fig:model-partitioning},
~\ref{fig:overview of healthcare},
~\ref{fig: Components of a Video Streaming pipeline},
~\ref{fig: Continuous Learning for video analytics},
~\ref{fig: Components of Autonomous Driving Pipeline},
~\ref{fig: Overview of AR/VR/MR pipeline} are made by Freepik from 
\href{https://www.flaticon.com} {https://www.flaticon.com}.
The views and conclusions contained in this document are those of the authors and should not be interpreted as representing the official views or policies, either expressed or implied, of the ARL, the Department of Defense or the U.S. Government. The U.S. Government is authorized to reproduce and distribute reprints for Government purposes notwithstanding any copyright notation herein.

\bibliographystyle{ACM-Reference-Format} %
\bibliography{main}

\end{document}